\documentclass[tc, manuscript]{copernicus}

\begin{document}
\nolinenumbers

\newcommand{\btheta}{\boldsymbol{\theta}}
\newcommand{\bpsi}{\boldsymbol{\psi}}
\newcommand{\bx}{\mathbf{x}}
\newcommand{\by}{\mathbf{y}}
\definecolor{myyel}{rgb}{0.8, 0.8, 0}
\definecolor{mypur}{rgb}{0.8, 0, 0.8}

\title{Climate-informed cryospheric reanalysis via hierarchical Bayesian data assimilation}

% Open question: Do we update notation to "mark" the observations as \mathbf{y}^o to differentiate these from the noisy observables \mathbf{y}

% \Author[affil]{given_name}{surname}

\Author[1][kristoffer.aalstad@geo.uio.no]{Kristoffer}{Aalstad} %% correspondence author
\Author[2]{Esteban}{Alonso-González}
\Author[3,4]{Joel}{Fiddes}
\Author[5]{Brian}{Groenke}
\Author[1]{Gregoire}{Guillet}
\Author[1]{Regine}{Hock}
\Author[6]{Bartłomiej}{Luks}
\Author[1]{Norbert}{Pirk}
\Author[1]{Sebastian}{Westermann}
\Author[1]{Yeliz}{A. Yılmaz}

\affil[1]{Department of Geosciences, University of Oslo, Oslo, Norway}
\affil[2]{Instituto Pirenaico de Ecología, Spanish Research Council (IPE-CSIC), Jaca, Spain}
\affil[3]{WSL Institute for Snow and Avalanche Research SLF, Davos, Switzerland}
\affil[4]{Mountain Futures GmbH, Klosters, Switzerland}
\affil[5]{Potsdam Institute for Climate Impact Research (PIK), Potsdam, Germany}
\affil[6]{Institute of Geophysics, Polish Academy of Sciences, Warsaw, Poland}

\runningtitle{Climate-informed cryospheric reanalysis}

\runningauthor{Aalstad et al.}

\received{}
\pubdiscuss{} %% only important for two-stage journals
\revised{}
\accepted{}
\published{}

%% These dates will be inserted by Copernicus Publications during the typesetting process.

\firstpage{1}

\maketitle

% By assimilating observations into models, cryospheric reanalysis can be used to infer the historical trajectory of essential climate variables, such as seasonal snow water equivalent and glacier mass balance. Existing cryospheric reanalyses typically treat each water year independently, with no pooling of information across years

\begin{abstract}
The cryosphere regulates global cycles of water, energy, and carbon, affecting societies and ecosystems. By assimilating observations into models, we can infer the past trajectory of essential climate variables, such as seasonal snow water equivalent and glacier mass balance, generating a cryospheric reanalysis. Existing cryospheric reanalyses typically treat each water year independently, with no pooling of information across years. With sparse and noisy observations, such `climate-assumed' reanalyses perform poorly, reverting to an assumed yet uncalibrated background climatology determined by fixed hyperparameters. Conversely, Bayesian calibration with static parameters completely pools information across water years, inferring climatological parameters without capturing inter-annual variability. Here, we propose a new reanalysis approach using partial pooling through hierarchical Bayesian data assimilation, which we coin `climate-informed' cryospheric reanalysis. This approach jointly infers the historical trajectory of state variables, annual parameters, and local climatological hyperparameters using a nested hybrid particle smoothing workflow. We test the approach through three experiments by assimilating in situ snow water equivalent, satellite-based fractional snow-covered area, and glacier-wide mass balance data into a temperature index model at eight study areas, from the Swiss Alps to the High Arctic. Hierarchical reanalysis with partial pooling outperforms annually independent cryospheric reanalysis (no pooling) and static parameter calibration (complete pooling) for all experiments during calibration and for all but one during validation. Overall, the climate-informed hierarchical reanalysis yields a considerable 51\% improvement in continuous ranked probability score relative to the prior, markedly exceeding that of annual reanalysis (38\%) and static calibration (19\%). Compared to a particle Markov chain Monte Carlo benchmark, our workflow reduces computational cost by several orders of magnitude through recycling and expectation maximization, with comparable or improved skill. The proposed workflow is modular and efficient, providing a foundation for future cryospheric applications of hierarchical Bayesian data assimilation.
\end{abstract}

\introduction  %

Snow, glaciers, and permafrost form the bulk of the terrestrial cryosphere and perform key climate services as refrigerators of the Earth system by regulating global cycles of energy, water, and carbon \citep{Euskirchen2013,Sturm2017}. Ongoing anthropogenic global warming is amplified in cold regions due to several positive feedback mechanisms involving these cryospheric components \citep{Meredith2019,Hock2019}. This amplification and the resulting depletion of the cryosphere have a direct impact on ecosystems, communities, and the broader coupled Earth system \citep{Arias2021}. In recognition of its importance, the cryosphere has been designated multiple Essential Climate Variables (ECVs) by the Global Climate Observing System \citep{GCOS245} that are targeted by major Earth observation efforts such as the Climate Change Initiative (CCI) of the European Space Agency \citep{Plummer2017}. Despite these ongoing efforts, the history, current state, and future fate of the cryosphere remain shrouded in considerable uncertainty, given the persistent grand challenges associated with modeling and observing cryospheric ECVs \citep{Dozier2016,Trofaier2017,Hock2017}. 

On the one hand, the evolution of cryospheric state variables can be simulated using numerical models of varying complexity, ranging from simpler models with a single vertical layer \citep{Hock2003}, intermediate complexity models with a few layers \citep{Essery2025}, to more complex multi-layered models \citep{Westermann2023}. Given sufficient meteorological forcing data and a model at the appropriate level of complexity \citep{Parker2020}, it is possible, with enough computational resources, to estimate state variables of interest at high spatio-temporal resolution. Nonetheless, like all geophysical models, cryospheric models are necessarily simplified representations of open natural systems and are, as such, subject to uncertainty arising from initial and boundary conditions, model structure, and internal parameters \citep{Gunther2019}. In particular, these models are highly sensitive to errors in atmospheric forcing fields, such as precipitation and air temperature, which can be considerable, especially in complex terrain \citep{Liu2022,Fang2023} and cold regions more generally \citep{Tang2023,Bin2025}. As such, atmospheric forcing is a major source of uncertainty when modeling the evolution of seasonal snowpacks \citep{AlonsoGonzalez2022}, the mass balance of glaciers \citep{Rounce2023}, and the thermal state of permafrost \citep{Willmes2025}. Leveraging information from the global observing system is key to constraining these sources of uncertainty and keeping models tethered to reality \citep{Gettelman2022}.

On the other hand, elements of the cryosphere can be partially observable across a range of scales, platforms, and modalities by leveraging the ever growing global climate observing system \citep{GCOS245}. Sparse ground-based observations allow us to monitor local cryospheric state variables such as seasonal snow mass \citep{Mortimer2025} and the mass balance of glaciers \citep{WGMS2025} with relatively high accuracy, but at a necessarily limited level of spatial representativeness and coverage. Airborne observations from crewed aircraft \citep{Painter2016,AlonsoGonzalez2026} and drones \citep{Pirk2022,AlonsoGonzalez2023,vanHove2026} provide a promising vantage point to improve spatial representativeness while maintaining high accuracy. For now, these airborne platforms typically remain limited to a small number of campaigns, providing snapshots in time rather than more continuous monitoring. The wider perspective offered by satellite remote sensing can help provide even better spatial representativeness with potentially global spatial and multi-decadal temporal coverage \citep{Plummer2017}, albeit often with a loss of accuracy due to the further removed space-borne vantage point. Over the last decades, such satellite remote sensing has matured into a vital tool for monitoring the state of the cryosphere \citep{Berthier2023,Gascoin2024}. Long satellite-based climate data records can now be retrieved for essential climate variables such as snow-covered area \citep{Riggs2017,Gascoin2019,Aalstad2020,MODIScci}, land surface temperature \citep{Dupuis2024}, and snow mass \citep{Pulliainen2020}. In addition, satellite remote sensing has enabled the delineation of glacier outlines \citep{RGI7,Maslov2025} and tracking changes in glacier mass \citep{Hugonnet2021,Glambie2025} at a global scale. Moreover, recent work has highlighted the potential utility of emerging cryospheric satellite retrievals such as estimates of snow depth \citep{Dunmire2024,Mazzolini2025} and wet snow lines \citep{Cluzet2024}. At the same time, all observations have in common that they are indirect, uncertain, unrepresentative, and incomplete \citep{Parker2017}. As such, in isolation, the current global climate observing system and existing geophysical models are merely necessary but insufficient tools to monitor essential climate variables in order to meet the latest requirements laid out by \citet{GCOS245}. 

Data assimilation (DA) emerged in the early days of operational numerical weather prediction \citep{Eliassen1954,Sasaki1958} from the need for objective \emph{analysis} (DA jargon for inference) that gets the best of both worlds by fusing observations and models \citep{Kalnay2024,Wikle2026}. Currently, the practice of DA has been extended to most, if not all, fields of Earth system science and can be formalized mathematically through the probabilistic framework of Bayesian inference \citep{Evensen2022}. Through this Bayesian lens, DA can be viewed as solving a dynamical inverse problem where we seek to infer plausible configurations of hidden state variables of interest that may have given rise to the noisy observations under a particular forward model \citep{SanzAlonso2023}. Modern DA methods can constrain uncertain geophysical models with noisy observations and provide gap-free probabilistic estimates of the (often unobserved) variables of interest to help reconstruct the past, nowcast the present, forecast the near future, and even constrain climate projections. Indeed, DA is an engine that helps power both numerical weather predictions and atmospheric reanalyses such as ERA5 \citep{Hersbach2020}, where the latter are perhaps the most used datasets in climate science \citep{Parker2016,Parker2017}. Reanalysis, short for retrospective analysis, involves using data assimilation to infer the historical trajectory of a system, originally the atmosphere, typically over several decades. Such reanalyses have now also become a vital source of training data for emerging machine learning-based weather prediction that rivals the predictive performance of traditional approaches \citep{Price2025,Moldovan2026,Abel2026}. 

% Kris comment to self: Many coauthors suggest cutting most of the ML stuff here.
The synergies between data assimilation \citep{Evensen2022} and probabilistic machine learning \citep{Murphy2023} run deep, with foundational Bayesian methods \citep{Robert2007} forming a natural bridge between the two fields \citep{Geer2021,Cheng2023,Bach2025}. In cryosphere-related applications, machine learning has been used to enhance DA by, for example, enabling efficient spatial information propagation \citep{Guidicelli2024}, rapidly and accurately emulating otherwise costly ensemble-based filtering \citep{Blandini2025}, and emulating models to make sampling more tractable \citep{Rounce2023,Keetz2025}. Similarly, Bayesian DA methods have been used to enhance cryosphere-related machine learning workflows by, for example, enabling uncertainty-aware estimates \citep{Pirk2024} and adaptive data collection to maximize information gain \citep{vanHove2026}. A common challenge in both DA and machine learning, as well as in combinations thereof, is tuning higher-level `hyperparameters' (i.e., parameters controlling parameters) that play a key, albeit uncertain, role  in probabilistic models. The hierarchical Bayesian approach, in which additional levels of inference are introduced, provides a principled way of solving the hyperparameter tuning problem with a promising track record in statistics \citep{Good1980,Robert2007,Gelman2013,McElreath2020}. These hierarchical methods are also well established in probabilistic machine learning \citep{MacKay1992,Neal1996}, through a Bayesian spectrum from maximum likelihood to full hierarchical Bayesian inference \citep{Murphy2023}. 

 A case can be made that, through hierarchical Bayesian models, statisticians have a long history of performing uncertainty-aware `deep modeling' \citep{Wikle2023}. In this view, revolutionary deep neural networks used in machine learning can be seen as scalable and more flexible plugin approximations of the ideal Bayesian modeling paradigm, albeit typically at the cost of little to no uncertainty quantification \citep{Murphy2023}. The dire need for uncertainty quantification via Bayesian inference in machine learning was recently emphasized by \citet{Papamarkou2024} and \citet{Bengio2025}, although scaling Bayesian methods to large neural networks remains a challenge. A similar challenge exists in climate science, where there is always a practical trade-off between uncertainty quantification, through ensemble size, and model complexity, through resolution and process representation. Therein, additional complexity is often prioritized to the detriment of uncertainty quantification \citep{Ferro2012}. 
 
 Recently, hierarchical Bayesian approaches have started to gain traction in the DA and wider state space modeling community \citep{Katzfuss2020,Lucini2021,Drovandi2022,Viani2023,PerezVieites2025}. These methods are still not widely adopted in large-scale DA and machine learning applications, as they can be computationally costly to implement in practice. At the same time, hierarchical methods hold considerable potential to improve model-data fusion across climate science, which remains mostly limited to `shallow' probabilistic modeling \citep{Wikle2026}. The highly data-intensive exercise of generating reanalyses for different parts of the Earth system \citep{Baatz2021} could greatly benefit from these techniques, especially for the sparsely observed cryosphere.

Cryospheric reanalysis involves tailoring DA-powered reanalysis workflows, originating in the numerical weather prediction community \citep[e.g.,][]{Hersbach2020}, to the cryospheric domain. Such \emph{retrospective analysis} (reanalysis) updates model hindcasts with observations using DA to retrospectively infer the historical trajectory of the state of a system \citep{Parker2016}. Our precise use of the technical term reanalysis from DA should not be confused with the more informal use of `reanalysis' (as in `analyze again') to refer to the reprocessing of historical glaciological observations \citep{Holmlund2005,Zemp2013,Andreassen2016}. The exact models, observations, and DA schemes used for reanalysis should generally be tailored to the target system \citep{Baatz2021,Abel2026}. The need for tailoring is evident in the cryospheric domain since global atmospheric reanalyses are highly uncertain in cold regions \citep{Liu2022,Fang2023,Bin2025} and often do not even assimilate existing data in these areas \citep{Orsolini2019}. Although reanalysis data are often criticized as not being observations due to their reliance on models, this is arguably a strength rather than a weakness since models help impose geophysical constraints while allowing for consistent, uncertainty-aware, and gap-free estimates that can be informed by multiple sensors through joint DA \citep{Parker2016,Parker2017}. Moreover, what are sometimes seen as superior `pure observational' alternatives, such as gridded observation products \citep{Lussana2018,Glambie2025}, also implicitly involve statistical models. Thus, such products are arguably just reanalyses without any physical constraints or the ability to leverage multimodal data \citep{Wikle2026}. On closer inspection, a superficially clear distinction between models and observations is usually much blurrier than it seems \citep{Parker2017}, with the rapidly evolving field of DA helping to blend the two in a statistically optimal manner via (re)analysis \citep{Evensen2022}. 

The practice of cryospheric reanalysis likely began in snow science with \citet{Martinec1981} and \citet{Cline1998}, who directly inserted fractional snow-covered area (FSCA) retrievals into snowmelt models to deterministically reconstruct peak snow mass. Accurately estimating snow mass, often recast as snow water equivalent (SWE) depth, is both crucial and a major challenge in snow science  \citep{Dozier2016,Gascoin2024}. On the one hand, modern variations \citep{Bair2023,Avanzi2023} on the deterministic method of \citet{Martinec1981} are seen as a promising step towards solving this challenge \citep{Dozier2016}. On the other hand, these methods are plagued by unquantified uncertainty in forcing data, snow models, and satellite retrievals \citep{Slater2013}. Probabilistic (i.e., Bayesian) snow reanalysis, starting with the seminal work of \citet{Kolberg2006} and \citet{Durand2008}, can be seen as uncertainty-aware generalizations of the deterministic approach \citep{Girotto2014}. Not only is the probabilistic approach uncertainty-aware, which can be key to generating a consistent reanalysis \citep{Parker2016}, but it also generally outperforms deterministic approaches in the few intercomparisons that exist \citep{Girotto2014,Yang2023}. Henceforth, by snow reanalysis and, more generally, cryospheric reanalysis, we implicitly refer to such a probabilistic approach.

Starting with \citet{Durand2008}, most snow reanalyses have relied on the use of efficient approximate Bayesian techniques from ensemble-based DA rather than more computationally intensive Markov Chain Monte Carlo \citep[MCMC;][]{Kolberg2006} sampling methods. Using real satellite data, this approach was first applied at the basin \citep{Girotto2014b} and mountain range scale \citep{Margulis2016} in the California Sierra Nevada. It has since been extended to parts of the Andes \citep{Cortes2017}, sites in the High Arctic \citep{Aalstad2018}, parts of the Swiss Alps \citep{Fiddes2019}, across the Lebanese mountains \citep{AlonsoGonzalez2021}, all of High Mountain Asia \citep{Liu2021}, the entire western US \citep{Fang2022}, and selected basins in western Canada \citep{Sun2025}. To begin with, these experiments were mostly carried out assimilating high resolution (30 m), albeit temporally scarce (ca. fortnightly) FSCA from Landsat \citep{Margulis2016}, but have since been extended to high resolution (20 m), more frequent (ca. weekly) Sentinel-2 data \citep{Aalstad2018}, coarser (500 m) but frequent (daily) MODIS data \citep{Fiddes2019}, and combinations thereof \citep{Fang2022}. A key aspect of snow reanalysis is the use of Bayesian smoothing schemes that can exploit the dynamic information in FSCA data by allowing observations in the ablation season to update the antecedent accumulation season \citep{Margulis2015,AlonsoGonzalez2022}. More simply stated, smoothers have the benefit of hindsight, which is key to successful reanalysis. This sets snow reanalysis problems apart from other snow DA problems \citep{Largeron2020} such as forward looking filtering in snow hydrological forecasting \citep{Blandini2025}. 

Recent snow reanalysis studies have explored the assimilation of remotely sensed snow depth data retrieved from crewed aircraft \citep{Margulis2019,AlonsoGonzalez2026}, drones \citep{AlonsoGonzalez2023}, space-borne radar \citep{Girotto2024}, and satellite-based laser altimetry \citep{Mazzolini2025}. Therein, snow depth assimilation was mostly performed independently without FSCA \citep{Margulis2019,AlonsoGonzalez2023,Girotto2024,AlonsoGonzalez2026}, but also jointly to show the added value of both modalities \citep{Mazzolini2025}, either using purely `1D' temporal DA \citep{Margulis2019,Girotto2024} as well as `4D' spatio-temporal DA \citep{AlonsoGonzalez2023,Mazzolini2025,AlonsoGonzalez2026} to handle spatial data gaps. Other satellite retrievals, such as wet snow  lines \citep{Cluzet2024} and skin temperature \citep{AlonsoGonzalez2023thermal}, remain largely untapped in snow reanalysis. The recent development of open source tools \citep{AlonsoGonzalez2022} helps facilitate efforts by the snow science community to make advances in methods, models, and assimilating emerging data streams.

Beyond snow, data assimilation methods are increasingly gaining traction across the broader terrestrial cryosphere. For example, \citet{Willmes2025} assimilated FSCA retrievals from Sentinel-2 into the complex terrestrial cryospheric model CryoGrid  \citep{Westermann2023} to constrain the ground thermal regime at a High Arctic permafrost site. \citet{Groenke2023, Groenke2024} applied ensemble-based Bayesian inverse modeling to infer the thermal evolution of permafrost over both decadal and centennial time scales using a variant of CryoGrid. \citet{Cao2025} tested the performance of ensemble-based assimilation of albedo and snow depth in CryoGrid to constrain glacier surface mass balance in large ensemble synthetic experiments. In an earlier reanalysis exercise, \citet{Navari2021} showed how albedo assimilation substantially improved the performance of surface mass balance simulations using the Crocus model along a transect on the Greenland Ice Sheet. A variational Bayesian approach was used to great effect by \citet{Brinkerhoff2025} to dynamically constrain an ice flow model and perform probabilistic projections of the fate of a large glacier.

A related line of work has used costlier Markov Chain Monte Carlo (MCMC) sampling for Bayesian calibration of static parameters in glacier mass balance models both regionally \citep{Rounce2020,Sjursen2025} and worldwide \citep{Rounce2023}. Applications of Bayesian inference to calibrate static parameters have also been explored when modeling glacier flow \citep{Berliner2008,Gopalan2018}. Such temporally global Bayesian model calibration with static parameters does not strictly constitute a reanalysis \citep{Parker2016,Abel2026}, but it provides a promising alternative problem formulation. This global approach to inference is called \emph{complete pooling} since all available data are pooled to update a set of static parameters \citep{Gelman2013,McElreath2020}. In contrast to complete pooling, the aforementioned cryospheric reanalysis examples \citep{Margulis2016,Navari2021} tend to use independent annual DA windows \citep{AlonsoGonzalez2022} across which annual (i.e., water year-level) parameters can vary.  This local approach to inference is called \emph{no pooling} since no information is pooled across DA windows when inferring annual parameters. Unlike complete pooling, inference via no pooling enables tracking of annual variations in parameters that reflect the dynamics of seasonal errors in the meteorological forcing and the process model. The primary limitation of no pooling is the absence of information sharing across windows, leading to especially poorly constrained parameters in years with little to no data.
On the one hand, the local no pooling method cannot, as the saying goes, see the forest for the trees. On the other hand, the global complete pooling method can only see the forest. 

Here, we seek to combine these approaches to see both the forest and the trees. Translating this analogy to our problem, the forest is the parameter climatology (i.e., what we expect), while the trees are the annual parameters reflecting seasonal realizations of meteorological forcing and model error (i.e., what we got). Throughout, we use the term climate to denote the statistical distribution of weather over a long but finite period \citep{Werndl2016,IPCC2021}. The ERA5 data used to force our models provide coarse estimates of historical weather, whereas the climate of interest here is the distribution over local weather realizations, which our hyperparameters help encode. Past studies have demonstrated the value of tapping into temporal patterns in snow reanalysis to learn parameters by transferring information across water years \citep{Kolberg2010,Fiddes2019,vonKaenel2025} using heuristic methods that we aim to formalize via hierarchical Bayesian DA. The apparently simple annual parameter filtering solution of letting the posterior parameters inform the prior for the next year requires careful tuning of the memory and noise hyperparameters in the parameter dynamics. Instead, here we propose a conceptually simple \emph{partial pooling} hierarchical Bayesian method where the local parameters in a given year are informed by both local data and the entire climatology via hyperparameters that are updated globally. The goal is for this approach to automatically identify the right compromise between no pooling and complete pooling for a given water year, resulting in a \emph{climate-informed} reanalysis. As such, the parameters will be pulled towards an inferred (rather than assumed) parameter climatology, especially in sparsely observed years, a phenomenon known as \emph{shrinkage} \citep{Robert2007,Gelman2013}. This partial pooling approach can also be extended to more complex formulations, including a hierarchical filtering variant where memory hyperparameters are also inferred. % blending timescales, analogous to weather and climate distinction

By introducing a climate-informed hierarchical cryospheric reanalysis framework to target both seasonal SWE and glacier mass balance, this study seeks to:
\begin{enumerate}
    \item Present the theory behind climate-informed cryospheric reanalysis via partial pooling.
    \item Provide a tractable workflow for hierarchical Bayesian DA based on hybrid nested particle smoothers. 
    \item Evaluate the added value of partial pooling compared to no pooling and complete pooling for four pairs of glacier and snow sites in distinct climatic regions from Svalbard to the Swiss Alps.
    \item Benchmark the performance of the proposed workflow using gold-standard particle MCMC. 
\end{enumerate}
To achieve this, we first present a detailed methodology (Section~\ref{sec:method}) to address objectives 1 and 2. We then present results (Section~\ref{sec:results}) from three experiments with a temperature index model assimilating in situ SWE data, satellite-based FSCA retrievals, and glacier-wide mass balance estimates across multiple sites. These results are used to demonstrate the benefits and benchmark the performance of the hierarchical DA workflow to meet objectives 3 and 4, respectively. In the discussion (Section~\ref{sec:discussion}), we contextualize the results while outlining remaining challenges and prospects for future research before concluding.

\section{Experimental design}
\label{sec:data}

\subsection{Study areas}

\begin{figure}[ht]
\includegraphics[width=\textwidth]{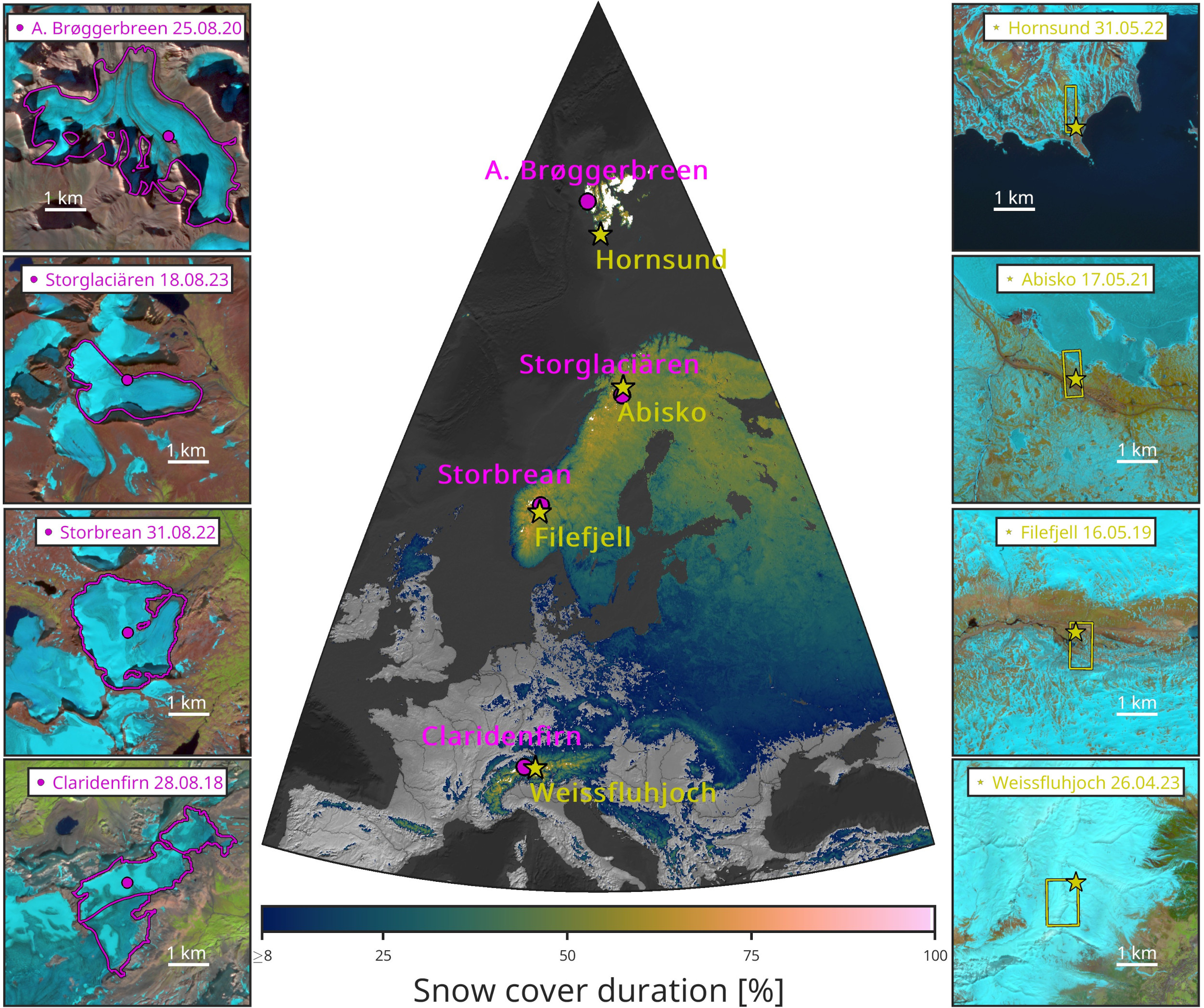}
\caption{Overview of the eight study areas. \textbf{Middle}: Climatological mean snow cover duration (in $\%$ of a water year) estimated from the Snow\_cci MODIS SCFG product \citep{MODIScci} at $0.01^\circ$ resolution showing pixels with $\geq8 \%$ (30 days) of mean snow cover duration across Europe north of $40^\circ\mathrm{N}$ along with the location of the four glacier (purple circles) and four seasonal snow (yellow stars) study areas in Tables~\ref{tab:snow} and~\ref{tab:glaciers} draped over a basemap made with Natural Earth. \textbf{Left}: Latitudinally (north-south) ordered Sentinel-2 $10$ m resolution false color images (where snow and ice are cyan) of the four glacier sites with outlines (purple line) and center coordinates (purple circle) defined by \citet{RGI7}. \textbf{Right}: Analogous false color images, but now for each of the four snow sites (yellow stars) with the encompassing Snow\_cci MODIS pixel (yellow rectangles).}
\label{fig:area}
\end{figure}

The study areas span roughly $30^\circ$ latitude from the Swiss Alps to Svalbard (Tables~\ref{tab:snow}-\ref{tab:glaciers}, Figure~\ref{fig:area}). These study areas were selected as they have some of the longest in situ records on snow mass and glacier mass balance, covering all or most of the ERA5 reanalysis \citep{Hersbach2020,Hersbach2023} period from 1940 to the present day. Unless stated otherwise, all mean quantities are computed based on the in situ glacier and snow site data for the current climate normal period 1991-2020.

\begin{table}[ht]
\caption{Overview of the four seasonal snow sites considered in this study where $\Delta \tau$ is the original typical sampling interval and $D_p$ denotes the climatological mean annual peak SWE (1991-2020).}  \label{tab:snow}
\centering
\begin{tabular}{l c c c c c c c}
\hline
Name   & Region & (Lat$[^\circ\mathrm{N}]$,Lon$[^\circ\mathrm{E}]$) & $z$  $ [\mathrm{m}\,\mathrm{asl}]$ & Water years & $\Delta \tau$ [days] & $D_p$ [mm] \\
 \hline
   Hornsund  & Svalbard & $(77.0010,15.5397)$ & $8$ & 1983-2024 & 5  & $132$ \\
   Abisko   & N. Sweden & $(68.3538,18.8166)$ & $393$ &  1955-2022 & 1  & $214$\\
   Filefjell & S. Norway & $(61.1779,8.1125)$ & $958$ &  1968-2022 & 1  & $402$\\
   Weissfluhjoch & Swiss Alps & $(46.8294,9.8093)$ & $2540$ & 1941-2024 & 14 & $883$ \\
 \hline
 \end{tabular}
 \end{table}

\subsubsection{Snow sites}\label{sec:snow}

The four snow sites were selected based on their exceptionally long in situ measurement records, as listed in the water years column of Table~\ref{tab:snow}. Although the SWE measurement techniques vary somewhat across the sites, for simplicity, we adopt a fixed moderate in situ SWE observation error standard deviation of $\sigma_D=0.01$ m used for data assimilation in the observation error covariance matrix.

Hornsund is the northernmost site located close to sea level near the Polish Polar Station Hornsund in SW Spitsbergen, Svalbard. The measurements consist of manual SWE surveys \citep{Luks2024}. The time series is fairly complete in terms of coverage, although there are a few gaps, including the entirety of the water year 1985. To the best of our knowledge, it provides the longest existing and still ongoing data record of in situ SWE measurements in the High Arctic. The Abisko site is located near the Abisko Scientific Research Station in northern Sweden. Snow depth measurements go back to 1913 \citep{Kohler2006}, but we use snow depth measurements from 1955 onward, where the coverage is more consistent. The data are distributed by the Swedish Meteorological and Hydrological Institute (SMHI) as station number 188800 and are available up to 2022. We convert the data into SWE using the empirical density model of \citet{Sturm2010} for the tundra snow class. 

The Filefjell site is located in the central mountains of southern Norway and is operated by the Norwegian Water and Energy Resources Directorate (NVE). This is a low alpine open site situated around the treeline in a valley. A snow pillow has been installed to measure SWE since 1967, and since the turn of the century, this has been supplemented by more modern snow scales \citep{Stranden2024}. The available SWE data span from water year 1968 to 2022, with a few gaps except for $4$ sequences of water years that are missing entirely, namely: 1977-1979, 1990-1991, 1993, and 2001-2004. The Weissfluhjoch site is operated by the WSL Institute for Snow and Avalanche Research (SLF) above Davos in the eastern Swiss Alps. There, manual SWE measurements have been conducted operationally since 1936 \citep{Marty2012}. We make use of these SWE measurements for the entire ERA5 reanalysis period from water year 1941 until 2024, with few gaps.

\begin{table}[ht]
\caption{Overview of the four glaciers considered in this study. The center coordinates and median (minimum,maximum) elevation $z$ for each glacier were obtained from RGI7 and $\overline{B}_a$ denotes the climatological mean annual mass balance (1991-2020).} \label{tab:glaciers}
\centering
\begin{tabular}{l c c c c c c c}
\hline
Name & Region & (Lat$[^\circ\mathrm{N}]$,Lon$[^\circ\mathrm{E}]$) & $z$  $ [\mathrm{m}\,\mathrm{asl}]$ & Water years & Area [km$^2$]  &  $\overline{B}_a$ [m]  \\
 \hline
    Austre Brøggerbreen  &  Svalbard & $(78.8876,11.8661 )$ & $275(61,702)$ & 1967-2024 & $9.81$ &  $-0.62$ \\
    Storglaciären &  N. Sweden &  $(67.9038,18.5615)$ & $1428(1178,1933)$ & 1946-2024  & $3.41$ &  $-0.24$ \\
    Storbrean &  S. Norway & $(61.5741,8.1410)$ & $1764(1401,2072)$ &  1949-2023 &  $5.21$ &  $-0.62$ \\
    Claridenfirn &  Swiss Alps & $(46.8436,8.8897)$ &$2880(2452,3215)$ &  1941-2024 & $5.18$ &  $-0.78$\\
 \hline
 \end{tabular} 
 \end{table}

\subsubsection{Glaciers}\label{sec:glacier}
 The glaciers (Table~\ref{tab:glaciers}) were selected not only in light of their relative proximity to the snow sites but also due to their long data records of glacier-wide mass balance that are available from the World Glacier Monitoring Service \citep{WGMS2025}.  The glaciers range in area from $3.41$ (Storglaciären) to $9.81$ km$^2$ (Austre Brøggerbreen) and in median elevation from 275 (Austre Brøggerbreen) to 2880 m a.s.l. (Claridenfirn) \citep{RGI7}. Specifically, we use glacier-wide winter and annual (sum of winter and summer) mass balance data derived from in situ stake measurements and/or geodetic methods over continuous periods of 58 (Austre Brøggerbreen) to 84 years (Claridenfirn). For simplicity and without loss of generality, we assume somewhat optimistic observation error standard deviations of $\sigma_W=0.2$ and $\sigma_A=0.4$ m w.e. for winter and annual balance, respectively, based loosely on \citet{Zemp2013}.

Austre Brøggerbreen, located on the Brøgger peninsula in north-western Svalbard, has Svalbard's longest continuous mass balance record starting in 1967 and running to the present day \citep{Hagen2003}. For years with missing winter data in \citet{WGMS2025}, we use estimates provided directly by the Norwegian Polar Institute \citep{NPI2023}. Storglaciären is situated just within the Arctic circle in northern Sweden, in the Kebnekaise massif. The mass balance record at Storglaciären began in 1946 and continues to this day, making this the longest continuous glacier-wide record in the world \citep{Holmlund2005}. Storbrean, previously known as Storbreen, located in the Jotunheimen mountain range in southern Norway, has the longest continuous glacier-wide mass-balance record in Norway \citep{Andreassen2016}.

Claridenfirn is located in the central Swiss Alps. Annual surveys began in 1914 and continue until present day. As such, Claridenfirn provides the longest record of uninterrupted glacier mass balance in the world \citep{Huss2021}. Unlike Storglaciären \citep{Holmlund2005}, the Claridenfirn historical record is mostly based on just two stakes \citep{Huss2021} and was thus not designed to systematically estimate glacier-wide mass balance. Nonetheless, with appropriate post-processing \citep{Zemp2013}, such data can help provide a unique long-term time series of glacier-wide mass balance estimates in the Alps. Herein, we employ the entire record of glacier-wide mass balance estimates provided in \citet{WGMS2025} from 1941 corresponding to the ERA5 reanalysis period.

\subsection{Fractional Snow-Covered Area (FSCA) retrievals}\label{sec:fsca}

To demonstrate the wider applicability of hierarchical cryospheric reanalysis, we also perform experiments assimilating globally available satellite data at the snow sites. For this, we adopt version 4 of the daily global canopy-corrected `snow on ground' Terra MODIS FSCA data from the ESA Snow\_cci \citep{MODIScci}, henceforth MODIS Snow\_cci. These MODIS-based data are available at a $0.01^\circ$ ($\simeq1$ km) spatial and daily (light permitting) resolution for water years 2000-2023 for all land areas except Antarctica. 

The FSCA retrieval algorithm in the MODIS Snow\_cci data is based on a version of the SCAmod method detailed in \citet{Metsamaki2015}. The retrievals are obtained from MOD021KM 1 km level 1B radiances \citep{MOD02} and therefore do not leverage the full resolution of the MODIS sensor \citep{Sun2025}. The MODIS Snow\_cci product includes an uncertainty layer, but despite this laudable effort to quantify uncertainty, we ignore this layer as we expect it to underestimate the actual error in the product \citep{Sun2025}. Instead, we assume a fixed observation error standard deviation of $\sigma_F=0.15$ for the MODIS Snow\_cci data based on previous studies evaluating \citep{Aalstad2020} and assimilating \citep{AlonsoGonzalez2021,Sun2025} related data. Although we did not explicitly model the effects of view angle on observation error, we omitted highly off-nadir retrievals with view zenith angles above $60^\circ$.

\subsection{Seasonal snow and glacier model}\label{sec:forward}

As our forward model, we employ a parsimonious degree-day model \citep{Hock2003} at a daily resolution. The model requires only daily mean near-surface air temperature $T$ [K] and daily precipitation $P$ [mm day$^{-1}$], along with the specification of $3$ uncertain parameters: the degree day factor $a$ [mm K$^{-1}$ day$^{-1}$], air temperature bias $b$ [K], and a snowfall scaling $c$ [-]. Such temperature index models continue to be employed to great effect both at the regional \citep{Lussana2018,Rounce2020,Avanzi2023} and global scales \citep{Elias2024,Rounce2023} to simulate the state of seasonal snow and glaciers. The relative ease of running these in ensemble mode, as well as their applicability at large scales and high resolutions, means that they are well suited as kernels for developing cryospheric DA methods \citep{Aalstad2026,Guillet2026,Yang2026}.

The main state variable in the model $D_k$ is the snow and ice water equivalent depth, with SI base units of meters water equivalent (m w.e.), where $k$ indexes days within a given water year. In snow science, this is referred to as snow water equivalent (SWE), whereas in glaciology, it is referred to as cumulative glacier mass balance. In contrast to seasonal snow sites, for glaciers, $D_k$ can be negative since the mass loss of firn and ice continues once seasonal snow has melted. This state variable evolves in discrete time according to the following finite difference equation
\begin{equation}
    D_{k+1}=\max\left(D_k+\dot{C}_{k+1}\Delta t -\dot{A}_{k+1}\Delta t,D_\mathrm{min}\right)  \, , \label{eq:TIM}
\end{equation}
where $D_\mathrm{min}$ is the minimum water equivalent, $\Delta t=1$ [day] is the daily time step, $\dot{A}_{k+1}$ [mm day$^{-1}$] denotes the ablation rate for day $k+1$ which is parametrized through
\begin{equation*}
    \dot{A}_{k+1} = a\max(T_{k+1}-T_m,0) \, , 
\end{equation*}
where the degree day factor $a$ [mm K$^{-1}$ day$^{-1}$] is an uncertain parameter, $T_m=273.15$ [K] is the freezing point of water, and $T_{k+1}$ is a bias corrected daily mean near surface air temperature given by
\begin{equation*}
    T_{k+1}=\widehat{T}_{k+1}-b \, ,
\end{equation*}
where $\widehat{T}_{k+1}$ [K] denotes the topographically downscaled daily mean near surface air temperature with a bias given by an uncertain parameter $b$ [K]. The daily accumulation rate $\dot{C}_{k+1}$ in \eqref{eq:TIM} is parametrized by 
\begin{equation*}
    \dot{C}_{k+1}=c s_{k+1}\dot{P}_{k+1} \, ,
\end{equation*}
where the snowfall scaling $c$ [-] is an uncertain parameter, $\dot{P}_{k+1}$ is the topographically downscaled daily total precipitation [mm day$^{-1}$], and $s_{k+1}$ is the snowfall fraction based on the linear transition
\begin{equation*}
    s_{k+1}= \min\left(\max\left(\frac{T_r-T_{k+1}}{T_r-T_s},0\right),1\right) \, ,
\end{equation*}
where $T_r=276.15$ [K] and $T_s=272.15$ [K] are assumed threshold air temperatures for pure rainfall and snowfall, respectively \citep{Aalstad2018}. The initial condition for each water year is defined as $D_0=0$ [mm] so the model computes the mass gain or loss (in water equivalent) relative to the start of the year. For seasonal snow $D_\mathrm{min}=0$, whereas for a glacier $-D_\mathrm{min}$ is proportional to the mass per unit area of the entire glacier, which, given our focus on the past, is assumed to be infinite here without loss of generality.

For glaciers, we use this temperature index model in a lumped mode for simplicity by approximating the glacier-wide climatic mass balance state variable through a single point simulation corresponding to the median glacier elevation. As such, glacier-wide mass balance observations can be assimilated directly as they correspond to a noisy version of the implicitly glacier-averaged mass balance state predictions from this lumped model. Moreover, the degree-day factor is increased by a factor of $1/0.7$ when the surface water equivalent is negative to parametrize the enhanced melt sensitivity of exposed ice \citep{Hock2003,Rounce2020}. 

For the experiments assimilating SWE data from the respective seasonal snow sites, we run the temperature index model implicitly at an idealized site scale corresponding to the spatial extent (e.g., the snow scale for Filefjell) of the SWE measurements at each site. As such, the observed SWE can also be assimilated directly as it corresponds to a noisy version of the state of the model. 

To enable seasonal snow experiments assimilating the $0.01^\circ$ MODIS Snow\_cci FSCA data, the forward model needs to predict FSCA. In line with previous snow reanalysis experiments assimilating FSCA \citep[e.g.,][]{Margulis2015,Aalstad2018,AlonsoGonzalez2021}, we use the probabilistic snow depletion curve of \citet{Liston2004} as our FSCA parametrization. This approach assumes that the subgrid peak SWE distribution is lognormal and that the ablation rate is uniform across the grid cell in question. In this way, both the FSCA and grid-cell mean SWE can be obtained analytically as a function of the accumulated melt depth, the peak grid-cell mean SWE, and the coefficient of variation parameter $v$ (standard deviation over mean). The first two variables can be obtained directly from the temperature index model, whereas the latter coefficient $v\geq0$ becomes an additional uncertain parameter that encodes the right-skewness, with higher $v$ indicating a more skewed peak subgrid snow distribution. 

\subsection{Meteorological forcing}

To force our experiments, we use the ERA5 atmospheric reanalysis product \citep{Hersbach2020}. This is the fifth generation global atmospheric reanalysis from the European Centre for Medium Range Weather Forecasts (ECMWF) available at hourly temporal and $0.25^\circ$ spatial resolution available from 1940 until present day in near real time. Specifically, we use the 2 m air temperature and total precipitation fields from the surface level available through \citet{Hersbach2023}. Despite the relatively coarse spatial resolution, ERA5 represents the current state-of-the-art for global atmospheric reanalysis in terms of resolution and skill, making it widely used across various applications \citep{AlonsoGonzalez2021,Fiddes2022,Pirk2024,Price2025}. 

To adjust the coarse $0.25^\circ$ ERA5 atmospheric forcing to the scale of our forward model, which is essentially the point scale, we apply established topographic downscaling routines \citep{Fiddes2022,Filhol2023}. In particular, for each study area, we apply inverse distance weighting with a cosine correction to longitude to horizontally interpolate the ERA5 surface fields from the nearest $4$ ERA5 centroids to the latitude-longitude coordinates given in Tables~\ref{tab:snow} and~\ref{tab:glaciers}. For the near-surface air temperature, we also apply a lapse rate correction to adjust from the interpolated elevation of the ERA5 centroids to the elevation of each study area, also given in the aforementioned tables, by assuming a standard lapse rate of $-6.9^\circ$C per km following \citet{Liston2006}.

\subsection{Experiments}\label{sec:experiments}

To test hierarchical cryospheric reanalysis based on partial pooling (PP), we carry out two sets of experiments at the snow sites and one set for the glaciers. For each set, we employ different approaches for inferring the uncertain parameters $\btheta$ in the reanalysis problem to benchmark the performance of the proposed hierarchical approach. The uncertain parameters defining the vector $\btheta$ are the degree-day factor $a$, temperature bias $b$, snowfall scaling factor $c$, and the subgrid coefficient of variation $v$, where the latter is only used in the snow depletion curve for seasonal snow sites.

 \begin{figure}[ht]
 \centering
\includegraphics[width=0.45\textwidth]{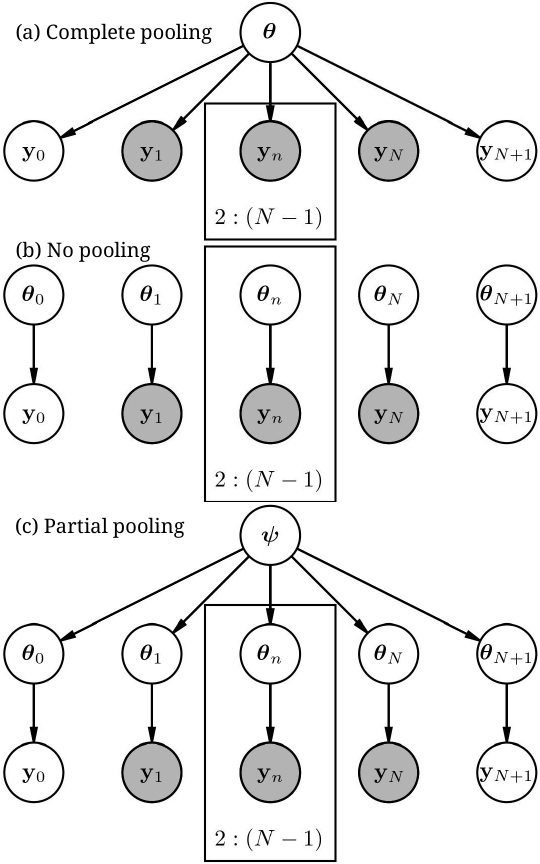}
\caption{Graphical models showing three formulations of a cryospheric reanalysis: (a) complete pooling (CP) where global static parameters $\btheta$ are inferred using observations $\by_n$ (gray nodes) across all observed water years indexed by $n=1,\dots,N$; (b) no pooling (NP) with local parameters $\btheta_n$ that vary across water years and are inferred independently using observations within a given water year $\by_n$; (c) partial pooling (PP) using climatological hyperparameters $\bpsi$ at the highest level to allow pooling information across water years via hierarchical inference while retaining local parameters $\btheta_n$. The structure of each graph, particularly the directed edges (arrows) between the nodes (circles), which can be either observed (shaded) or latent (unshaded) variables, encodes the entire dependence structure of each probabilistic model formulation \citep{Murphy2023}. The unobserved nodes $\by_0$ and $\by_{N+1}$ show that, unlike no pooling, complete (via $\btheta$) and partial (via $\bpsi$) pooling enable calibrated predictions in unobserved years.}
\label{fig:pgm}
\end{figure}

The first approach to reanalysis we test in each set of experiments is complete pooling (CP) where we infer `global' static parameters $\btheta$ using observations $\by_n$ across all observed years $n=1,\dots,N$ by targeting the `global' posterior $p(\btheta\mid\by_{1:N})$. The second approach is no pooling (NP), where we infer `local' water year parameters $\btheta_n$ independently for each water year using all observations within the given water year $\by_n$  by targeting the `local' posterior $p(\btheta_n\mid\by_{n})$. These approaches serve as useful benchmarks not only because they are widely used for cryospheric reanalysis \citep{Margulis2016} and calibration \citep{Rounce2023}, but also because they are on opposite ends of a temporal spectrum that partial pooling aims to straddle. The third approach is partial pooling (PP) where we also infer local water year parameters $\btheta_n$ while leveraging information from all available observations $\by_{1:N}$ via climatological hyperparameters $\bpsi$ by targeting a climatologically calibrated local posterior $p(\btheta_n\mid\by_{1:N})$ obtained via marginalization as described in Section~\ref{sec:dynamics}. 

The three pooling approaches are shown schematically in Figure~\ref{fig:pgm}. In addition to reporting the posterior predictive performance of these three approaches, we also show their prior predictive performance. The prior predictive distribution will generally differ between the hierarchical (PP) and non-hierarchical (CP, NP) approaches since the hierarchical prior considers uncertainty in the hyperparameters $\bpsi$. For both non-hierarchical approaches, we use the same default prior by fixing the typically implicit hyperparameters $\bpsi=\bpsi_0$ to the hyperprior mean given in Table~\ref{tab:prior} as a best guess.

We implement partial pooling via a hybrid nested particle smoothing method detailed in Section~\ref{sec:IES} through to Section~\ref{sec:AMIS}. This method was motivated by considerable gains in computational efficiency that may enable tractability at scale. To help evaluate the trade offs between inferential performance and computational efficiency, we test two other methods for the partial pooling approach. The first is type-II maximum a posteriori (MAP) using an `optimal' point estimate for the hyperparameters $\bpsi$. This skips half of the workflow in the nested particle method, increasing efficiency at the cost of performance as the hyperposterior $p(\bpsi\mid\by_{1:N})$ is replaced by a plug-in approximation \citep{Murphy2023}. The second method is PMCMC (Appendix~\ref{app:PMCMC}), which serves as a costly gold-standard benchmark \citep{Law2012}. We refer to these additional partial pooling benchmarking approaches with the acronyms MAP and PMCMC, respectively, reserving the acronym PP for the hybrid nested particle smoothing method. All the partial pooling approaches have the same prior, so we only report the prior performance for PP. Finally, all approaches from CP to PMCMC use the same hybrid data assimilation scheme, abbreviated MAGPIES (Section~\ref{sec:MAGPIES}), to infer the conditional parameter posterior $p(\btheta\mid\by,\bpsi)$.

The five approaches outlined above (CP, NP, PP, MAP, and PMCMC) are used across three sets of experiments assimilating SWE, FSCA, and glacier-wide mass balance data, respectively. Our SWE experiment conducts seasonal snow reanalyses by assimilating in situ SWE data for the four snow sites. To make the results more relatable to what can be expected in terms of SWE retrievals from upcoming satellite missions such as NISAR, we assimilate SWE data at most every $10$ days while using all the higher frequency data for evaluation (Table~\ref{tab:snow}). We only assimilate SWE data prior to water year $2000$, treating the 20$^\text{th}$ century as the calibration period. All the available SWE data from water year $2000$ onward are withheld from the assimilation as independent validation data, treating the $21^\text{st}$ century as the validation period. While this experiment can help demonstrate the potential of hierarchical snow reanalysis, it provides an upper bound on typical reanalysis performance since high quality SWE data are scarce and mostly in situ. There are currently no gridded global high resolution SWE data  \citep{Pulliainen2020} for global snow reanalysis, but this may change with InSAR data assimilation \citep{Margulis2026}. 

Our FSCA experiment conducts seasonal snow reanalyses by assimilating MODIS FSCA data for the four snow sites. We assimilate all the MODIS FSCA data, treating the available Snow\_cci MODIS-era $2000-2023$ as the calibration period for this experiment. No MODIS data are available for water years prior to $2000$, so we treat the $20^\text{th}$ century and $2024$ as the validation period. Performance is evaluated using independent in situ SWE data in both the validation and calibration periods. Since the FSCA reanalyses are not constrained by any data from pre-2000, this experiment allows us to test the added value offered by the hierarchical partial pooling approach for transferring information from the MODIS satellite sensor into the pre-MODIS era. If successful, this helps pave the way toward extending satellite-based cryospheric reanalyses into the pre-satellite era. Moreover, unlike SWE, FSCA data assimilation has the potential to be scaled up to global mountain snow reanalysis. 

Our glacier experiment conducts glacier reanalyses by assimilating glacier-wide mass balance data for the four glaciers. We assimilate all the available glacier-wide mass balance data for each glacier prior to the balance year $2000$, treating the 20$^\text{th}$ century as the calibration period. All the mass balance data from water year $2000$ onward are withheld from the assimilation, so we treat the $21^\text{st}$ century as the validation period. 

\begin{table}[ht]
\caption{Overview of the three sets of experiments distinguished by their name, assimilation data, evaluation data, calibration period, and validation period. The calibration and validation periods are given by their maximal extent across all sites and may start later or end earlier at a given site depending on data availability. The glacier experiment is run for four glaciers (Table~\ref{tab:glaciers}), while the SWE and FSCA experiments are run for four seasonal snow sites (Table~\ref{tab:snow}). For each experiment, we run three approaches to cryospheric reanalysis (CP, NP, PP) with two additional hierarchical benchmarking approaches (MAP, PMCMC). $^\dagger$Winter and annual glacier-wide mass balance data.} \label{tab:experiments}
\centering
\begin{tabular}{l c c c c}
\hline
Name & Assimilation & Evaluation & Calibration & Validation   \\
 \hline
 SWE & In situ SWE & In situ SWE & 1941-99 & 2000-24  \\
 FSCA & MODIS FSCA & In situ SWE & 2000-23 & 1941-99, 2024 \\
 Glacier & Mass balance$^\dagger$  & Mass balance$^\dagger$ & 1941-99 & 2000-24 \\
 \hline
 \end{tabular} 
 \end{table}

\subsection{Evaluation}

To evaluate the performance of the respective approaches to cryospheric reanalysis across the three experiments in Table~\ref{tab:experiments}, we rely on the Continuous Ranked Probability Score \citep[CRPS;][]{Gneiting2005}. The CRPS is used for probabilistic verification to evaluate the performance of the entire approximate posterior distribution rather than just a point estimate. For the reference `ground truth', we use the in situ SWE data for the snow sites and the glacier-wide mass balance for the glaciers. We compute CRPS separately for the calibration (train) and validation (test) periods to help gauge the performance of the reanalyses in both observed and unobserved periods. 

Following \citet{Hersbach2000}, the CRPS is defined as 
\begin{equation}
    \mathrm{CRPS}\left(P,x^\star \right)=\int \left[P(x)-H(x-x^\star)\right]^2 \, \mathrm{d}x \, , \label{eq:CRPS}
\end{equation}
where $P(x)=\int_{-\infty}^{x} p(x')\, \mathrm{d}x' $ is the cumulative probability distribution of the uncertain variable or parameter $x$ of interest with (prior or posterior) density $p(x)$, $x^\star$ is the reference truth value, and $H(x-x^\star)$ is the Heaviside function which is $1$ if $x\geq x^\star$ and $0$ otherwise. The CRPS is a non-negative score with an optimal value of $0$ indicating perfect agreement with the reference. It reduces to absolute error in the special case of degenerate (deterministic) predictions. In the more general non-degenerate case, CRPS measures both the accuracy (central goodness of fit) and the precision (sharpness) of the distribution $p(x)$ relative to the reference $x^\star$, making it a natural choice for evaluating cryospheric data assimilation experiments \citep[e.g.,][]{AlonsoGonzalez2023,Cao2025}. In practice, in line with these previous studies, we estimate the CRPS by assuming that the relevant marginal predictive distributions obtained via ensemble-based DA are Gaussian. As such, CRPS is estimated using the mean and standard deviation of the ensemble in the simple analytical expression for the Gaussian CRPS given by \citet{Gneiting2005}. Since a CRPS value is obtained for each (scalar) reference truth observation $x^\star$, we report the mean CRPS over all reference values as a summary statistic for each study area in a given experiment using a particular reanalysis approach for each study area. Moreover, we calculate the percentage improvement (PI, \%) in mean CRPS for each method compared to that of the (hierarchical) prior $\mathrm{CRPS}_{\mathrm{pri}}$ via $\mathrm{PI}=\left(1-\mathrm{CRPS}/\mathrm{CRPS}_{\mathrm{pri}}\right)\times10^2$
and average this across all sites to obtain the mean PI for each method in each experiment.

\section{Method}    
\label{sec:method}

Here, we present a new hierarchical Bayesian data assimilation method to generate climate-informed cryospheric reanalyses via temporal partial pooling. Recall that the task of reanalysis in Earth system science seeks to reconstruct the historical trajectory of the state of a system by combining  uncertain dynamical models with noisy and sparse observations \citep{Hersbach2020,Baatz2021,Fang2022,Abel2026}. 

We describe the new framework in detail, starting from the hierarchical Bayesian framing \citep{Robert2007,Gelman2013} of the reanalysis problem and progressing through the steps of our workflow. In theory, it is possible to solve the hierarchical reanalysis problem via joint inference of annual parameters and climatological hyperparameters, for example via standalone particle \citep{Chopin2020} or Markov Chain Monte Carlo (MCMC) \citep{Robert2004} methods. However, this approach is inefficient and often prohibitively expensive for large state spaces \citep{Sarkka2023} typical of geophysical data assimilation \citep{Carrassi2018,vanLeeuwen2019}. Nested methods present a more promising avenue, where PMCMC approaches are the state-of-the-art \citep{Andrieu2010}. Nonetheless, even PMCMC is generally intractable for larger scale multi-decadal cryospheric reanalysis applications.

To address this, we suggest a hybrid workflow based on nested particle methods \citep{Chopin2013}. In broad terms, this workflow involves annual parameter inference via hybrid particle ensemble Kalman smoothing (MAGPIES in Section~\ref{sec:MAGPIES}), nested inside outer climatological hyperparameter inference that is guided by expectation maximization (EM in Section~\ref{sec:EM}), which also refines each MAGPIES generation and is ultimately powered by adaptive particle smoothing (via AMIS in Section~\ref{sec:AMIS}), amortized by efficiently recycling particles from MAGPIES without model reruns. This nested particle smoother workflow is displayed schematically in Figure~\ref{fig:workflow}. Related hierarchical Bayesian problems and nested methods have recently been introduced in the data assimilation \citep{Katzfuss2020,Lucini2021,Drovandi2022} and broader Bayesian literature \citep{Chopin2013,Rainforth2018,Viani2023,PerezVieites2025}. Our work is inspired by these developments while presenting an original and opportunistic fusion of methods tailored to the cryospheric reanalysis problem. That said, the nested structure is modular: any scheme providing conditional posterior approximations and evidence estimates may serve in the inner loop, and any scheme that can approximate the hyperposterior via the evidence may serve in the outer loop.

\subsection{Cryospheric reanalysis as Bayesian inversion} \label{sec:reanalysis}

To formalize key concepts and notation, we cast cryospheric reanalysis as a Bayesian inverse problem \citep{SanzAlonso2023}. As such, we conceive of the data $\by$ that we are given as noisy samples from a forward (data generating) model 
\begin{equation}
    \by=\mathcal{G}(\btheta)+\boldsymbol{\epsilon} \, , \label{eq:forward}
\end{equation}
where $\by$ is an observation vector, $\mathcal{G}(\cdot)$ is the forward model, $\btheta$ is a vector of uncertain parameters, and $\boldsymbol{\epsilon}$ is a noise term representing unknown observation error. By only considering observation (and representation) error, we are adopting a strong constraint forcing formulation \citep{Evensen2022}. Therein, the forward model is assumed to be perfect, and all uncertainty is due to initial and boundary conditions, forcing, and internal model parameters, which can all be lumped into $\btheta$. This approach is widely used in practical cryospheric DA \citep{AlonsoGonzalez2022,Cao2025,Willmes2025,Guillet2026} and beyond \citep{Pirk2022,Keetz2025}, including reanalysis products \citep{Margulis2016,Hersbach2020} where it is often implicit. In cryospheric DA, this formulation can be justified by forcing being a dominant source of uncertainty \citep{Gunther2019,Fang2023,Tang2023}. The reanalysis problem of identifying historical model trajectories that match observations can, in principle, be solved by somehow \emph{inverting} the forward model in \eqref{eq:forward} to find the uncertain parameters $\boldsymbol{\theta}$. In practice, this inverse problem is challenging due to the non-linearity of the forward model and the fact that it is ill-posed with a non-unique solution due to the uncertain noise term $\boldsymbol{\epsilon}$ \citep{SanzAlonso2023}.

To invert \eqref{eq:forward}, we first note that here $\mathcal{G}(\cdot)$ is implicitly a composition of three deterministic operators
\begin{equation*}
\by=\mathcal{H}\left(\mathcal{M}\left(\mathcal{T}\left(\btheta \right)\right)\right)+\boldsymbol{\epsilon} \, ,
\end{equation*}
where $\mathcal{T}(\cdot)$ is a transformation operator, $\mathcal{M}(\cdot)$ is a dynamic model operator, and $\mathcal{H}(\cdot)$ is the observation operator. The transformation operator $\boldsymbol{\varphi}=\mathcal{T}(\btheta)$ maps transformed uncertain parameters $\btheta$ in an unbounded parameter space to the model parameters $\boldsymbol{\varphi}$ in the potentially bounded space of the geophysical model. It is analogous to the inverse-link and activation functions used in statistics and machine learning \citep{Gelman2013,Murphy2023}. The motivation for this is that most optimization and inference algorithms are designed to operate in unbounded spaces. As such, it is easier to explicitly target the unbounded transformed parameters $\btheta$ rather than their corresponding bounded model parameter counterparts $\boldsymbol{\varphi}$ that we recover implicitly. This approach can require keeping track of  Jacobian terms \citep{Gelman2013}, but not with the transformations used in this study described in Section~\ref{sec:prior}. Analogous transformations have been used successfully in many cryospheric DA applications \citep{Aalstad2018,AlonsoGonzalez2022,Groenke2023,Cao2025}. The next operator is the dynamic model operator $\mathbf{x}=\mathcal{M}(\boldsymbol{\varphi})$, taking the model parameters and (implicitly) forcing as input to compute the model state trajectory vector $\mathbf{x}$ for a given time period. In this study, $\mathcal{M}(\cdot)$ is the temperature index model detailed in Section~\ref{sec:forward} that is run one water year at a time. Thus, $\mathbf{x}$ contains the trajectory of daily water equivalent $D_k$ for a given water year. Finally, $\widehat{\mathbf{y}}=\mathcal{H}(\mathbf{x})$ is the observation operator that maps from the state to a vector of noise-free predicted observations $\widehat{\mathbf{y}}$. These predicted observations correspond to noise-free versions of the generally sparse observations in $\mathbf{y}$. As such, the observation operator maps the relevant entries in $\mathbf{x}$ to the observation space. In the case of direct observations, such as SWE and mass balance, this is done simply by selecting the corresponding state variables for times (and locations) with observations. For indirect observations, this operator generally involves additional modeling, such as the snow depletion curve used to predict FSCA from SWE, described in Section~\ref{sec:forward}.

In light of the layered forward model operator $\mathcal{G}$ and the inverse problem associated with \eqref{eq:forward}, we can begin to create a path to generate a reanalysis. The ill-posed nature of the inverse problem implies that there is no uniquely optimal parameter vector $\boldsymbol{\theta}$. Instead of optimizing parameters, we cast reanalysis as a dynamical Bayesian inverse problem \citep{SanzAlonso2023} corresponding to data assimilation \citep{Evensen2022}. Thereby, inversion boils down to a well-posed problem of inferring the posterior $p(\btheta\mid \by)$, the probability density of parameters given the data, via Bayes' rule
\begin{equation}
    p(\btheta\mid \by) = \frac{p(\by\mid \btheta) p(\btheta)}{p(\by)} \, , \label{eq:Bayes}
\end{equation}
where $p(\btheta)$ is the prior probability density encoding available information about the parameters before considering the data, $p(\by\mid\btheta)$ is the likelihood which loosely speaking tells us how well the data $\by$ fits our model with parameters $\btheta$, and $p(\by)$ is the evidence that is a normalizing constant here but vital for higher levels of inference discussed in Section~\ref{sec:hierarchical}. In this Bayesian approach probability should be seen as a general measure of uncertainty rather than merely a relative frequency \citep{MacKay2003,Robert2007}. As such, the posterior distribution $p(\btheta\mid\by)$ encodes what has been learned about the parameters $\btheta$ after considering the data $\by$ and is thus the Bayesian solution to the inverse problem. 

In practical reanalysis, we are not directly interested in the transformed parameters $\btheta$ per se, but primarily the state $\mathbf{x}$ and possibly the model parameters $\boldsymbol{\varphi}$. Nonetheless, we can obtain quantities of interest from the posterior distribution using posterior expectations of the form
\begin{equation}
    \mathrm{E}\left[u(\btheta)\mid\by \right]=\int u(\btheta) p(\btheta \mid \by) \, \mathrm{d}\btheta \, , \label{eq:postex}
\end{equation}
where $u(\btheta)$ is some function of the transformed parameters $\btheta$. For example, if $u(\btheta)=\mathcal{M}(\mathcal{T}(\btheta))=\mathbf{x}$, we obtain the posterior mean of the entire annual state trajectory $\mathbf{x}$. This formulation implicitly solves a Bayesian smoothing problem since the state for a given day $k$ is conditioned on \emph{all} observations in the given water year, as opposed to only up to day $k$ for filtering \citep{AlonsoGonzalez2022}. Through different choices of $u(\boldsymbol{\theta})$, this expectation is readily extended to the posterior variance and other summary statistics. We may analogously compute posterior expectations of some choice of utility (or loss) function to find optimal point estimates \citep{Sarkka2023}. In practical reanalysis with ensemble-based DA, expectations of the form \eqref{eq:postex} are straightforward to estimate in post-processing using Monte Carlo integration described in Section~\ref{sec:PIES}. To summarize, the posterior $p(\btheta\mid\by)$ is combined with corresponding historical model predictions $\mathbf{x}=\mathcal{M}(\mathcal{T}(\btheta))$ to obtain a cryospheric reanalysis. In practice, the posterior is approximated by an ensemble of particles, as described in Sections~\ref{sec:IES}-\ref{sec:AMIS}. Before digging into how to solve the problem computationally, we need to fully specify our probabilistic data generating model, starting with the prior and likelihood in \eqref{eq:Bayes}.

\subsection{Prior and likelihood}\label{sec:prior}

To fully specify a probabilistic model for cryospheric reanalysis, we need to define the prior $p(\btheta)$ and likelihood in $p(\by\mid \btheta)$ in \eqref{eq:Bayes}. The evidence $p(\by)$ is obtained from these (Section~\ref{sec:hierarchical}), so specifying the prior and likelihood suffices to infer the posterior. Up to now, our notation has been implicit, as we have not included all the assumptions and background information on which our probabilistic model is conditioned. Any so-called hyperparameters controlling these distributions are implicit; the rule being that variables which are not considered uncertain (never appearing to the left of the conditioning symbol $\mid$) are suppressed for brevity \citep{Robert2007,Gelman2013}. 

More explicitly, the prior, likelihood, evidence, and posterior in $\eqref{eq:Bayes}$ become $p(\btheta\mid\bpsi,\mathcal{P})$, $p(\by\mid\btheta,\bpsi,\mathcal{P})$, $p(\by\mid\bpsi,\mathcal{P})$, and $p(\btheta\mid\by,\bpsi,\mathcal{P})$, respectively, where $\bpsi$ denotes the hyperparameters and $\mathcal{P}$ denotes the probabilistic model. As shown in Figure~\ref{fig:pgm} we are using and comparing three different probabilistic models (CP, NP, and PP) in this study (see Section~\ref{sec:experiments}). For economy, we keep conditioning on $\mathcal{P}$ implicit, letting the target posteriors defined by \eqref{eq:CP} for CP, \eqref{eq:NP} for NP, and \eqref{eq:nparpost} for PP be indicators of these respective probabilistic models. The set of hyperparameters $\bpsi$ controlling the prior becomes key in hierarchical reanalysis, so we make this conditioning explicit wherever needed. 

For our prior distributions on the model parameters $\boldsymbol{\varphi}$, we use a logit-normal distribution for the physically double bounded parameters $a,c,v$ and a normal distribution for the effectively unbounded parameter $b$. The logit-normal distribution has been widely used to impose double bounds on parameters in DA \citep{Aalstad2018,Keetz2025}, with the key property that the logit transform of the logit-normally distributed uncertain variable of interest is normally distributed. In practice, logit-normal variables are generated by applying the generalized logit transform $\mathrm{glt}(\cdot)$ to the desired median, adding standard Gaussian noise scaled by the associated standard deviation $\sigma$, and then applying the corresponding generalized expit transform $\mathrm{get}(\cdot)$. These transforms are generalized as they scale the unit interval $(0,1)$, allowing for general double bounds. The use of the logit-normal prior and the associated generalized logit transform permits inference on unbounded transformed parameters $\btheta=[\alpha,\beta,\gamma,\nu]$, where $\alpha=\mathrm{glt}(a)$ and so on, with transforms given in Table~\ref{tab:prior}. The corresponding model parameters can then be recovered after inference by applying the corresponding inverse transform $a=\mathrm{get}(\alpha)$ and so on, where the specific bounds given by the support column in Table~\ref{tab:prior} are implicit in the transform.

\begin{figure}[ht]
\includegraphics[width=\textwidth]{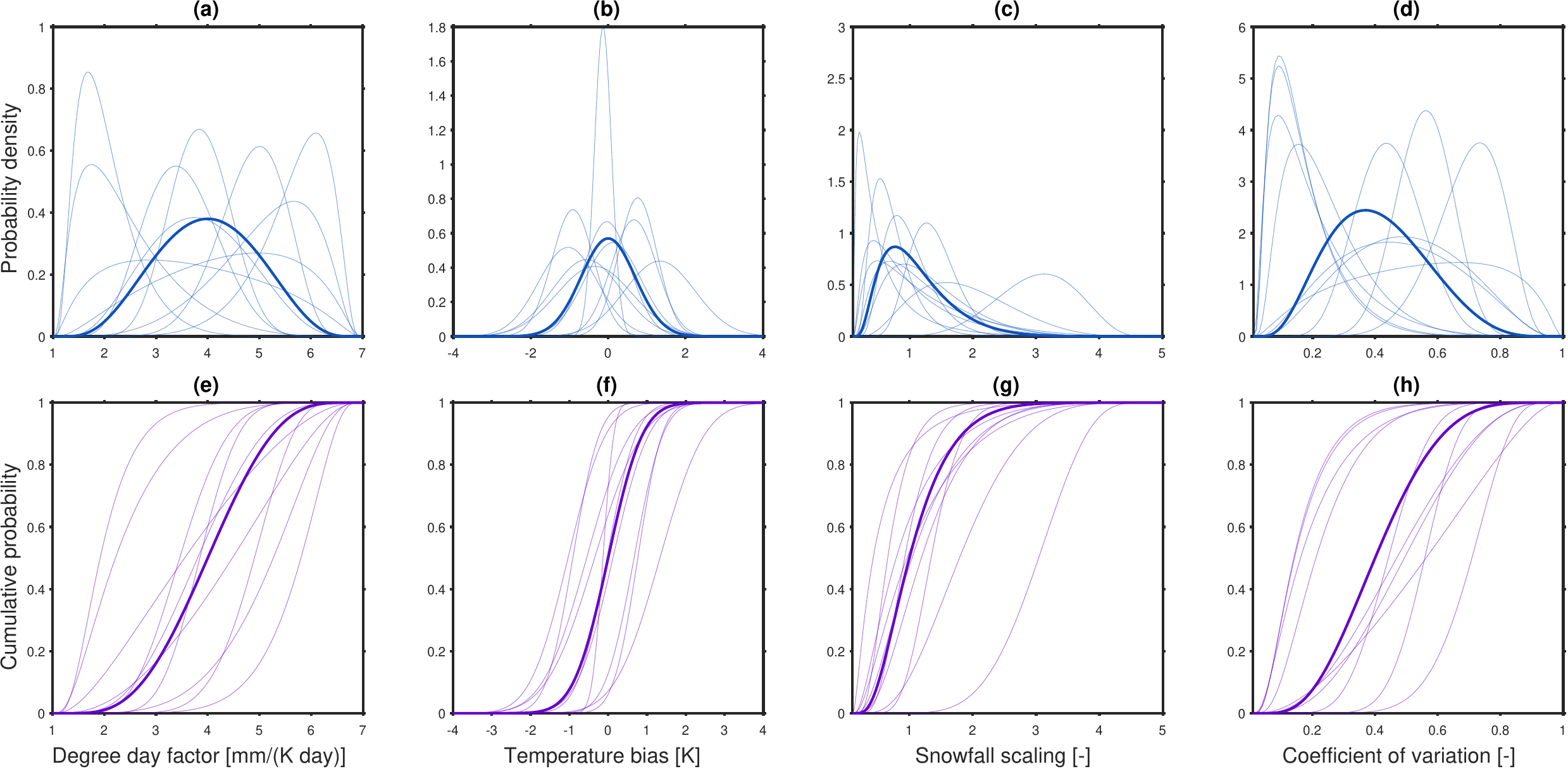}
\caption{Marginal prior densities (top) and corresponding cumulative distributions (bottom) for the degree day factor $a$ (panels a, e), temperature bias $b$ (panels b, f), snowfall scaling $c$ (panels c, g), and coefficient of variation $v$  (panels d, h). The default prior (thick solid line) used for non-hierarchical inference $p(\btheta|\bpsi_0)$ has fixed hyperparameters $\bpsi_0$ set to the hyperprior mean. The conditional priors (thin solid lines) used in hierarchical inference $p(\btheta|\bpsi_i)$ have uncertain hyperparameters $\bpsi_i\sim p(\bpsi)$ sampled from the hyperprior.}
\label{fig:prior} 
\end{figure}

For simplicity, we assume a conditionally independent prior $p(\btheta\mid\bpsi)$ on the unbounded parameters $\btheta$ (and thus also $\boldsymbol{\varphi}$), so that the joint prior factorizes into the product of marginal priors given $\boldsymbol{\psi}$. These marginal priors are fully specified by the information given in Table~\ref{tab:prior}, which includes the hyperparameters $\bpsi$ in the form of prior location $\mu$ and scale $\sigma$ for each transformed parameter. The conservative bounds given by the support are considered known based on physical constraints and previous work \citep{Aalstad2018,Rounce2020}, so these are held fixed across all experiments. The uncertain hyperparameters $\bpsi$ play a key role in hierarchical inference. The location and scale hyperparameters $\mu$ and $\sigma$ define the mean and standard deviation for the respective marginal priors of the transformed parameters $\btheta$. In Figure~\ref{fig:prior}, we show how the respective conditional priors $p(\btheta\mid\bpsi)$ vary by sampling different hyperparameters from the hyperprior $p(\bpsi)$. These hyperparameters are often fixed to some best guess $\bpsi_0$ in general non-hierarchical settings such as CP and NP that only seek to infer $\btheta$. This ignores the fact that we are generally uncertain about these hyperparameters, making priors overconfident and limiting information transfer between years. In hierarchical inference via partial pooling, shown in Figure~\ref{fig:pgm}, we also infer these hyperparameters $\bpsi$ that define the climatology of the lower level parameters $\btheta$. Since we have $N_\theta=4$ ($3$) types of snow (glacier) parameters $\btheta$ that each have a location and scale hyperparameter, this leaves us with a total of $N_\psi=8$ ($6$) hyperparameters. Unlike the parameters $\btheta$ that can vary from year to year, the climatological hyperparameters $\bpsi$ are assumed to be time invariant.

For the hyperprior $p(\bpsi)$ (i.e., the prior on the hyperparameters), we also assume conditional independence, enabling factorization into a product of marginal hyperpriors. These marginal hyperpriors are fully specified by the information given in Table~\ref{tab:prior}. Here we use normal (Gaussian) hyperpriors for the location hyperparameters and center these on typical values for the associated model parameters from related work \citep{Aalstad2018,Rounce2020,Mazzolini2025}, namely $a=4$ [mm K$^{-1}$ day$^{-1}$], $b=0$ [K], $c=1$ [-], and $v=0.4$ [-] each with a standard scale in transformed space of $s=1$. To impose positivity and avoid numerical instability, we assume logit-normal hyperpriors with support $(0,1.4)$ on the scale hyperparameters, centering them on a value of $\eta=\mathrm{glt}(0.7)=0$ in the transformed parameter ($\tau$) space with a unit scale of $\chi=1$. Although these hyperpriors are certainly highly uncertain and challenging to specify, the uncertainty on higher level probabilities is less consequential for inference \citep{Good1980}.

For the observation error model, we follow the standard approach in DA \citep{Carrassi2018} and state space modeling more generally \citep{Sarkka2023} by assuming zero-mean additive white Gaussian noise. This results in an observation error of the form $\boldsymbol{\epsilon}\sim\mathrm{N}(\mathbf{0},\mathbf{R})$ with observation error covariance $\mathbf{R}$. This typical choice can be justified from maximum entropy \citep{Jaynes2003} and central limit theorem arguments \citep{Chopin2020}, but it is possible to relax this assumption with more complex \citep{vanHove2026} and robust error models \citep{Guillet2026}. Adopting this error model implies a Gaussian likelihood \citep{AlonsoGonzalez2022}
\begin{equation}
    p(\by\mid \btheta) =\det(2\pi \mathbf{R})^{-1/2}\exp\left(-\frac{1}{2}\left[\by-\widehat{\by}\right]^\mathrm{T}\mathbf{R}^{-1}\left[\by-\widehat{\by}\right]\right) \, , \label{eq:likelihood}
\end{equation}
where $\widehat{\by}=\mathcal{H}(\mathcal{M}(\mathcal{T}(\btheta)))$ are the noise-free predicted observations given transformed parameters $\btheta$. We use a diagonal observation error covariance matrix $\mathbf{R}=\sigma_y^2\mathbf{I}$ where $\sigma_y$ is the observation error standard deviation specified for the respective observation types in Sections~\ref{sec:snow},\ref{sec:glacier}, and \ref{sec:fsca}. This assumes that the observation errors are conditionally independent in time, since $\by$ implicitly contains temporally distributed observations for a particular point or grid cell in space. We restrict our attention to purely temporal (i.e., embarrassingly parallel in space) DA, but it is possible to generalize further to spatio-temporal DA \citep{AlonsoGonzalez2023}. Having formally defined the Gaussian likelihood in \eqref{eq:likelihood}, we emphasize that evaluating the likelihood is often the main computational bottleneck in Bayesian inverse problems in general and cryospheric reanalysis in particular. The reason being that each likelihood evaluation requires running the entire forward model to compute the residual $\boldsymbol{\epsilon}=\by-\widehat{\by}$ between the observations $\by$ and predicted observation $\widehat{\by}(\btheta)$ for parameter setting $\btheta$.

\begin{table}[ht]
\caption{Specification of prior \mbox{$p(\btheta\mid\bpsi)$} and hyperprior $p(\bpsi)$ distributions for the unbounded parameters \mbox{$\btheta=[\alpha,\beta,\gamma,\nu]$} and hyperparameters \mbox{$\bpsi=[\boldsymbol{\mu},\boldsymbol{\tau}]=[\mu_\alpha,\mu_\beta,\mu_\gamma,\mu_\nu,\tau_\alpha,\tau_\beta,\tau_\gamma,\tau_\nu]$}, respectively. The model parameters \mbox{$\boldsymbol{\varphi}=[a,b,c,v]$} correspond to the unbounded parameters $\btheta$ that we infer through the specified transformations. So (e.g.) the prior of $a$ denoted $p(a)$ is a double bounded Logit-Normal (LogitN) with support $(1,7)$  whereby the generalized logit transform (glt), and \mbox{$\alpha=\mathrm{glt}(a)$} follows a normal distribution \mbox{$p(\alpha)=p(a)|\partial_\alpha a|$} with uncertain mean $\mu_\alpha$ and standard deviation $\sigma_\alpha$ hyperparameters. Similarly for (e.g.) $\sigma_\alpha$ the hyperprior $p(\sigma_\alpha)$ is a double bounded Logit-Normal (LogitN) with support $(0,1.4)$ so \mbox{$\tau_\alpha=\mathrm{glt}(\sigma_\alpha)$} follows a normal distribution $p(\tau_\alpha)$ with fixed mean $\eta_\alpha$ and standard deviation $\chi_\alpha$.}
\centering
\begin{tabular}{l c c c c c c c}
\hline
Parameter name   & Symbol & Transform & Family & Support & Location & Scale \\
 \hline
   Degree day factor  & $a$  & $\alpha=\mathrm{glt}(a)$ & LogitN & $(1,7)$ & $\mu_\alpha$ & $\sigma_\alpha$ \\
   Temperature bias  & $b$ & $\beta=b$ & Normal & $\mathbb{R}$ & $\mu_\beta$ & $\sigma_\beta$ \\
   Snowfall scaling  & $c$ & $\gamma=\mathrm{glt}(c)$ & LogitN & $(0.1,5)$ & $\mu_\gamma$ & $\sigma_\gamma$ \\
   Subgrid CV  & $v$ & $\nu=\mathrm{glt}(v)$ & LogitN & $(0,1)$ & $\mu_\nu$ & $\sigma_\nu$ \\
   \hline
Hyperparameter   & Symbol & Transform & Family & Support & Location & Scale \\
 \hline
    Mean of $\alpha$ & $\mu_\alpha$ & None & Normal & $\mathbb{R}$ & $m_\alpha=\mathrm{glt}(4)$ & $s_\alpha=1$ \\
    Mean of $\beta$ & $\mu_\beta$ & None & Normal & $\mathbb{R}$ & $m_\beta=0$ & $s_\beta=1$ \\
    Mean of $\gamma$ & $\mu_\gamma$ & None & Normal & $\mathbb{R}$ & $m_\gamma=\mathrm{glt}(1)$ & $s_\gamma=1$ \\
    Mean of $\nu$ & $\mu_\nu$ & None & Normal & $\mathbb{R}$ & $m_\nu=\mathrm{glt}(0.4)$ & $s_\nu=1$ \\
    Std. dev. of $\alpha$ & $\sigma_\alpha$ & $\tau_\alpha=\mathrm{glt}(\sigma_\alpha)$ & LogitN & $(0,1.4)$ & $\eta_\alpha=\mathrm{glt}(0.7)$ & $\chi_\alpha=1$ \\
    Std. dev. of $\beta$ & $\sigma_\beta$ & $\tau_\beta=\mathrm{glt}(\sigma_\beta)$ & LogitN & $(0,1.4)$ & $\eta_\beta=\mathrm{glt}(0.7)$ & $\chi_\beta=1$ \\
    Std. dev. of $\gamma$ & $\sigma_\gamma$ & $\tau_\gamma=\mathrm{glt}(\sigma_\gamma)$ & LogitN & $(0,1.4)$ & $\eta_\gamma=\mathrm{glt}(0.7)$ & $\chi_\gamma=1$ \\
    Std. dev. of $\nu$ & $\sigma_\nu$ & $\tau_\nu=\mathrm{glt}(\sigma_\nu)$ & LogitN & $(0,1.4)$ & $\eta_\nu=\mathrm{glt}(0.7)$ & $\chi_\nu=1$ 
 \\
 \hline
 \end{tabular} \label{tab:prior}
 \end{table}

\subsection{Hierarchical reanalysis} \label{sec:hierarchical}
In hierarchical reanalysis, we seek to infer the joint posterior over parameters $\btheta$ and hyperparameters $\bpsi$  given the observations $\by$ via Bayes' rule
\begin{equation}
p(\btheta,\bpsi\mid \by)=\frac{p(\by\mid \btheta,\bpsi)p(\btheta,\bpsi)}{p(\by)} \, , \label{eq:jpost}
\end{equation}
where $p(\by\mid \btheta,\bpsi)$ is the likelihood, $p(\btheta,\bpsi)$ is the joint prior, and we refer to $p(\by)$ as the hyperevidence for reasons that will soon become apparent. For now, the time indexing of the parameters, $\btheta=\btheta_{1:N}=\left[\btheta_1,\dots,\btheta_n\dots,\btheta_N\right]^\mathrm{T}$, and observations $\by=\by_{1:N}=\left[\by_1,\dots,\by_n\dots,\by_N\right]^\mathrm{T}$, is kept implicit for economy. The joint posterior can be decomposed into
\begin{equation}
p(\btheta,\bpsi\mid \by) = p(\btheta\mid \by,\bpsi)p(\bpsi\mid \by) \, , \label{eq:jdecomp}
\end{equation}
where
\begin{equation}
    p(\btheta\mid \by,\bpsi) = p(\by\mid \btheta)p(\btheta\mid \bpsi)/p(\by\mid \bpsi) \, , \label{eq:cposttheta}
\end{equation}
is the usual (conditional) posterior of interest in which the hyperparameters $\bpsi$ are fixed and often implicit. Here $\bpsi$ does not include observation hyperparameters (such as the noise variance $\sigma_y^2$), whereby observations are conditionally independent of $\bpsi$ given $\btheta$, so $p(\by\mid \btheta,\bpsi)=p(\by\mid \btheta)$ in \eqref{eq:cposttheta}. We recognize the conditional posterior in \eqref{eq:cposttheta} as the usual target of non-hierarchical methods where the hyperparameters $\bpsi$ are fixed. This conditional posterior can be approximated using any of the usual approaches to non-hierarchical data assimilation, such as particle or ensemble Kalman methods.

A new challenge is to obtain the marginal posterior for the hyperparameters, $p(\bpsi\mid \by)$, which is a key feature of the hierarchical approach. Applying Bayes' rule
\begin{equation}
 p(\bpsi\mid \by) = p(\by\mid \bpsi)p(\bpsi)/p(\by) \, , \label{eq:postpsi}
\end{equation}
where $p(\by\mid \bpsi)$ is the familiar evidence (marginal likelihood) defined by the generally intractable integral \citep{MacKay2003,Llorente2023} 
\begin{equation}
    p(\by\mid \bpsi)=\int p(\btheta,\by \mid \bpsi) \, \mathrm{d}\btheta = \int p(\by \mid \btheta)p(\btheta \mid \bpsi) \, \mathrm{d}\btheta \, , \label{eq:evidence}
\end{equation}
in which the fixed hyperparameters tend to be implicit in non-hierarchical settings, $p(\bpsi)$ is the prior for the hyperparameters called the hyperprior, and $p(\by)$ is the evidence (or marginal likelihood) after integrating out the hyperparameters that we refer to as the hyperevidence to continue the `hyper' theme at the second level of inference. Comparing \eqref{eq:cposttheta} and \eqref{eq:postpsi}, it is clear that at the second (hyperparameter) level of inference the evidence plays the same role as the likelihood does at the usual first (parameter) level of inference. Moreover, the evidence can be estimated (point-wise) as a by-product while performing the usual first level of parameter inference in \eqref{eq:cposttheta} while keeping the hyperparameters fixed. These insights suggest a nested approach to hierarchical inference. 
% Explain why direct joint inference (as opposed to the nested approach) is not feasible

Although the hyperparameters may be of interest in certain settings, the joint posterior in \eqref{eq:jpost} is just a means to an end when our final target is the marginal posterior for the parameters and the corresponding posterior state estimates. Using the sum (marginalization) rule, this marginal posterior is obtained through
\begin{equation*}
p(\btheta\mid \by)=\int p(\btheta,\bpsi\mid \by) \, \mathrm{d}\bpsi \, ,
\end{equation*}
where we can insert \eqref{eq:jdecomp} into the integrand to obtain
\begin{equation}
p(\btheta\mid \by)=\int p(\btheta\mid \by,\bpsi) p(\bpsi\mid \by) \, \mathrm{d}\bpsi \, , \label{eq:mtarget}
\end{equation}
so the final target marginal posterior for the parameters $p(\btheta\mid \by)$ can be viewed as an expectation of the usual (conditional) posterior $p(\btheta\mid \by,\bpsi)$ with respect to the marginal posterior of the hyperparameters $p(\bpsi\mid \by)$. As such, the final target we seek is a weighted average of the usual (conditional) posteriors, with weights given by the marginal hyperparameter posteriors. Thus, this target marginal posterior is a kind of (infinite) weighted model average across conditional posteriors, where the weight of each `model' is defined by the choice of hyperparameters. This kind of weighted model average interpretation becomes particularly pertinent when adopting particle methods for the second (hyper) level of inference. It also reaffirms the aforementioned nested approach to hierarchical inference. 

\subsection{Inferential dynamics} \label{sec:dynamics} % Might actually have to be an appendix

So far, the inferential dynamics of the hierarchical reanalysis have been hidden by using implicit time indices $n=1,\dots,N$. To formalize these dynamics, we reintroduce the time index in the target marginal posterior \eqref{eq:mtarget}, obtaining
\begin{equation}
p(\btheta_{1:N}\mid \by_{1:N})=\int p(\btheta_{1:N}\mid \by_{1:N},\bpsi) p(\bpsi\mid \by_{1:N}) \, \mathrm{d}\bpsi \, . \label{eq:mdtarget}
\end{equation}
Ultimately, we wish to perform posterior predictions for a given water year $n$ conditioned on \emph{everything all at once} $\by_{1:N}$, which requires the marginal annual posterior
\begin{equation}
p(\btheta_n\mid \by_{1:N})=\int p(\btheta_{1:N}\mid \by_{1:N}) \, \mathrm{d}\btheta_{-n} \, , \label{eq:amdtarget}
\end{equation}
where $\btheta_{-n}$ contains the parameter vectors for all years except $n$. Inserting \eqref{eq:mdtarget} in \eqref{eq:amdtarget} yields
\begin{equation}
p(\btheta_n\mid \by_{1:N})= \int p(\bpsi\mid \by_{1:N}) \int p(\btheta_{1:N}\mid \by_{1:N},\bpsi) \, \mathrm{d}\btheta_{-n} \, \mathrm{d}\bpsi \, , \label{eq:amdetarget}
\end{equation}
where we have pulled $p(\bpsi\mid \by_{1:N})$ defined in \eqref{eq:postpsi} out of the inner multiple integral  since it does not depend on $\btheta_{-n}$. To simplify the integral in \eqref{eq:amdetarget}, we first expand the inner integrand, which is the usual (conditional) posterior of interest for all parameters $\btheta_{1:N}$ using \eqref{eq:cposttheta} 
\begin{equation}
p(\btheta_{1:N}\mid \by_{1:N},\bpsi) = p(\by_{1:N}\mid \btheta_{1:N})p(\btheta_{1:N}\mid \bpsi)/p(\by_{1:N}\mid \bpsi) \, , \label{eq:cdposttheta}
\end{equation}
this can be expanded further since the annual parameter vectors in the prior $p(\btheta_{1:N}\mid \bpsi)$ are conditionally independent $p(\btheta_n\mid \bpsi,\btheta_{-n})=p(\btheta_n\mid \bpsi)$, so we get the factorization
\begin{equation*}
p(\btheta_{1:N}\mid \bpsi)=\prod_{n=1}^N p(\btheta_n\mid \bpsi) \, ,
\end{equation*}
similarly the annual observation vectors in the likelihood $p(\by_{1:N}\mid \btheta_{1:N})$ are also conditionally independent $p(\by_n\mid \btheta_n,\btheta_{-n},\by_{-n})=p(\by_n\mid \btheta_n)$, where $\by_{-n}$ contains the observations for all years except $n$, resulting in the factorization
\begin{equation*}
p(\by_{1:N}\mid \btheta_{1:N})=\prod_{n=1}^N p(\by_n\mid \btheta_n) \, ,
\end{equation*}
by combining these factorizations, the numerator in \eqref{eq:cdposttheta} becomes
\begin{equation}
p(\by_{1:N}\mid \btheta_{1:N})p(\btheta_{1:N}\mid \bpsi)=\prod_{n=1}^N p(\by_n\mid \btheta_n)p(\btheta_n\mid \bpsi) \, . \label{eq:jfac}
\end{equation}
The evidence in the denominator of \eqref{eq:cdposttheta} is given by
\begin{equation}
 p(\by_{1:N}\mid \bpsi)=\int  p(\by_{1:N},\btheta_{1:N}\mid \bpsi) \,  \mathrm{d}\btheta_{1:N} \, , \label{eq:evidef}
\end{equation}
 noting that $p(\by_{1:N},\btheta_{1:N}\mid \bpsi)=p(\by_{1:N}\mid \btheta_{1:N})p(\btheta_{1:N}\mid \bpsi)$, we may insert the factorization in \eqref{eq:jfac} to convert the multiple integral into a product of integrals of the form 
\begin{equation}
p(\by_{1:N}\mid \bpsi)=\prod_{n=1}^N \int 
 p(\by_n\mid \btheta_n)p(\btheta_n\mid \bpsi) \,  \mathrm{d}\btheta_{n} \, , \label{eq:mipi}
\end{equation}
where we recognize the integral as the annual evidence
\begin{equation}
p(\by_n\mid \bpsi) = \int  p(\by_n\mid \btheta_n)p(\btheta_n\mid \bpsi) \,  \mathrm{d}\btheta_{n} \, , \label{eq:annualevi}
\end{equation}
such that \eqref{eq:mipi} can be written as
\begin{equation}
p(\by_{1:N}\mid \bpsi) = \prod_{n=1}^N p(\by_n\mid \bpsi ) \, , \label{eq:jevi}
\end{equation}
whereby the evidence also factorizes into a product of conditionally independent annual evidence terms. This evidence factorization is a special case that is a result of the aforementioned conditional independence properties of both the prior and the likelihood. We may now insert \eqref{eq:jfac} and \eqref{eq:jevi} into \eqref{eq:cdposttheta} and merge the product operators to obtain the following expression for the conditional posterior 
\begin{equation}
   p(\btheta_{1:N}\mid \by_{1:N},\bpsi) = \prod_{n=1}^N p(\by_n\mid \btheta_n)p(\btheta_n\mid \bpsi)/p(\by_n\mid \bpsi ) \, , \label{eq:cpfacl}
\end{equation}
using the definition of the annual conditional posterior
\begin{equation}
p(\btheta_n\mid \by_n,\bpsi)=p(\by_n\mid \btheta_n)p(\btheta_n\mid \bpsi)/p(\by_n\mid \bpsi ) \, , \label{eq:actpos}
\end{equation}
then the posterior in \eqref{eq:cpfacl} factorizes to a product of annual conditional posteriors
\begin{equation}
p(\btheta_{1:N}\mid \by_{1:N},\bpsi) = \prod_{n=1}^N p(\btheta_n\mid \by_n,\bpsi) \, .\label{eq:cpfac}
\end{equation}

Having shown how both the evidence and the conditional posterior factorize into their annual counterparts, we can now proceed to tackle the nested integral in \eqref{eq:amdetarget} that defines our final target distribution, namely the marginal annual posterior $p(\btheta_n\mid \by_{1:N})$ conditioned on everything all at once. 
Inserting \eqref{eq:cpfac} into \eqref{eq:amdetarget}
\begin{equation}
p(\btheta_n\mid \by_{1:N})= \int 
 p(\bpsi\mid \by_{1:N}) \int \left( \, \prod_{k=1}^N p(\btheta_k\mid \by_k,\bpsi) \right) \, \mathrm{d}\btheta_{-n} \, \mathrm{d}\bpsi \, , \label{eq:famdetarget}
\end{equation}
where, due to the independence of the annual conditional posteriors, the inner multiple integral can be factorized as follows
\begin{equation}
\int \left(  \, \prod_{k=1}^N p(\btheta_k\mid \by_k,\bpsi) \right) \mathrm{d}\btheta_{-n}  = \prod_{k=1}^{N} \left[\left(1-\delta_{kn}\right) \int p(\btheta_k\mid \by_k,\bpsi)  \, \mathrm{d}\btheta_k+\delta_{kn}p(\btheta_n\mid \by_n,\bpsi)\right] \, , \label{eq:kron}
\end{equation}
where $\delta_{kn}$ is the Kronecker delta, which is $1$ for $k=n$ and $0$ otherwise. The annual conditional posteriors necessarily integrate to one over their support 
\begin{equation*}
1= \int p(\btheta_k\mid \by_k,\bpsi)  \, \mathrm{d}\btheta_k \, ,
\end{equation*}
for all $k=1,\dots,N$ whereby \eqref{eq:kron} simplifies considerably to
\begin{equation*}
    \int \left( \,  \prod_{k=1}^N p(\btheta_k\mid \by_k,\bpsi) \right) \mathrm{d}\btheta_{-n}  = p(\btheta_n\mid \by_n,\bpsi) \, ,
\end{equation*}
such that \eqref{eq:famdetarget} simplifies to the following 
\begin{equation}
    p(\btheta_n\mid \by_{1:N}) = \int p(\bpsi\mid \by_{1:N})p(\btheta_n\mid \by_n,\bpsi) \, \mathrm{d}\bpsi \, , \label{eq:nparpost}
\end{equation}
which can be seen as an expectation of the annual conditional parameter posterior with respect to the marginal hyperparameter posterior. Applying explicit indices in \eqref{eq:postpsi} and using the joint evidence factorization in \eqref{eq:jevi}, the latter is given by
\begin{equation}
    p(\bpsi\mid \by_{1:N})=\left(\prod_{n=1}^N p(\by_n\mid \bpsi)\right)p(\bpsi)/p(\by_{1:N}) \, , \label{eq:hypost}
\end{equation}
where, analogously to the evidence in \eqref{eq:evidef}, the hyperevidence $p(\by_{1:N})$ is obtained by marginalization
\begin{equation*}
p(\by_{1:N})=\int p(\by_{1:N},\bpsi) \, \mathrm{d}\bpsi \, ,
\end{equation*}
noting that $p(\by_{1:N},\bpsi)=p(\by_{1:N}\mid \bpsi)p(\bpsi)$ then
\begin{equation*}
p(\by_{1:N})=\int p(\by_{1:N}\mid \bpsi)p(\bpsi) \, \mathrm{d}\bpsi \, ,
\end{equation*}
and inserting for \eqref{eq:jevi}, we obtain
\begin{equation*}
p(\by_{1:N})=\int \left(\prod_{n=1}^N p(\by_n\mid \bpsi ) \right) p(\bpsi)  \, \mathrm{d}\bpsi \, ,
\end{equation*}
where we recognize the integrand as the numerator in \eqref{eq:hypost}. As such, the hyperevidence serves as a normalizing constant at the second (hyperparameter) level of inference. This is analogous to the role played by the evidence at the first (parameter) level of inference.

In summary, the inferential dynamics described by these equations suggest a \emph{nested} workflow to obtain our final target distributions in the form of the individual annual parameter posteriors in \eqref{eq:nparpost}. An outer loop involves estimating an approximation for the hyper-posterior $p(\bpsi\mid \by)$ in \eqref{eq:hypost} over the hyperparameters $\bpsi$. A nested inner loop involves estimating an approximation for the \emph{conditional} annual parameter posteriors $p(\btheta_n\mid \by_n,\bpsi)$ in  \eqref{eq:actpos}. Finally, these nested estimates are used together in \eqref{eq:nparpost} to marginalize out the hyperparameter and obtain the target distributions of interest $p(\btheta_n\mid \by_{1:N})$. Although naive implementations of such nested loops with subsequent marginalization are expensive \citep{Rainforth2018}, costly likelihood function evaluations can be reused to increase efficiency and accuracy.  

As alluded to in Section~\ref{sec:experiments} and visualized in Fig.~\ref{fig:pgm}, there are alternative ways to cast the probabilistic dynamics in the cryospheric reanalysis problem. The approach outlined above using \eqref{eq:nparpost} corresponds to a novel \emph{partial pooling} (PP) approach to hierarchical reanalysis that allows us to pool climatological information across years \emph{and} fit the particular `weather' of the year in question. This approach straddles two more extreme and simpler approaches, namely the farsighted complete pooling (CP) approach with global static parameters and the shortsighted no pooling (NP) approach with local annually varying parameters. For global inference via CP, the target posterior is 
\begin{equation}
p(\btheta\mid\by_{1:N},\bpsi_0)=\left(\prod_{n=1}^Np(\by_n\mid\btheta)\right)p(\btheta\mid\bpsi_0)/p(\mathbf{y}_{1:N}\mid\bpsi_0) \, ,  \label{eq:CP}
\end{equation}
where we have made the typically implicit conditioning on a fixed hyperparameter configuration $\bpsi_0$ explicit to emphasize that this is a non-hierarchical method. This CP posterior is straightforward to estimate by concatenating all data assimilation windows (water years) into a single large window to update time invariant global parameters. On the other extreme, for the local inference via NP, the target posterior is simply
\begin{equation}
p(\btheta_n\mid\by_n,\bpsi_0)=p(\by_n\mid\btheta_n)p(\btheta_n\mid\bpsi_0)/p(\mathbf{y}_{n}\mid\bpsi_0) \, , \label{eq:NP}
\end{equation}
which is computed independently for all water years $n$. This has been the standard approach to cryospheric reanalysis \citep{Margulis2016,AlonsoGonzalez2021,Cao2025} and amounts to treating the reanalysis problem in each water year independently. For years with data, this tends to perform better than CP, as the uncertain parameters $\btheta_n$, particularly those related to forcing, can vary considerably from year to year. For both CP and NP, the target posteriors are approximated using a single generation of the MAGPIES scheme described in Section~\ref{sec:MAGPIES}, using the hyperprior mean in Table~\ref{tab:prior} as the hyperparameters $\bpsi_0$. To enable a fairer comparison with PP, which uses an outer hyperparameter loop, the non-hierarchical CP and NP approaches use a larger ensemble size of $N_e=10^3$ compared to $N_e=10^2$ for PP.

% Question: Why is nesting even needed, why don't we do a "joint" inference to begin with?

%\textbf{Summarize key equations in workflow}: 1st obtain "hypost" \eqref{eq:hypost} then use this to weight the individual annual parameter posteriors \eqref{eq:nparpost}, in practice these are obtained at the same time. 

\subsection{Nested particle smoothing workflow}

% High level overview, also describe how obtaining hyperposterior is hardest part; from this we can get target marginal posterior of interest relatively cheaply (explain how)

Before diving into the various algorithms used in our nested particle smoothing approach, we provide a high level overview of the overall workflow. Knowledge of the method up to and including this nested workflow section is sufficient to follow the results and discussion. On the one hand, readers not interested in the details surrounding these algorithms and their implementation may thus consider skipping the remainder of the method from Section~\ref{sec:IES} onward. On the other hand, readers curious about details in the form of the various interconnected schemes from EM to AdaPBS, via MAGPIES and the Laplace approximation, and how the workflow can be benchmarked via PMCMC may find Sections~\ref{sec:IES}-\ref{sec:AMIS} and the appendices instructive. 

\begin{figure}[ht]
\includegraphics[width=\textwidth]{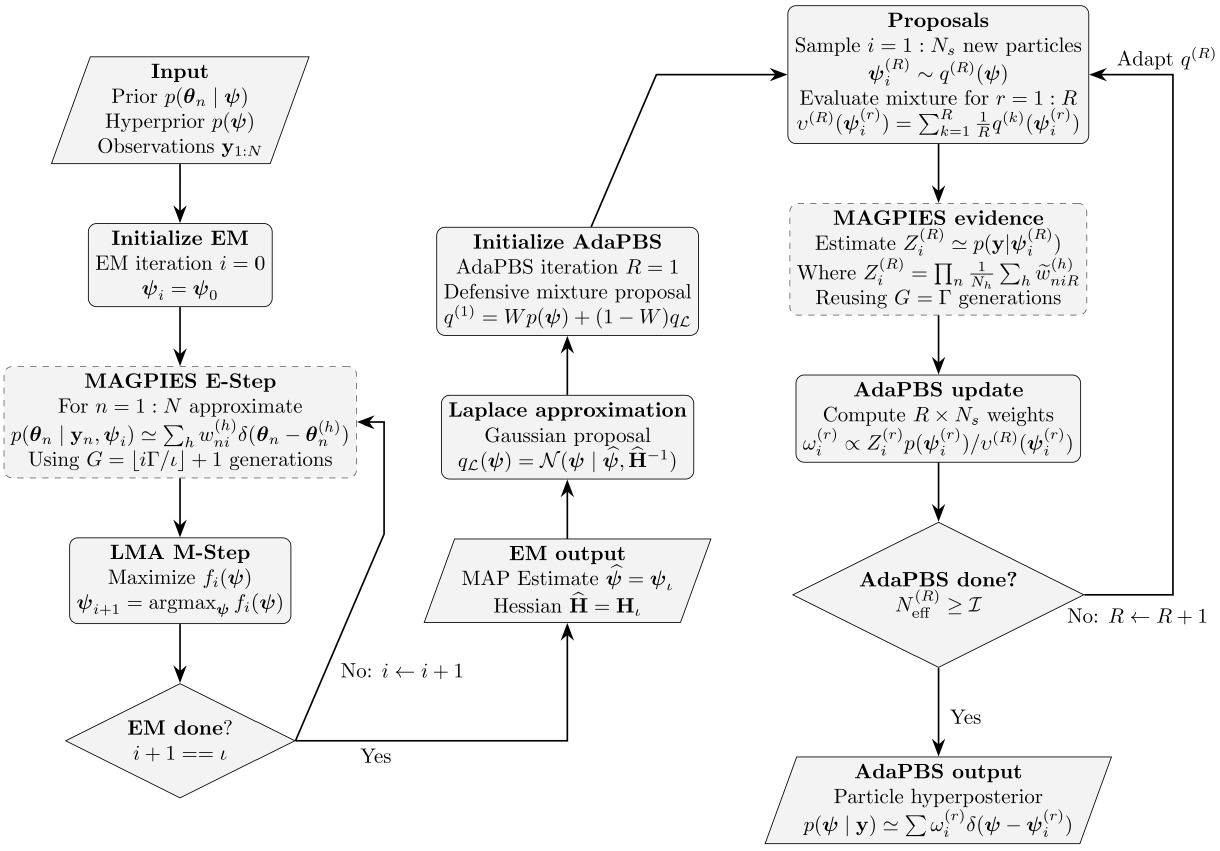}
\caption{Flowchart showing the workflow for the hybrid nested particle smoother method used to conduct tractable hierarchical Bayesian cryospheric reanalysis. The workflow starts with Expectation Maximization (EM) combining a hybrid particle smoother (MAGPIES) to estimate conditional posteriors and gradient-based optimization (LMA) to estimate optimal hyperparameters $\boldsymbol{\widehat{\psi}}$. The EM output is then used to form a Laplace approximation of the hyperposterior. This in turn is used together with the hyperprior to serve as an initial proposal for an amortized Adaptive Particle Batch Smoother (AdaPBS) which provides a particle approximation of the hyperposterior while reusing information provided by MAGPIES in the EM iterations. This hyperposterior approximation can then be used for marginal posterior prediction.}
\label{fig:workflow}
\end{figure}
% hyperposterior is the most expensive part

The ensuing overview of the nested workflow follows, step by step, the flowchart in Figure~\ref{fig:workflow}. The workflow begins by specifying the input in the form of a (conditional) prior $p(\btheta\mid\bpsi)$ and hyperprior $p(\bpsi)$, as outlined in Section~\ref{sec:prior}, as well as the observations $\by_{1:N}$ for all water years $n=1:N$. Next, expectation maximization (EM) is initialized by setting the EM iteration counter $i=0$ and providing an initial guess for the hyperparameters $\bpsi_0$ in the form of the hyperprior mean. Subsequently, we enter into the main loop of the EM algorithm (Section~\ref{sec:EM}), which seeks to provide an optimal point estimate of the hyperparameters in the form of a type-II MAP estimate. The EM consists of cycling through two steps: the E-step inferring the conditional posterior $p(\btheta\mid\by,\bpsi_i)$ and the M-step optimizing hyperparameters $\bpsi_{i+1}$. The E-step is carried out by running the Multiple Adaptively Guided Particle-adjusted Iterative Ensemble Smoother (MAGPIES; Section~\ref{sec:MAGPIES}), a hybrid iterative particle-ensemble Kalman scheme, to infer the conditional posterior. This E-step is the only part of the workflow that involves running a costly ensemble of forward model runs, so MAGPIES was chosen with efficiency in mind. After the E-step, the M-step finds a new hyperparameter estimate $\bpsi_{i+1}$ via optimization based on the Levenberg-Marquardt algorithm (LMA; Section~\ref{app:grad}). The EM continues for a total of $\iota$ iterations, after which it terminates and outputs a final type-II MAP estimate $\boldsymbol{\widehat{\psi}}=\boldsymbol{\psi}_\iota$ as well as the Hessian $\widehat{\mathbf{H}}=\mathbf{H}_\iota$ as a useful by-product of optimization. These two outputs then serve as mean ($\boldsymbol{\widehat{\psi}}$) and covariance ($\widehat{\mathbf{H}}^{-1}$) for an initial Laplace approximation (Section~\ref{sec:Laplace}) of the hyperposterior.

The second branch of the workflow in Figure~\ref{fig:workflow} consists of an amortized Adaptive Particle Batch Smoother (AdaPBS; Section~\ref{sec:AMIS}) to infer the hyperposterior. This instantiation of the AdaPBS scheme is initialized by setting the iteration counter to $R=1$ and defining an initial defensive mixture proposal using the hyperprior and the Laplace approximation. Note that the role of the EM scheme in accelerating AdaPBS is twofold: it both provides a better starting point \emph{and} amortizes inference using the existing MAGPIES generations to provide recycled samples that can be re-used to obtain evidence estimates within AdaPBS at negligible cost. As such, thanks to the recycling of MAGPIES, no additional forward model runs are needed within the amortized AdaPBS loop. Since this loop is effectively free, it can continue iterating until a desired particle diversity threshold is reached, at which point it terminates and returns a particle approximation of the hyperposterior $p(\bpsi\mid\by)$. This hyperposterior can then be combined with MAGPIES-based conditional posterior approximations $p(\btheta\mid\by,\bpsi)$ to marginalize and obtain the marginal posterior $p(\btheta\mid\by)$ of interest.

We also perform independent Particle Markov Chain Monte Carlo (PMCMC; Appendix~\ref{app:PMCMC}) runs in which MAGPIES is used without recycling to obtain improved evidence estimates, serving as a gold-standard yet costlier benchmark against which we compare the performance of the nested particle workflow. The task of estimating the hyperposterior $p(\bpsi\mid\by)$ is generally much more expensive than obtaining the non-hierarchical (typically implicitly) conditional posterior $p(\btheta\mid\by,\bpsi)$, which is relatively straightforward to estimate with ensemble-based DA schemes such as MAGPIES. Our nested workflow is designed primarily to provide an efficient way to estimate this hyperposterior. As an even more computationally efficient benchmark, we use the MAP estimate $\boldsymbol{\widehat{\psi}}=\boldsymbol{\psi}_\iota$ as a cruder plug-in approximation of the hyperposterior. 

The nested workflow in Figure~\ref{fig:workflow} is nested since it involves a particle approximation of the higher level hyperposterior based on a particle approximation of the lower level conditional posterior. It is nonetheless not brute force nested Monte Carlo \citep{Rainforth2018}, as it is guided by EM and leverages recycling via MAGPIES. Thereby, it only involves running on the order of $\Gamma \times N_e \times (N_a+1)$ forward model runs with a small ensemble size $N_e\simeq 10^2$, a handful of ensemble smoother iterations $N_a=4$, and an even lower number of MAGPIES generations $\Gamma = 2$. This is in contrast to a more naive nested particle approach that would require on the order of $N_e^2$ forward model runs, generally with a much larger ensemble size, typically with $N_e\geq 10^3$. An even higher cost is incurred in our intensive PMCMC benchmarking implementation, which requires on the order of $10^8$ forward model runs. The primary contribution of the presented workflow is to provide a tractable approach to hierarchical inference in cryospheric reanalysis. We benchmark this approach by comparing it to non-hierarchical approaches (CP, NP), as well as more intensive (PMCMC) and cruder (MAP) hierarchical approaches. In the remaining method, we present the various schemes used in the workflow and for benchmarking in more detail.  
% Only conceptually nested but not strictly nested since we amortize the hyperposterior AMIS?

\subsection{Iterative Ensemble Smoother (IES)} \label{sec:IES}

At the first parameter level of inference targeting $p(\btheta\mid \by,\bpsi)$, we employ a variant of the particle-adjusted iterative ensemble smoother (PIES) scheme proposed by \citet{Pirk2022}. This scheme is a particle-based importance sampling adjustment of the ensemble Kalman-based iterative ensemble smoother (IES) inspired by related hybrid methods \citep{Papadakis2010,Stordal2015}. Before describing PIES, we outline the IES scheme that we use as a first key calibration step to obtain the proposal distribution in the PIES scheme. 

%and highlight a few modifications compared to the original derivation in \citet{Pirk2022}. 
 Among the many variants of IES schemes \citep{Evensen2022}, we use the ensemble smoother with multiple data assimilation (ES-MDA) proposed by \citet{Emerick2013} due to its ease of implementation, relatively low cost, and robust performance in previous cryospheric data assimilation studies \citep{Aalstad2018,AlonsoGonzalez2022,AlonsoGonzalez2023,AlonsoGonzalez2026,Mazzolini2025}. The ES-MDA scheme is initialized by generating an ensemble of $j=1,\dots,N_e$ parameter vectors from the prior $\btheta_{ni}^{(0j)} \sim p(\btheta\mid \bpsi_i)$, which we store in the $N_\theta \times N_e$ prior parameter ensemble matrix $\boldsymbol{\Theta}_{ni}^{(0)}=\left[ \btheta_{ni}^{(0j)}\right]_{j=1}^{N_e}$ where the superscript $(\cdot)^{(0)}$ is used to denote the prior ensemble. Based on the prior parameter ensemble, we obtain a prior ensemble of predicted observations from the forward model $\widehat{\by}_{ni}^{(0j)}=\mathcal{G}(\btheta_{ni}^{0j})$ that we store in the $N_o\times N_e$ prior predicted observation ensemble matrix $\widehat{\mathbf{Y}}_{ni}^{(0)}=\left[ \widehat{\mathbf{y}}_{ni}^{(0j)}\right]_{j=1}^{N_e}$. Here we emphasize that this prior ensemble is conditional on the choice of hyperparameters $\bpsi_i$ as made explicit through the use of the index $i$ reserved for the hyperparameter configuration. Moreover, we keep the year index $n$ explicit throughout, emphasizing that parameter inference is for a particular DA window corresponding to a water year.

Starting with the prior ensemble, our variant of the ES-MDA cycles through update, resample, and prediction steps for $N_a$ sequential MDA iterations indexed by $\ell=0,\dots,(N_a-1)$. The first step of each iteration updates the parameter ensemble 
\begin{equation*}
    \widetilde{\boldsymbol{\Theta}}_{ni}^{(\ell+1)}= \boldsymbol{\Theta}_{ni}^{(\ell)}+\mathbf{K}_{ni}^{(\ell)}\left(\mathbf{Y}_n^{(\ell)}-\widehat{\mathbf{Y}}_{ni}^{(\ell)}\right) \, ,
\end{equation*}
using the inflated ensemble Kalman gain \citep{Evensen2022}
\begin{equation*}
\mathbf{K}_{ni}^{(\ell)}=\frac{1}{N_e}\boldsymbol{\Theta}_{ni}^{(\ell)'}\widehat{\mathbf{Y}}_{ni}^{(\ell)'\mathrm{T}}\left(\frac{1}{N_e}\widehat{\mathbf{Y}}_{ni}^{(\ell)'}\widehat{\mathbf{Y}}_{ni}^{(\ell)'\mathrm{T}}+\alpha \mathbf{R}\right)^{-1} \, ,
\end{equation*}
predicted observation matrix $\widehat{\mathbf{Y}}_{ni}^{(\ell)}$, and inflated perturbed observation matrix \citep{Emerick2013,vanLeeuwen2020}
\begin{equation*}
\mathbf{Y}_n^{(\ell)}=\by_n\mathbf{1}_{N_e}^{\mathrm{T}}+\sqrt{\alpha}\mathbf{L}_\mathbf{R}\mathbf{Z}_\by^{(\ell)} \, ,
\end{equation*}
where $(\cdot)^{'}$ denotes anomalies from the ensemble mean, $(\cdot)^{\mathrm{T}}$ denotes the transpose, $\mathbf{1}_{N_e}$ is a $N_e \times 1$ column vector of ones, $\alpha=N_a$ is the fixed inflation coefficient, and the lower triangular matrix $\mathbf{L}_\mathbf{R}$ is obtained from the Cholesky decomposition of the observation error covariance matrix $\mathbf{R}=\mathbf{L}_\mathbf{R}\mathbf{L}_\mathbf{R}^\mathrm{T}$. In our case with a diagonal $\mathbf{R}$, this $\mathbf{L}_\mathbf{R}$ is just a diagonal matrix containing the observation error standard deviations. The $N_o\times N_e$ matrix $\mathbf{Z}_\by$ contains independent samples from a standard normal distribution. In the second step of each iteration, we generate a \emph{resampled} updated parameter ensemble matrix $\boldsymbol{\Theta}_{ni}^{(\ell+1)}$ from the multivariate normal distribution $\mathcal{N}(\widetilde{\boldsymbol{\mu}}_{ni}^{(\ell+1)},\widetilde{\boldsymbol{\Sigma}}_{ni}^{(\ell+1)})$
, where $\widetilde{\boldsymbol{\mu}}_{ni}^{(\ell+1)}=\frac{1}{N_e}\widetilde{\boldsymbol{\Theta}}_{ni}^{(\ell+1)}\mathbf{1}_{N_e}$ and $\widetilde{\boldsymbol{\Sigma}}^{(\ell+1)}=\frac{1}{N_e}\widetilde{\boldsymbol{\Theta}}_{ni}^{(\ell+1)'}\widetilde{\boldsymbol{\Theta}}_{ni}^{(\ell+1)'\mathrm{T}}$ are the updated ensemble mean and covariance, respectively. This resampling is achieved through
\begin{equation}
    \boldsymbol{\Theta}_{ni}^{(\ell+1)} =\widetilde{\boldsymbol{\mu}}_{ni}^{(\ell+1)}+\mathbf{L}_{\boldsymbol{\Sigma}}^{(\ell+1)}\mathbf{Z}_{\btheta}^{(\ell+1)} \, , \label{eq:cholies}
\end{equation}
where the lower triangular matrix $\mathbf{L}_{\boldsymbol{\Sigma}}^{(\ell+1)}$ is obtained from a Cholesky decomposition of the updated ensemble covariance matrix $\widetilde{\boldsymbol{\Sigma}}_{ni}^{(\ell+1)}=\mathbf{L}_{\boldsymbol{\Sigma}}^{(\ell+1)}\mathbf{L}_{\boldsymbol{\Sigma}}^{(\ell+1)\mathrm{T}}$, and the $N_\theta \times N_e$ matrix $\mathbf{Z}_{\btheta}^{(\ell+1)}$ contains independent samples from a standard normal distribution. In the third and final step of each iteration, we perform the prediction step to obtain the updated predicted observations from the forward model $\widehat{\by}_{ni}^{((\ell+1)j)}=\mathcal{G}(\btheta_{ni}^{((\ell+1)j)})$ that we store in the $N_o\times N_e$ updated predicted observation ensemble matrix $\widehat{\by}_{ni}^{(\ell+1)}$. Once we have cycled through $N_a$ iterations of these three steps, we end up with $\boldsymbol{\Theta}_{ni}^{(N_a)}$, along with the associated ensemble of states and predicted observations that represent the posterior estimate for water year $n$ and hyperparameter setting $i$ from this version of the ES-MDA scheme. 

We highlight that, although the ES-MDA assimilates the same data several times via `multiple data assimilation' (MDA), it does so in a controlled manner with observation errors that are inflated to ensure Bayesian consistency via the constraint $\sum_{\ell=1}^{N_a}1/\alpha=1$. This is a special case of the more general technique of likelihood tempering \citep{Murphy2023} that makes the ES-MDA more robust than the non-iterative ES for nonlinear models \citep{Aalstad2018,AlonsoGonzalez2022,Pirk2022}. Here we use $N_a=4$ MDA iterations, which have been found to provide robust performance in many nonlinear data assimilation problems while maintaining computational tractability \citep{Emerick2013,Aalstad2018,Pirk2022,Keetz2025}. The resampling step is not strictly necessary and is rather unconventional in the IES, but it ensures the validity of the subsequent importance sampling described in Section~\ref{sec:PIES} while maintaining robust performance \citep{AlGhattas2024}.

\subsection{Particle-adjusted Iterative Ensemble Smoother (PIES)}\label{sec:PIES}

The PIES scheme \citep{Pirk2022} uses the IES to build a more optimal proposal distribution for importance sampling \citep{vanLeeuwen2019,Rainforth2020}. To clarify, we provide a quick recap of importance sampling \citep{Chopin2020,Sarkka2023,Murphy2023} that is also at the core of the nested particle methods used in our hierarchical inference workflow. Keeping indices implicit, following \eqref{eq:postex}, the expectation of a possibly vector-valued function $u(\btheta)$ with respect to the conditional parameter posterior $p(\btheta\mid \by, \bpsi)$ is defined as
\begin{equation}
    \mathbb{E}[u(\btheta)\mid \by, \bpsi]=\int u(\btheta) p(\btheta\mid \by, \bpsi) \, \mathrm{d}\btheta \, , \label{eq:expectation}
\end{equation}
where once more the expectation of $u(\btheta)=\btheta$ corresponds to the conditional posterior mean and so on (Section~\ref{sec:reanalysis}). Now, if we could generate an ensemble of $N_e$ independent samples from the posterior $\btheta^{(j)}\sim p(\btheta\mid \by, \bpsi)$, it would be straightforward to estimate such expectations via direct Monte Carlo integration
\begin{equation*}
\mathbb{E}[u(\btheta)\mid \by, \bpsi] \simeq \frac{1}{N_e} \sum_{j=1}^{N_e} u(\btheta^{(j)})  \, ,
\end{equation*}
due to the law of large numbers and the central limit theorem, this unbiased estimate will converge almost surely, with a Monte Carlo variance inversely proportional to $N_e$ \citep{Robert2004}. The catch is that, except for special cases, we generally cannot directly generate independent samples from the intractable target posterior distribution of interest. In fact, at least with Monte Carlo (ensemble) methods, obtaining such samples is the main goal and key challenge of inference \citep{MacKay2003}.

Importance sampling helps us generalize Monte Carlo integration to cases with intractable target distributions. Defining a simpler proposal distribution $q(\btheta)$ with at least the same support as the target posterior, we use $q(\btheta)/q(\btheta)=1$ to write \eqref{eq:expectation} as
\begin{equation*}
    \mathbb{E}[u(\btheta)\mid \by, \bpsi]=\int \frac{u(\btheta) p(\btheta\mid \by, \bpsi)}{q(\btheta)}q(\btheta) \, \mathrm{d}\btheta \, ,
\end{equation*}
so we can use $N_e$ independent samples from the simpler proposal distribution $\btheta^{(j)}\sim q(\btheta)$ and Monte Carlo integration to obtain the importance sampling estimate
\begin{equation}
    \mathbb{E}[u(\btheta)\mid \by, \bpsi] \simeq \frac{1}{N_e} \sum_{j=1}^{N_e} \frac{u(\btheta^{(j)})p(\btheta^{(j)} \mid \by,\bpsi)}{q(\btheta^{(j)})} \, ,  \label{eq:IS}
\end{equation}
which appears tractable since we now only need to evaluate the posterior point-wise $p(\btheta^{(j)} \mid \by,\bpsi)$ for each of the $N_e$ samples. Note that the optimal proposal is the posterior, for which we recover direct Monte Carlo sampling \citep{MacKay2003}. An obstacle remains in that we are typically only able to perform exact point-wise evaluations of the unnormalized posterior $p(\by\mid \btheta)p(\btheta \mid \bpsi)$ (i.e., the joint $p(\btheta,\by \mid \bpsi)$), since the evidence $p(\by \mid \bpsi)$ in \eqref{eq:evidence} involves an intractable integral. As a workaround, we use importance sampling to estimate the evidence itself by inserting $q(\btheta)/q(\btheta)$ into the integrand of \eqref{eq:evidence} 
\begin{equation*}
    p(\by\mid \bpsi) = \int \frac{p(\by \mid \btheta)p(\btheta \mid \bpsi)}{q(\btheta)}q(\btheta) \, \mathrm{d}\btheta \, , 
\end{equation*}
and use $\btheta^{(j)}\sim q(\btheta)$ to obtain the importance sampling estimate of the evidence
\begin{equation}
p(\by\mid \bpsi) \simeq \widehat{Z} =\frac{1}{N_e}\sum_{j=1}^{N_e}  \frac{p(\by \mid \btheta^{(j)})p(\btheta^{(j)} \mid \bpsi)}{q(\btheta^{(j)})} = \frac{1}{N_e}\sum_{j=1}^{N_e} \widetilde{w}^{(j)} \, , \label{eq:ISevi}
\end{equation}
where we introduce $\widehat{Z}$ as a convenient shorthand for this evidence estimate and define the unnormalized weights $\widetilde{w}^{(j)}$ as. Using the evidence approximation $\widehat{Z}\simeq p(\by \mid \bpsi)$, we can obtain a self-normalized importance sampling estimate \citep{Rainforth2020} of the posterior expectation by inserting for $p(\by \mid \btheta)p(\btheta)/\widehat{Z} \simeq p(\btheta \mid \by)$ to approximate the point-wise posterior in \eqref{eq:IS}
\begin{equation}
    \mathbb{E}[u(\btheta)\mid \by, \bpsi] \simeq \frac{1}{N_e} \sum_{j=1}^{N_e} \frac{u(\btheta^{(j)})p(\by \mid \btheta^{(j)})p(\btheta^{(j)} \mid \bpsi)}{\widehat{Z} q(\btheta^{(j)})} = \sum_{j=1}^{N_e} w^{(j)}u(\btheta^{(j)}) \, ,  \label{eq:SNIS}
\end{equation}
where we have used \eqref{eq:ISevi} to define the normalized weights
\begin{equation}
    w^{(j)}=\frac{\widetilde{w}^{(j)}}{\sum_{k=1}^{N_e}\widetilde{w}^{(k)}} \, , \label{eq:normw}
\end{equation}
which sum to one and can thus be seen as representing the probability mass associated with each sample $\btheta^{(j)}$. This corresponds to a particle approximation of the posterior
\begin{equation}
    p(\btheta\mid \by, \bpsi) \simeq \sum_{j=1}^{N_e} w^{(j)}\delta(\btheta-\btheta^{(j)}) \label{eq:parapp} \, ,
\end{equation}
where $\delta(\cdot)$ is the Dirac delta with the key properties that $\int \delta(\btheta-\btheta^{(j)})\,\mathrm{d}\btheta=1$ and $\int u(\btheta)\delta(\btheta-\btheta^{(j)})\,\mathrm{d}\btheta=u(\btheta^{(j)})$. As such, the continuous posterior density is approximated as a discrete set of particles at the sampled locations $\btheta^{(j)}$ in parameter space, whose probability mass is given by their weights. We use the word particle to refer to the set consisting of a sampled parameter vector $\btheta^{(j)}$, the corresponding model state $\mathbf{x}^{(j)}=\mathcal{M}(\btheta^{(j)})$, predicted observations $\widehat{\by}^{(j)}=\mathcal{H}(\mathbf{x}^{(j)})=\mathcal{G}(\btheta^{(j)})$, and the associated weight $w^{(j)}$ obtained via importance sampling using an ensemble of $j=1,\dots,N_e$ particles. 

Building on the idea of importance sampling as the foundation of particle methods, the particle adjustment in the PIES scheme is importance sampling using the IES to construct the proposal distribution $q(\btheta)$. The idea is simple: the posterior approximation from the IES will generally be closer to the optimal proposal (i.e., the posterior) than basic importance sampling with the prior as the proposal. To obtain a proposal from the IES, we can use the resampled ensemble members $\btheta_{ni}^{(N_a j)}\sim q_{ni}^{(N_a)}$ and the corresponding normal distribution $q_{ni}^{(N_a)}(\btheta)=\mathcal{N}(\btheta \mid \widetilde{\boldsymbol{\mu}}_{ni}^{(N_a)},\widetilde{\boldsymbol{\Sigma}}_{ni}^{(N_a)})$ from the final MDA step $\ell=N_a$ of the IES \eqref{eq:cholies}. With this choice of proposal, the unnormalized PIES weights are 
\begin{equation}
    \widetilde{w}_{n}^{(j)}(\bpsi_i)=\frac{p(\by_{n}\mid \btheta_{ni}^{(N_a j)})p(\btheta_{ni}^{(N_a j)}\mid \bpsi_i)}{q_{ni}^{(N_a)}(\btheta_{ni}^{(N_a j)})} \, , \label{eq:PIES}
\end{equation}
which are then normalized analogously to \eqref{eq:normw} to obtain the PIES  particle approximation \eqref{eq:parapp} of the conditional parameter posterior. The water year $n$ and hyperparameter $i$ indices in \eqref{eq:PIES} are a reminder that we will use such a particle update in a manner that is nested within the inferential dynamics described in Section~\ref{sec:dynamics}.

\subsection{Multiple Adaptively Guided PIES (MAGPIES)} \label{sec:MAGPIES}

The PIES scheme provides a robust choice for the first parameter level of inference, achieving good performance while maintaining an affordable cost of $(N_a+1)N_e$ parallelizable forward model runs. We expect PIES to improve the IES at no extra cost \citep{Papadakis2010} while outperforming basic importance sampling methods \citep{Pirk2022}. However, using PIES with a proposal based solely on the final IES iteration is inefficient in the hierarchical Bayesian inference setting for two reasons. 

Firstly, the proposal in PIES only makes explicit use of the $N_e$ particles resampled from the final iteration of IES. Recall that IES generates a history of $(N_a+1)\times N_e$ particles sampled either from the prior or one of the $N_a$ subsequent MDA iterations. There is generally no reason, especially with $N_a=4$, to expect that only the $N_e$ particles from the final iteration of IES are informative. As such, it is wasteful to discard the history of $N_a N_e$ particles and the information contained in their forward model runs and the associated likelihood evaluations. 

Secondly, the hyperparameter setting $\bpsi_i$ enters the proposal, obtained by running the IES with prior hyperparameters $\bpsi_i$, and the prior for which we are computing the weights in \eqref{eq:PIES}. With PIES, we thus need to run a new IES with $(N_a+1)N_e$ costly forward model runs to generate a new proposal for each new setting of $\bpsi_i$. Theoretically, this approach is promising, in that an IES proposal obtained with hyperparameters $\bpsi_i$ will be closer to the optimal proposal. In practice, however, this becomes expensive when sampling multiple hyperparameter settings in hierarchical inference. Instead, we can just reuse existing historical proposals constructed using similar hyperparameter settings to the $\bpsi_i$ under consideration and only rerun the IES when strictly needed. Thereby, approximate hierarchical inference is possible at essentially no extra cost \citep{Viani2023}.

We tackle these sources of inefficiency in the PIES scheme for hierarchical inference via the novel Multiple Adaptively Guided Particle-adjusted Iterative Ensemble Smoother (MAGPIES) scheme. This is a simple extension of the PIES scheme that uses the entire history of particles generated from multiple IES runs that are adaptively guided towards promising hyperparameter configurations. The MAGPIES scheme is based on Adaptive Multiple Importance Sampling \citep[AMIS;][]{Cornuet2012}, as also used by the AdaPBS scheme \citep{Aalstad2026} in Section~\ref{sec:AMIS}, leveraging multiple proposals that are adaptively guided towards the target distribution. This falls under the umbrella of adaptive importance sampling \citep{Bugallo2017}, but is distinguished by its use of a deterministic mixture proposal \citep[DMP;][]{Owen2000} that avoids wasting particles and lowers Monte Carlo variance. Unlike the original AMIS scheme, which adapts the DMP via importance sampling, the MAGPIES scheme uses multiple IES runs with guided hyperparameter configurations to adapt the DMP. By leveraging IES, MAGPIES makes implicit use of the likelihood tempering procedure that has been shown to be advantageous for hierarchical inference \citep{Viani2023,Llorente2023}, while maintaining a tractable computational cost.

Building on PIES in \eqref{eq:PIES}, the unnormalized weights in the MAGPIES scheme are 
\begin{equation}
\widetilde{w}_{ng}^{(\ell j)}(\bpsi_i)=\frac{p(\by_{n}\mid \btheta_{ng}^{(\ell j)})p(\btheta_{ng}^{(\ell j)}\mid \bpsi_i)}{\upsilon_{nG}(\btheta_{ng}^{(\ell j)})} \, ,\label{eq:MAGPIES}
\end{equation}
where the ensemble of parameter vectors $\btheta_{ng}^{(\ell j)}$ constitutes the entire currently available history of particles, which are recast as samples from a multiple IES-based DMP $\btheta_{ng}^{(\ell j)}\sim \upsilon_{n G}(\btheta)$ of the form
\begin{equation}
    \upsilon_{nG}(\btheta)=\sum_{\kappa=1}^{G}\sum_{k=0}^{N_a}\frac{1}{ (N_a+1)G}q_{n\kappa}^{(k)}\left(\btheta \mid \bpsi_\kappa^\star \right) \, , \label{eq:MAGP}
\end{equation}
where the dummy indices $\kappa$ and $k$ correspond to $g$ and $\ell$, respectively, in 
\begin{equation*}
    q_{ng}^{(\ell)}\left(\btheta \mid \bpsi_g^\star \right)=\mathcal{N}\left(\btheta \mid \widetilde{\boldsymbol{\mu}}_{ng}^{(\ell)},\widetilde{\boldsymbol{\Sigma}}_{ng}^{(\ell)}\right) \, ,
\end{equation*}
which are Gaussian proposals from successive generations $g$ of running the IES with iterations $\ell$ and hyperparameters $\bpsi_g^\star$. The subscript $G$ in the DMP $\upsilon_{nG}$ defined in \eqref{eq:MAGP} denotes the current total number of generations $g=1,\dots,G$ of IES, each based on a sequentially adapted hyperparameter setting $\bpsi_g^\star$ and $\ell=0,\dots,N_a$ MDA iterations that together make up this current mixture proposal. In effect, this proposal is just an adaptive mixture of $(N_a+1)G$ Gaussians. Note that we can reuse all generations of sampled particles $\btheta_{ng}^{(\ell j)}$ and costly likelihood evaluations in \eqref{eq:MAGPIES}. As such, for each water year, MAGPIES can provide  a history of $N_h=N_e(N_a+1)G$ particle weights without having to rerun any additional ensembles of forward model runs, by instead leveraging those that have already been carried out in the existing generations of IES. Crucially, this means that we can freely re-evaluate the MAGPIES weights in \eqref{eq:MAGPIES} as a function of  arbitrary new hyperparameters $\bpsi$ allowing us to perform approximate hyperparameter optimization and inference without having to rerun the forward model. Reruns of the forward model only occur if we decide to update the current total number of generations that the IES scheme has been run, i.e., by setting $G\leftarrow G+1$. This can be done adaptively using the effective sample size for the weights obtained using MAGPIES as part of the hyperparameter inference described in the following sections. In practice, as sketched in Figure~\ref{fig:workflow}, we run MAGPIES for a total of $\Gamma=2$ generations. The first generation ($G=1$) uses $\bpsi_1^\star=\bpsi_0=\mathbf{m}$ set to the hyperprior mean in an IES run to build the proposal \eqref{eq:MAGP} and recycles this proposal in MAGPIES \eqref{eq:MAGPIES} for $10$ EM iterations (Section~\ref{sec:EM}). The second generation ($G=2$) then uses $\bpsi_2^\star=\bpsi_{10}$, obtained from the tenth EM iteration, in a new IES run to extend the proposal \eqref{eq:MAGP} and recycles this proposal in MAGPIES \eqref{eq:MAGPIES} for another $10$ EM iterations, for a total of $\iota=20$ EM iterations. This $\Gamma=2$ generation MAGPIES construction is then reused in its entirety, without new model runs, to infer the hyperparameters via an amortized AdaPBS (Section~\ref{sec:AMIS}).

To reduce the proliferation of subscripts and superscripts at the hyperparameter level of inference, we will collapse multiple indices, namely the particle $j=1,\dots,N_e$, multiple data assimilation $\ell=0,\dots,N_a$, and generation $g=1,\dots,G$ indices, into a single particle history index $h=1,\dots,N_h$. This single index runs over the entire current history of available $N_h=(N_a+1)G N_e$ particles, allowing us to write $\btheta_{n}^{(h)}$ in lieu of $\btheta_{ng}^{(\ell j)}$, where we can map between this single particle history index and the aforementioned multiple indices via $h(j,\ell,g)=(g-1)(N_a+1)N_e+\ell N_e+j$.

Adopting this new particle history index, the MAGPIES annual conditional posterior is estimated using the particle approximation
\begin{equation} 
p(\btheta_n\mid \by_n,\bpsi_i)\simeq \sum_{h=1}^{N_h} w_{ni}^{(h)}\delta(\btheta_n-\btheta_n^{(h)}) \, , \label{eq:MAGpost}
\end{equation}
where the weights in \eqref{eq:MAGPIES} are normalized in the usual way outlined in \eqref{eq:normw} via
\begin{equation}
w_{ni}^{(h)}=\frac{\widetilde{w}_{ni}^{(h)}}{\sum_{k=1}^{N_h} \widetilde{w}_{ni}^{(k)}} \, , \label{eq:MAGnorm}
\end{equation}
in which we have inserted for $\widetilde{w}_{ni}^{(h)}=\widetilde{w}_{ng}^{(\ell j)}(\bpsi_i)$ in \eqref{eq:MAGPIES} using the single particle history index $h=h(j,\ell,g)$. Moreover,
 the annual evidence in \eqref{eq:annualevi} can be approximated via
\begin{equation}
    p(\by_n\mid \bpsi_i) \simeq \widehat{Z}_{ni} = \frac{1}{N_h}\sum_{k=1}^{N_h} \widetilde{w}_{ni}^{(k)} \, , \label{eq:MAGevi} 
\end{equation}
which is already used implicitly in the normalization step \eqref{eq:MAGnorm}. We re-emphasize that these MAGPIES approximations of the annual conditional posterior and the evidence can be computed for any new setting of the hyperparameters $\bpsi_i$ at essentially no cost. Since $\bpsi_i$ only enters through the closed-form prior, no additional forward model runs are required: we can store and reuse the entire currently available particle history $\btheta_{n}^{(h)}=\btheta_{ng}^{(\ell j)}$ and the associated costly likelihood evaluations.  The fact that these MAGPIES weights are functions of the hyperparameter vector $\bpsi_i$ is symbolized implicitly through the index $i$ in $\widetilde{w}_{ni}^{(h)}$. 

\subsection{Expectation maximization (EM)} \label{sec:EM}

To accelerate hierarchical inference, we use the expectation maximization (EM) algorithm \citep{Dempster1977,Neal1998} to rapidly obtain a maximum a posteriori (MAP-II) estimate \citep{Murphy2023} of the hyperparameters $\boldsymbol{\widehat{\psi}}=\mathrm{argmax}_{\bpsi}\,p(\bpsi\mid \by)$. The EM algorithm is an iterative approach for maximum (marginal) likelihood estimation in general state space modeling \citep{Sarkka2023}, including geophysical data assimilation \citep{Pulido2018,Lucini2021}, which we have adapted to MAP-II estimation. This algorithm has the advantage that it does not require direct access to the typically intractable function that is being optimized, in this case $p(\bpsi\mid \by)$, while still ensuring convergence, often with relatively few iterations. Herein, we outline how we implement the EM to obtain the MAP-II estimate $\boldsymbol{\widehat{\psi}}$ and refer to \citet{Neal1998} for a detailed justification of EM.

As the name implies, the EM algorithm proceeds by iteratively cycling between two steps: the expectation step (E-step) followed by the maximization step (M-step). These two steps are carried out sequentially over a set of $\iota$ iterations indexed by $i=0,\dots,\iota$ starting with an initial guess for the hyperparameters for which we use the hyperprior mean vector $\bpsi_0=\mathbf{m}$. The result of the M-step in the last iteration is then taken as the MAP-II hyperparameter estimate $\boldsymbol{\widehat{\psi}}=\bpsi_{\iota}$. 

It turns out that we are already performing the E-step in typical data assimilation, since this involves inferring the usual parameter posterior  $p(\btheta\mid \by,\bpsi_i)$ conditioned on the current estimate of the hyperparameters $\bpsi_i$. As such, the E-step amounts to running the usual ensemble-based data assimilation to infer the (local) parameter posterior for each year. Using the factorization in \eqref{eq:cpfac}, we define the following particle approximation $q_i(\btheta\mid \bpsi_i)$ of the parameter posterior 
\begin{equation}
p(\btheta\mid \by,\bpsi_i) \simeq q_i(\btheta\mid \bpsi_i) = \prod_{n=1}^N \sum_{h=1}^{N_h} w_{ni}^{(h)}\delta \left( \btheta_n-\btheta_{n}^{(h)}\right) \, , \label{eq:parq}
\end{equation}
where the year indexing on $p(\btheta\mid \by,\bpsi_i)=p(\btheta_{1:N}\mid \by_{1:N},\bpsi_i)$ is implicit for economy and
\begin{equation*}
p(\btheta_n\mid \by_n,\bpsi_i)\simeq  q_{ni}(\btheta,\bpsi_i)=\sum_{h=1}^{N_h} w_{ni}^{(h)}\delta \left( \btheta_n-\btheta_{n}^{(h)}\right) \, ,
\end{equation*}
is the particle approximation of the conditional posterior for year $n$ with hyperparameter $\bpsi_i$. Note that here the MAGPIES weights $w_{ni}^{(h)}$ from \eqref{eq:MAGnorm} of the $h=1,\dots,N_h$ historical particles at locations $\btheta_{n}^{(h)}$ in parameter space depend on the water year $n$ and hyperparameter $i$ under consideration. Without loss of generality, these sum-to-one weights $\sum_{h=1}^{N_h}w_{ni}^{(h)}=1 $ will either be varying $w_{ni}^{(h)}\in[0,1]$ or can be made equal $w_{ni}^{(h)}=1/N_h$ through a resampling step. Other approximations  $q_i(\btheta\mid \bpsi_i)$ for the conditional parameter posterior $p(\btheta\mid \by,\bpsi_i)$ could be used. Herein, we restrict our attention to the MAGPIES particle approximation from \eqref{eq:MAGpost} using \eqref{eq:MAGnorm} as a robust, accurate, and computationally tractable choice for our moderately dimensional nonlinear problem based on other `particle EM' applications \citep{Lucini2021,Sarkka2023}. 

The subsequent M-step is considerably more involved and is where the hyperparameter optimization occurs. As will become clear, for this MAP-II application of the EM algorithm, it is instructive to define the objective function to be maximized as
\begin{equation}
    f_i(\bpsi)=F(q_i,\bpsi)+\log(p(\bpsi)) \, \label{eq:fdef} \, ,
\end{equation}
where the negative variational free energy \citep{Parr2022} is given by
\begin{equation*}
    F(q_i,\bpsi)=\int q_i(\btheta\mid \bpsi_i)\log\left(\frac{p(\btheta,\by\mid \bpsi)}{q_i(\btheta\mid \bpsi_i)}\right) \, \mathrm{d}\btheta , %\label{eq:F}
\end{equation*}
which can be written as
\begin{equation}
     F(q_i,\bpsi) = \mathcal{Q}(\bpsi,\bpsi_i)+H(q_i) \, , \label{eq:Fdef}
\end{equation}
where the entropy term
\begin{equation}
    H(q_i)=-\int q_i(\btheta\mid \bpsi_i)\log(q_i(\btheta\mid \bpsi_i)) \, \mathrm{d}\btheta \, , \label{eq:entropy}
\end{equation}
is independent of $\bpsi$ and the $\mathcal{Q}$ function is defined as the expectation
\begin{equation}
    \mathcal{Q}(\bpsi,\bpsi_i)=\int q_i(\btheta\mid \bpsi_i)\log\left(p(\btheta,\by\mid \bpsi)\right) \, \mathrm{d}\btheta \, . \label{eq:Qdef} 
\end{equation}
We note in passing that by inserting $p(\btheta,\by\mid \bpsi)=p(\btheta\mid \by,\bpsi)p(\by\mid \bpsi)$ into \eqref{eq:Fdef}, $F$ can also be reformulated as
\begin{equation}
F(q_i,\bpsi)=-D_\mathrm{KL}\left(q_i(\btheta\mid \bpsi_i)\mid \mid p(\btheta\mid \by,\bpsi)\right)+\log\left(p(\by\mid \bpsi)\right) \, , \label{eq:ELBO}
\end{equation}
where the Kullback-Leibler divergence (relative entropy) term is given by
\begin{equation}
D_\mathrm{KL}\left(q_i(\btheta\mid \bpsi_i)\mid \mid p(\btheta\mid \by,\bpsi)\right)=\int q_i(\btheta\mid \bpsi_i)\log\left(\frac{q(\btheta\mid \bpsi_i)}{p(\btheta\mid \by,\bpsi)}\right) \, \mathrm{d}\btheta \, , \label{eq:DKL}
\end{equation}
which is an asymmetric non-negative measure of the `distance' between two distributions that is only zero when the distributions are identical. As such, the first term on the right hand side of \eqref{eq:ELBO} is always $\leq0$ implying that $F(q_i,\bpsi)\leq \log\left(p(\by\mid \bpsi)\right)$ where $p(\by\mid \bpsi)$ is the evidence term \eqref{eq:evidef}. The negative variational free energy is thus also known as the \emph{evidence lower bound} (ELBO for short) and forms the basis of variational Bayesian inference \citep{Parr2022,Murphy2023}. Therein, the ELBO is used as the objective when optimizing the variational parameters (which need not be $\bpsi$) of a variational distribution $q$ that is used to approximate the posterior. In our case, we use the ELBO to note that $f_i$, the sum of $F(q_i,\bpsi)$ (where $q_i$ is parametrized by $\bpsi_i$) and $\log(p(\bpsi))$, provides a tractable lower bound on the log hyperposterior $\log(p(\bpsi\mid \by))$ where the tightness of the bound will depend on the veracity of the particle approximation $q_i(\btheta\mid \bpsi_i)$ to the true posterior in \eqref{eq:parq}. Defining the energy function for the hyperposterior as 
\begin{equation}
    \phi(\bpsi)=-\log(p(\bpsi\mid \by)) \, , \label{eq:energy}
\end{equation}
using the lower bound from \eqref{eq:ELBO} in \eqref{eq:fdef} and the product rule of probability, we obtain
\begin{equation}
   \phi(\bpsi) \leq  -f_i(\bpsi) + \log\left(p(\mathbf{y})\right)  \, ,
\end{equation}
to justify the approximation $-f_i(\bpsi)+\log\left(p(\mathbf{y})\right)\simeq \phi$ when the bound is tight. Thereby, using that $p(\by)$ is constant with respect to $\bpsi$, minimizing $\phi$ amounts to maximizing $f_i$ so that
\begin{equation}
\frac{\partial \phi }{\partial \bpsi } \simeq-\frac{\partial f_i}{\partial \bpsi} \, , \label{eq:phig}
\end{equation}
where using the definition of $f_i$ in \eqref{eq:fdef} and that the entropy \eqref{eq:entropy} is independent of $\bpsi$
\begin{equation}
\frac{\partial \phi}{\partial \bpsi} \simeq  -\frac{\partial\mathcal{Q}}{\partial \bpsi}-\frac{\partial \log(p(\bpsi))}{\partial \bpsi} \, , \label{eq:Fisheri}
\end{equation}
which relates the gradient of the energy function to the gradient of the expectation $\mathcal{Q}$ and the gradient of the (log) hyperprior \citep{Sarkka2023}. Now we are in a position to use the tractable expression on the right hand side of \eqref{eq:Fisheri} as a guide to gradient-based minimization of the energy function for the maximization step to obtain the updated MAP-II estimate $\boldsymbol{\widehat{\psi}}\simeq \bpsi_{i+1}=\mathrm{argmax}_{\bpsi}\, f_i(\bpsi)$ for the next iteration $i+1$. In practice, we perform this minimization using a second-order gradient-based optimization method based on the Levenberg-Marquardt algorithm \citep{Pujol2007} with an analytical gradient vector and Hessian matrix as described in~\ref{app:grad}.

Wrapping up  the EM algorithm \citep{Neal1998}, we emphasize that even if $q_i(\btheta\mid \bpsi_i)$ were the true posterior and not an approximation, it is conditioned on the current $\bpsi_i$ and not the updated $\bpsi_{i+1}$, so after the M-step the divergence in \eqref{eq:DKL} is $D_\mathrm{KL}\left(q_i(\btheta\mid \bpsi_i)\mid \mid p(\btheta\mid \by,\bpsi_{i+1})\right)$. This implies that even if $q_i$ in \eqref{eq:parq} was not an approximation, we generally still have a non-zero divergence and thus only a lower bound on the MAP-II hyperparameter unless $\bpsi_{i}=\bpsi_{i+1}$. This is why new iterations $i\leftarrow i+1$ with new expectation steps, where we update $q_{i}$ by conditioning on updated hyperparameters $\bpsi_i$, are also a key part of the EM algorithm. The flip side of the divergence argument is that if $\bpsi_i$ does not change noticeably from one EM iteration to the next, then this is a sign that the algorithm has converged to a (local) maximum. While this suggests using an adaptive stopping criterion, it is easier to plan computational loads by using a fixed yet relatively large maximum number of $\iota=20$ iterations to help ensure convergence for the problem at hand.
% Link to 1053-5888/08/$25.00©2008IEEE variational inference and EM paper. As well as the Lindsten paper on particle EM

\subsection{Laplace approximation (LA)} \label{sec:Laplace} 

The MAP-II Hessian matrix $\widehat{\mathbf{H}}$ turns out to be a useful estimate of the hyperposterior precision (inverse covariance) matrix that can be paired with the MAP-II point estimate $\boldsymbol{\widehat{\psi}}$ of the mean to form a Laplace approximation \citep{MacKay2003} of the hyperposterior, which we denote as $q_\mathcal{L}(\bpsi)$. Specifically, this Laplace approximation of the hyperposterior $p(\bpsi\mid \by)\simeq  q_\mathcal{L}(\bpsi)=\mathcal{N}(\bpsi\mid \boldsymbol{\widehat{\psi}},\widehat{\mathbf{H}}^{-1})$ is a multivariate Gaussian (normal) distribution with mean $\boldsymbol{\widehat{\psi}}$ and precision $\widehat{\mathbf{H}}$ of the form
\begin{equation}
    q_\mathcal{L}(\bpsi)=\sqrt{\frac{\det(\widehat{\mathbf{H}})}{(2\pi)^{N_\psi}}}\exp\left(-\frac{1}{2}\left[\bpsi-\boldsymbol{\widehat{\psi}}\right]^\mathrm{T} \widehat{\mathbf{H}}\left[\bpsi-\boldsymbol{\widehat{\psi}}\right]\right) \, , \label{eq:precnorm}
\end{equation}
where the (Hessian-based) precision $\widehat{\mathbf{H}}$ matrix is just the inverse of the usual Gaussian covariance matrix. Other than the negligible cost of computing $\widehat{\mathbf{H}}$ once the MAP-II estimate has been $\boldsymbol{\widehat{\psi}}$ obtained, this Laplace approximation turns out to be an essentially free by-product of our instantiation of the EM algorithm. Herein, we employ this Laplace approximation $q_\mathcal{L}(\bpsi)$ as a proposal distribution when forming a particle approximation of the hyperposterior through importance sampling. As such, the proposed EM-powered particle Laplace approximation approach provides a map to guide the design of the proposal when applying importance sampling to the hyperparameters $\bpsi$.
Note that \citet{Schillings2020} established the theoretically robust properties of such a hybrid approach and its superiority compared to basic importance sampling for inverse problems.

Having obtained the Laplace approximation proposal $q_\mathcal{L}(\bpsi)$, this hybrid approach proceeds by first sampling from the proposal $\bpsi_i \sim q_\mathcal{L}(\bpsi)$. To do so, we use a Cholesky decomposition of the precision matrix $\widehat{\mathbf{H}}=\mathbf{L}_{\widehat{\mathbf{H}}}\mathbf{L}_{\widehat{\mathbf{H}}}^\mathrm{T}$ \citep{Rue2001} and solve the shifted linear system  $\mathbf{L}_{\widehat{\mathbf{H}}}^\mathrm{T}\left(\bpsi_i-\boldsymbol{\widehat{\psi}}\right)=\mathbf{z}_i$ for $\bpsi_i$, where the $N_\psi \times 1$ vector $\mathbf{z}_i$ contains independent samples from a standard normal distribution. We use a triangular solver for $i=1,\dots,N_s$ simultaneously by using an $N_\psi\times N_s$ matrix of standard normal samples $\mathbf{Z}$.

Using the Laplace approximation directly as the proposal, we could obtain a particle approximation of the hyperposterior by computing the unnormalized weights
\begin{equation*}
    \widetilde{\omega}_i = \frac{p(\by\mid \bpsi_i)p(\bpsi_i)}{q_\mathcal{L}(\bpsi_i)} \, ,
\end{equation*}
where the Laplace proposal $q_\mathcal{L}$ is given by \eqref{eq:precnorm} and the joint evidence $p(\by\mid \bpsi_i)$ is approximated by combining the factorization in \eqref{eq:jevi} with the MAGPIES estimate \eqref{eq:MAGevi} via
\begin{equation}
    p(\by \mid \bpsi_i)\simeq \widehat{Z}_{i} = \prod_{n=1}^N \widehat{Z}_{ni} \, , \label{eq:MAGPIESjevi}
\end{equation}
and the hyperprior $p(\bpsi_i)$ is known in closed form. Note that we have chosen to use the distinct symbols $\omega$ and $w$ to denote the outer hyperparameter weights and the inner parameter weights, respectively, to help distinguish these two nested stages of particle-based inference. The self-normalized weights are obtained in the usual way via 
\begin{equation*}
    \omega_i = \frac{\widetilde{\omega}_i}{\sum_{k=1}^{N_s}\widetilde{\omega}_k} \, , 
\end{equation*}
which provides a particle Laplace approximation of the hyperposterior of the form
\begin{equation*}
    p(\bpsi\mid \by) \simeq \sum_{i=1}^{N_s} \omega_i \delta\left(\bpsi-\bpsi_i\right) \, .
\end{equation*}
As usual, these hyperposterior weights are first obtained in log-space to ensure numerical stability \citep{Chopin2020}. In practice, this particle Laplace approximation serves as part of a key first step in an adaptive importance sampling strategy \citep{Aalstad2026,Cornuet2012} described in Section~\ref{sec:AMIS}. Notably, using the MAP-II guided Laplace approximation in the proposal, rather than just the more basic prior proposal, already in the first step of adaptive importance sampling can speed up convergence to the desired level of particle diversity in the outer hyperparameter ensemble and thus reduce computational costs substantially.

\subsection{Adaptive Particle Batch Smoother (AdaPBS)} \label{sec:AMIS}

Together, the particle EM and Laplace approximation provide a guided initial proposal that is useful for importance sampling across annual windows at the hyperparameter level of inference. At the same time, the possibility of degenerate hyperparameter inference remains high even with this proposal. This is because the Laplace posterior approximation used as the proposal can be biased and overconfident. Therefore, the importance sampling estimator for the hyperposterior may have considerable variance that can result in particle degeneracy. As a first step to alleviate this, we use a defensive mixture approach \citep{Hesterberg1995,Willmes2025} by sampling from the proposal 
\begin{equation}
    q^{(1)}(\bpsi)=W p(\bpsi)+(1-W)q_\mathcal{L}(\bpsi) \, , \label{eq:defensive}
\end{equation}
which is a mixture of the hyperprior $p(\bpsi)$ and the Laplace approximation $q_\mathcal{L}(\bpsi)$, where, for simplicity, we set equal mixture weights $W=0.5$. This mitigates the risk of an overconfident proposal \citep{Bugallo2017} while leveraging the target-aware Laplace approximation. 

To hedge against the fact that this may still produce a degenerate particle approximation for the hyperparameter posterior, this defensive mixture proposal is simply used as a first step in an AdaPBS scheme \citep{Aalstad2026} to infer hyperparameters across water years. The AdaPBS uses AMIS \citep{Cornuet2012} to build a DMP similar to MAGPIES except that now the DMP is updated via importance sampling rather than the IES. This variant of AdaPBS differs slightly from that in \citet{Aalstad2026}, in that we use it to target climatological hyperparameters via the evidence rather than annual parameters via the likelihood. The overall algorithm is nonetheless mostly identical to that in \citet{Aalstad2026}, so we do not repeat it here, but include it in Appendix~\ref{app:ada} for completeness. 

\section{Results}
\label{sec:results}

 \begin{table}[ht]
\caption{Mean Continuous Ranked Probability Score (CRPS) [mm w.e.] for each study area (rows) and method (columns) in three experiments assimilating in situ SWE, MODIS fractional snow-covered area (FSCA), and glacier-wide mass balance, respectively. The CRPS generalizes absolute error to handle probabilistic estimates with $0$ mm w.e. indicating a perfect score. Values are reported separately for the calibration (train) and validation (test) periods with the validation in brackets. For each row, \textbf{bold} and \underline{underlined} values indicate the best performing pooling approach among complete (CP), no (NP), and partial (PP) pooling, for the calibration and validation period, respectively. Best means either the lowest site-level mean CRPS or, across sites, the highest mean percentage improvement in CRPS (PI, in \%). The Prior column reports the results for the hierarchical prior that serves as a baseline for improvement throughout. Results from the additional partial pooling inference schemes (MAP and PMCMC) are provided alongside for benchmarking the nested particle smoothing in PP.} \label{tab:crps}
\centering
\small  
\setlength{\tabcolsep}{4pt} % Adjust column padding if it's too wide
\begin{tabular}{l c c c c | c c}
\hline In situ SWE & Prior & CP & NP & PP & MAP & PMCMC\\ \hline 
Hornsund & $\mathrm{136}(\mathrm{112})$ & $\mathrm{42}(\mathrm{35})$ & $\mathbf{15}(\mathrm{86})$ & $\mathbf{15}(\underline{26})$ & $\mathrm{15}(\mathrm{25})$ & $\mathrm{15}(\mathrm{29})$ \\ 
Abisko & $\mathrm{46}(\mathrm{36})$ & $\mathrm{23}(\underline{19})$ & $\mathrm{10}(\underline{19})$ & $\mathbf{9}(\underline{19})$ & $\mathrm{10}(\mathrm{20})$ & $\mathrm{9}(\mathrm{33})$ \\ 
Filefjell & $\mathrm{63}(\mathrm{53})$ & $\mathrm{26}(\mathrm{32})$ & $\mathrm{14}(\mathrm{29})$ & $\mathbf{13}(\underline{23})$ & $\mathrm{14}(\mathrm{23})$ & $\mathrm{13}(\mathrm{88})$ \\ 
Weissfluhjoch & $\mathrm{126}(\mathrm{119})$ & $\mathrm{74}(\mathrm{106})$ & $\mathrm{24}(\mathrm{78})$ & $\mathbf{22}(\underline{70})$ & $\mathrm{23}(\mathrm{72})$ & $\mathrm{22}(\mathrm{68})$ \\ 
Mean PI (\%) & $\mathrm{0}(\mathrm{0})$ & $\mathrm{54}(\mathrm{42})$ & $\mathrm{81}(\mathrm{38})$ & $\mathbf{83}(\underline{55})$ & $\mathrm{82}(\mathrm{54})$ & $\mathrm{83}(\mathrm{14})$ \\ 
\hline MODIS FSCA & Prior & CP & NP & PP & MAP & PMCMC\\ \hline 
Hornsund & $\mathrm{99}(\mathrm{117})$ & $\mathrm{85}(\mathrm{146})$ & $\mathbf{45}(\mathrm{114})$ & $\mathrm{57}(\underline{94})$ & $\mathrm{57}(\mathrm{97})$ & $\mathrm{40}(\mathrm{94})$ \\ 
Abisko & $\mathrm{36}(\mathrm{46})$ & $\mathrm{36}(\mathrm{55})$ & $\mathrm{35}(\underline{25})$ & $\mathbf{26}(\mathrm{44})$ & $\mathrm{38}(\mathrm{51})$ & $\mathrm{23}(\mathrm{55})$ \\ 
Filefjell & $\mathrm{57}(\mathrm{73})$ & $\mathrm{64}(\mathrm{110})$ & $\mathrm{35}(\underline{33})$ & $\mathbf{26}(\mathrm{63})$ & $\mathrm{32}(\mathrm{69})$ & $\mathrm{33}(\mathrm{99})$ \\ 
Weissfluhjoch & $\mathrm{122}(\mathrm{134})$ & $\mathrm{181}(\mathrm{254})$ & $\mathrm{148}(\mathrm{83})$ & $\mathbf{93}(\underline{81})$ & $\mathrm{99}(\mathrm{78})$ & $\mathrm{103}(\mathrm{132})$ \\ 
Mean PI (\%) & $\mathrm{0}(\mathrm{0})$ & $\mathrm{-12}(\mathrm{-46})$ & $\mathrm{19}(\underline{35})$ & $\mathbf{37}(\mathrm{19})$ & $\mathrm{25}(\mathrm{13})$ & $\mathrm{38}(\mathrm{-8})$ \\ 
\hline Mass balance & Prior & CP & NP & PP & MAP & PMCMC\\ \hline 
A. Brøggerbreen & $\mathrm{227}(\mathrm{302})$ & $\mathrm{154}(\mathrm{345})$ & $\mathrm{134}(\underline{265})$ & $\mathbf{109}(\mathrm{336})$ & $\mathrm{105}(\mathrm{323})$ & $\mathrm{87}(\mathrm{313})$ \\ 
Storglaciären & $\mathrm{693}(\mathrm{965})$ & $\mathrm{315}(\mathrm{347})$ & $\mathbf{134}(\mathrm{1128})$ & $\mathrm{139}(\underline{294})$ & $\mathrm{131}(\mathrm{308})$ & $\mathrm{107}(\mathrm{296})$ \\ 
Storbrean & $\mathrm{1803}(\mathrm{2100})$ & $\mathrm{591}(\mathrm{635})$ & $\mathrm{183}(\mathrm{2669})$ & $\mathbf{122}(\underline{496})$ & $\mathrm{125}(\mathrm{502})$ & $\mathrm{121}(\mathrm{468})$ \\ 
Claridenfirn & $\mathrm{427}(\mathrm{486})$ & $\mathrm{352}(\mathrm{467})$ & $\mathrm{271}(\mathrm{460})$ & $\mathbf{120}(\underline{458})$ & $\mathrm{125}(\mathrm{450})$ & $\mathrm{108}(\mathrm{463})$ \\ 
Mean PI (\%) & $\mathrm{0}(\mathrm{0})$ & $\mathrm{43}(\mathrm{31})$ & $\mathrm{62}(\mathrm{-7})$ & $\mathbf{74}(\underline{35})$ & $\mathrm{75}(\mathrm{36})$ & $\mathrm{79}(\mathrm{37})$ \\ 
\hline
 \end{tabular} 
 \end{table}

\subsection{In situ snow water equivalent data assimilation}
\label{sec:resS}

\begin{figure}[ht]
\includegraphics[width=\textwidth]{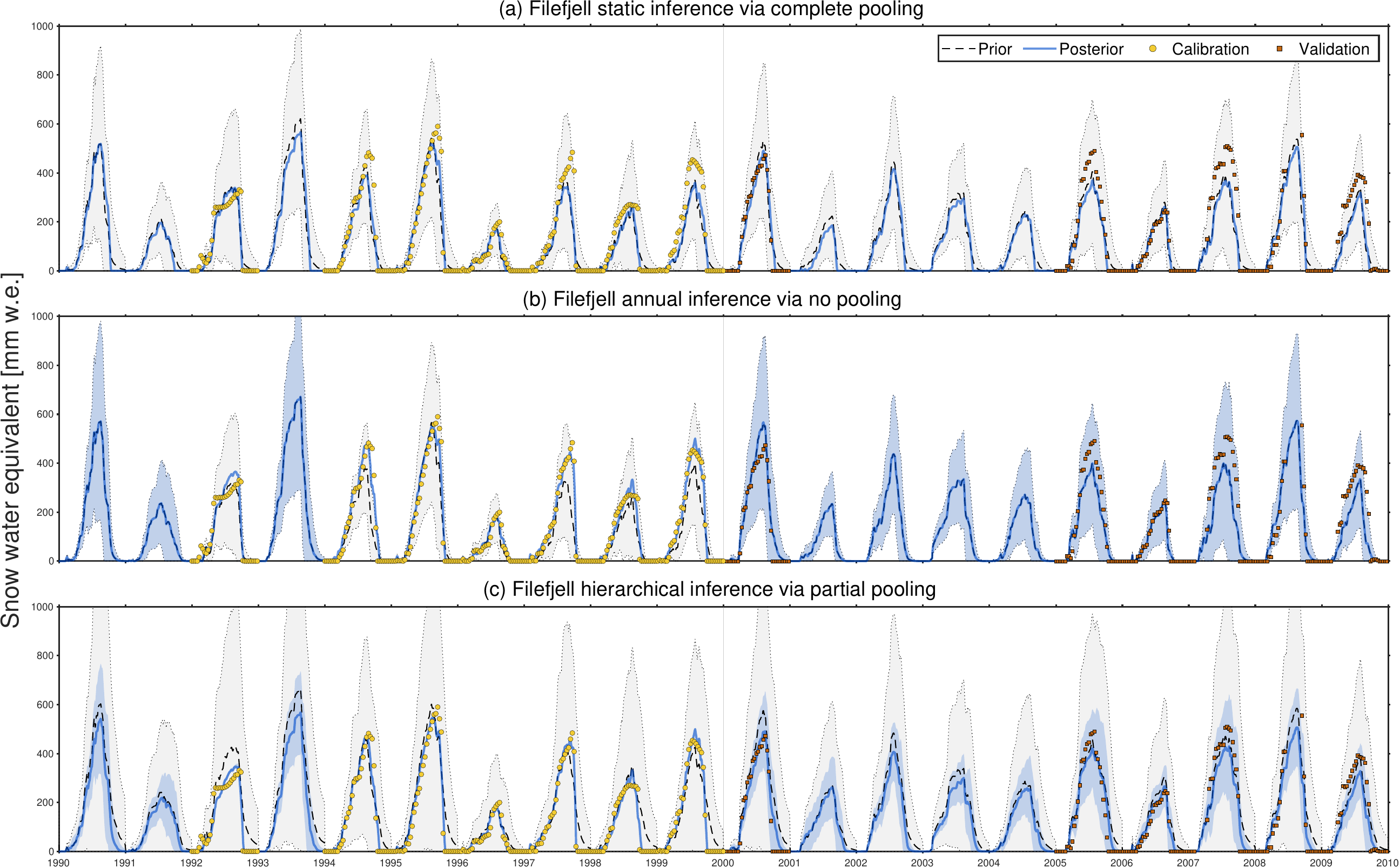}
\caption{Daily SWE reanalysis results from the in situ SWE assimilation experiment at Filefjell using three pooling approaches zoomed in on water years 1990-2009 shown on the $x$-axis. The posterior mean (blue solid line) $\pm 1$ standard deviation (blue shading) is shown together with the prior mean (black dashed line) $\pm 1$ standard deviation (gray shading, dotted bounds). Yellow circles and orange squares show observations in the calibration (before 2000) and validation (2000 onward) periods, separated by a gray line. The prior is conditional (fixed hyperparameters) in (a)\&(b) and hierarchical in (c). The prior and posterior are identical for the no pooling validation period in (b). A full 1968-2022 SWE reanalysis time series is provided as Figure~S3 in the Supplement.}
\label{fig:S_SWE_FF}
\end{figure}

We start with the results from the first experiment using in situ SWE data assimilation at the four seasonal snow sites outlined in Table~\ref{tab:snow}. To demonstrate the effects of hierarchical inference for snow reanalysis, we focus on an exemplary site and time period, namely Filefjell 1990-2009 through the time series shown in Figure~\ref{fig:S_SWE_FF}. Full versions of the corresponding time series for all the snow sites are provided via Figures~S1--S4 in the Supplement, with qualitatively identical results across all sites. Panel (a) of Figure~\ref{fig:S_SWE_FF} shows that the posterior SWE estimates (blue) from the static inference approach using complete pooling (CP) perform relatively well compared to the broad prior SWE estimates (gray shading) across all years, both in the calibration (before 2000) and validation periods. At the same time, this CP approach is generally overconfident in that the posterior spread is far from encompassing the observations, especially for high SWE water years, both in the calibration period (e.g., 1994, 1995) and validation period (e.g., 2005, 2007). The CP inference also struggles more to capture lower snow years, such as 1998 in the calibration period and 2009 in the validation period. Even if the CP gives constrained estimates that can be more accurate than those of the prior, it is generally overconfident in both the calibration and validation periods. As such, the CP approach is too aggressive in trying to see the forest for the trees: it learns an overconfident set of parameters that are a compromise across years in the climatology while failing to capture the particularities of individual water years.

Switching to the other extreme in panel (b) of Figure~\ref{fig:S_SWE_FF}, annual inference via no pooling (NP) performs much better in the calibration period, in which the posterior closely matches the noisy SWE data that are assimilated. In the validation period, as well as in years without data (e.g., 1990-1991, 1993), however, the NP approach simply reverts to the conditional prior and forgets what has been learned about the parameters in years with data. As such, even if the probabilistic predictions from the NP approach are both accurate and well calibrated in the calibration period, they are completely unconstrained in the validation period. In summary, the NP approach is myopic in that it cannot see the forest for the trees: it can infer the weather in observed water years where data are available, but it cannot learn the climatology by pooling information across years. 

Bridging these two extreme approaches to inference in panel (c) of Figure~\ref{fig:S_SWE_FF}, hierarchical inference via partial pooling (PP) performs well in both the calibration and validation periods. Just as with NP, the posterior with the PP approach is able to closely match the assimilated SWE data during the calibration period while accounting for observation error and the strong constraint assumption. At the same time, PP learns a well-calibrated parameter climatology across the calibration period so that it can constrain SWE estimates in the validation period relative to the prior, in contrast to NP. Furthermore, unlike the CP approach, the PP approach is not overconfident in the validation period, where the posterior spread generally encompasses the independent SWE observations. To make these findings more quantitative, considering the posterior CRPS in Table~\ref{tab:crps} for Filefjell, PP clearly performs best; while NP is close in the calibration period, the PP CRPS is markedly ($21\%$) lower (i.e., better) than the second best performing approach (NP) in the validation period. These results show that the PP approach can see both the trees and the forest by inferring the parameters in individual water years and learning a well-calibrated parameter climatology.

\begin{figure}[ht]
\includegraphics[width=\textwidth]{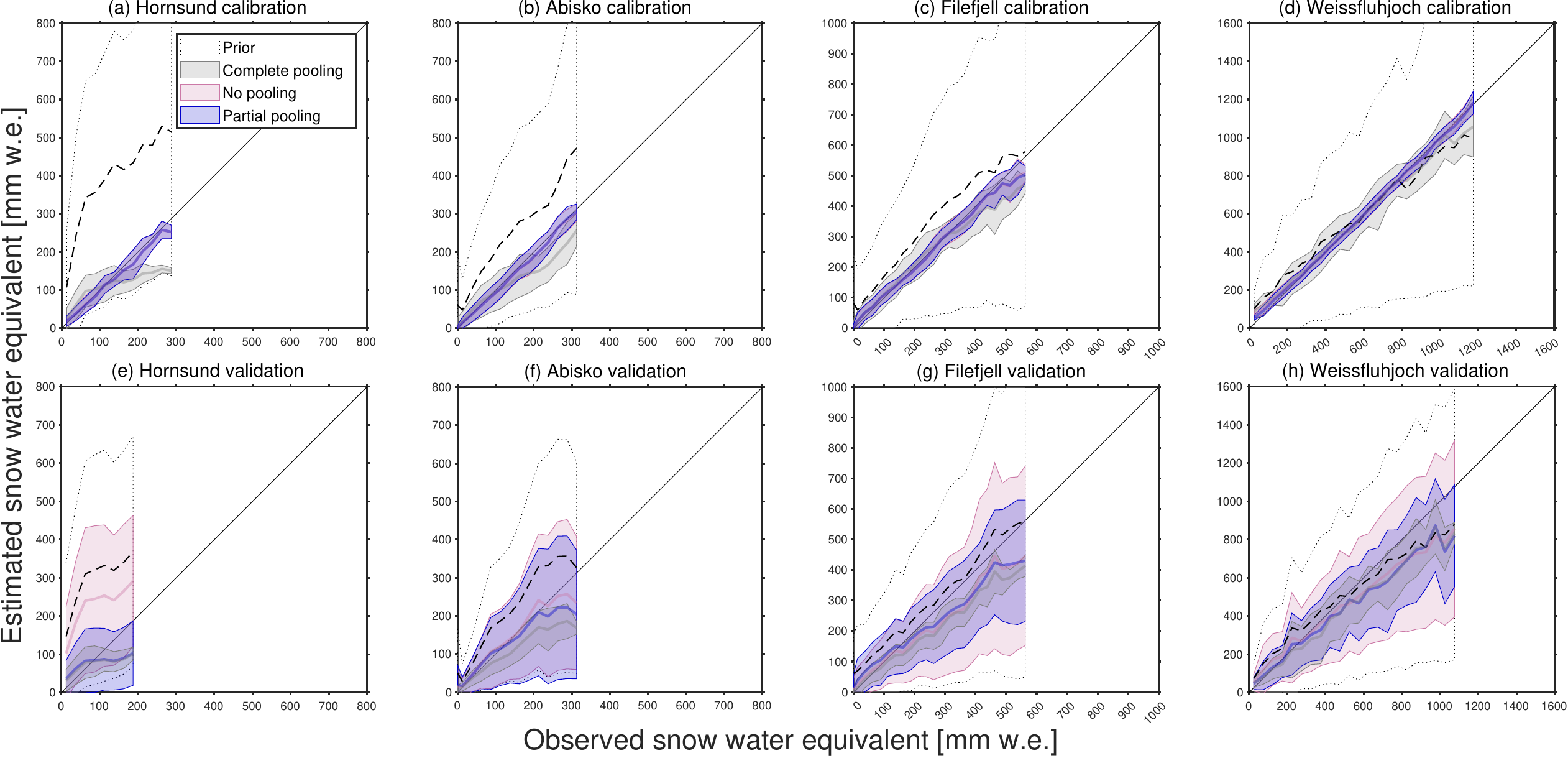}
\caption{Binned scatter plot showing estimates of SWE sampled from the hierarchical prior (dashed black line, transparent dotted envelope) as well as from the complete pooling (gray line and shading), no pooling (pink line and shading), and partial pooling (blue line and shading) posteriors for each snow study site (columns) in the calibration (top) and validation (bottom) periods. The central lines and envelopes show the sample mean $\pm1$ standard deviation in each observation bin. Bins with fewer than $5$ observations are not shown. The thin solid diagonal line is the 1:1 line. The axes are square and equal for symmetry, covering most of the estimated dynamic range of the broad prior.}
\label{fig:S_scatter}
\end{figure}

Taking into account the in situ SWE assimilation experiment as a whole, the binned scatter plots in Figure~\ref{fig:S_scatter} show the performance of the respective approaches for both the calibration and validation periods at all snow sites. As expected, the posterior from all pooling approaches clearly outperforms the hierarchical prior in both periods, where the latter systematically exhibits a much larger spread across the board and has a strong positive bias for Hornsund and Abisko. From the top row of plots, with the exception of a few very high SWE measurements, both the PP and NP approaches are able to accurately and precisely infer the observed SWE across all sites in the calibration period. The static CP approach, in contrast, performs considerably worse and deviates to varying degrees from the 1:1 line for higher observed SWE values. These findings hold for both the mean of the posterior mean and standard deviation (shading) within each bin for all sites. For the validation period, the story changes, and it is generally the PP approach that performs best by tracking the 1:1 line relatively closely while maintaining well calibrated uncertainty. The other two pooling schemes perform worse overall in the validation period, in which NP and CP tend to be underconfident and overconfident, respectively. This pattern is especially evident for Hornsund in panel (e), where the wide conditional prior that NP reverts to in the calibration was already strongly positively biased, while CP overfits to a narrow posterior with a negative bias. 

A major advantage of the PP approach becomes apparent when looking at the CRPS across all sites for the SWE experiment in Table~\ref{tab:crps}. Recall that the CRPS is a probabilistic measure of both the accuracy and the precision of the ensemble-based predictions as a whole. Here we see that the PP approach comes out on top with the lowest CRPS for all sites in both the calibration and validation periods, tying at one of the four sites in each period. Although the differences to NP are small (i.e., a few \%) or near zero for most sites in the calibration period, the gains over CP are considerable. For example, at Weissfluhjoch, the errors obtained by the PP and NP approaches in terms of mean CRPS are both reduced by more than a factor of $3$ compared to CP. Averaging across all snow sites in the calibration period, PP provided an 83\% improvement in CRPS relative to the hierarchical prior, closely followed by no pooling with an 81\% improvement, followed by CP lagging behind at a 54\% improvement.

In the validation period, the improvements obtained by PP are even more stark where the CP and NP approaches are only able to match it in one case (Abisko) and at the remaining sites the PP approach improves on the CRPS of the runner up by at least 10\% (NP, Weissfluhjoch) and at most 25\% (CP, Hornsund). The gains obtained by adopting hierarchical inference over annual inference are especially pronounced in the validation period for sites where the prior and thus NP perform poorly, such as at Hornsund where PP improves on the NP CRPS by 70\%. The advantages of PP in the validation period are readily apparent for the time series across all sites (Figs.~S1--S4 in the Supplement), where the PP posterior is able to encompass most of the independent observations while tending to provide sharper predictions than the prior (and thus NP), with generally much better calibrated uncertainty than the overconfident CP approach. Averaging across all snow sites in the validation period, PP provided a 55\% improvement in CRPS relative to the hierarchical prior, followed by CP with a 42\% improvement, and NP with a 38\% improvement.

\subsection{Satellite-based fractional snow-covered area data assimilation}
\label{sec:resF}

% Kris is here :)

The results for the satellite-based FSCA data assimilation will be presented in the same way as the in situ SWE experiment. At the outset, we expect these two experiments to show similar qualitative results. Differences are nonetheless inevitable due to the unique characteristics of the assimilated $0.01^\circ$ ($\simeq 1$ km) MODIS FSCA data, which are indirectly related to and not at the same scale as the independent in situ SWE data used for evaluation. Furthermore, in this MODIS data assimilation experiment, the order of the calibration and validation periods is reversed, as the calibration period must necessarily focus on the available Snow\_cci MODIS data that begins in the water year 2000. 

\begin{figure}[ht]
\includegraphics[width=\textwidth]{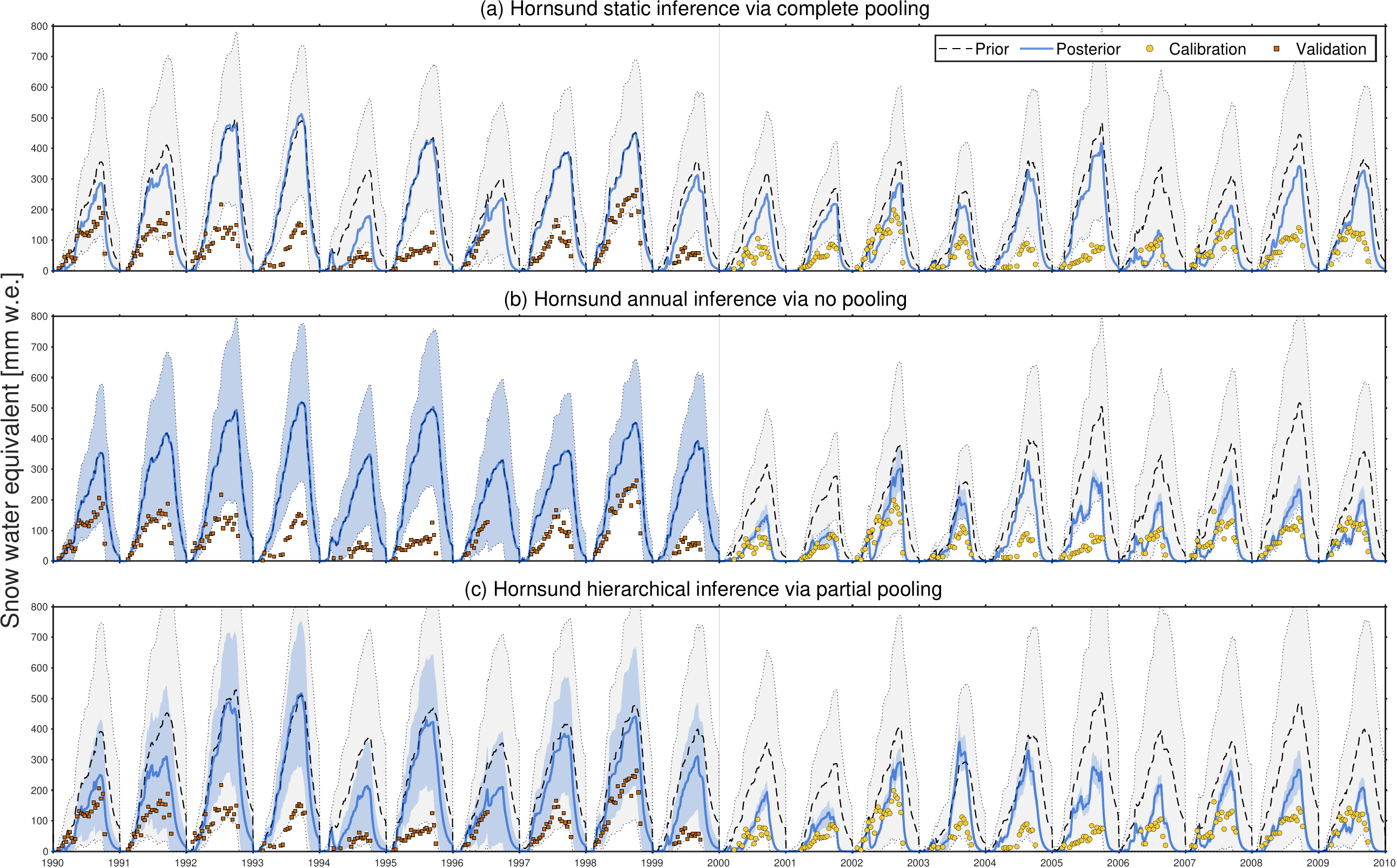}
\caption{As in Figure~\ref{fig:S_SWE_FF} but for the MODIS FSCA assimilation experiment at Hornsund, where the calibration and validation periods are now 2000--2023 and 1983--1999, 2024, respectively, and the plotted in situ SWE evaluation data are fully independent. The full 1983--2024 MODIS-based SWE reanalysis time series is provided as Figure~S5 in the Supplement.}
\label{fig:F_SWE_HS}
\end{figure}

For the FSCA assimilation experiment, we use the time series from Hornsund in Figure~\ref{fig:F_SWE_HS}, as this is among the sites where the benefits of the partial pooling were most striking. We found similar, albeit less pronounced, performance gains by going hierarchical both visually in Figures~S5-S8 and quantitatively in terms of CRPS in Table~\ref{tab:crps} for most (all but Hornsund) sites in the calibration period and half the sites (Hornsund, Weissfluhjoch) in the validation period. By inspecting Figure~\ref{fig:F_SWE_HS}, we see that both the wide conditional prior in panels (a)\&(b) and the even wider hierarchical prior in panel (c) generate far too much SWE in all years compared to the observations. The posterior obtained from static inference via the complete pooling (CP) approach in panel (a) of Figure~\ref{fig:F_SWE_HS} is able to partly correct this positive bias. At the same time, the CP posterior is generally far too confident in both the validation and calibration periods. Moreover, despite performing reasonably well in typical snow years such as 2007, the CP approach struggles to capture both low (2001) and high (2002) snow years in the calibration period. In the validation period before 2000, the CP approach provides sharp, albeit overconfident, predictions and retains much of the positive bias of the prior. The posterior from the annual inference via no pooling (NP) approach performs considerably better on average across the calibration period, where it is able to capture the aforementioned low and high snow years reasonably well while maintaining better calibrated uncertainty. Nonetheless, as expected, the NP approach does not learn the climatology: in the validation period it simply reverts to the prior as in the previous experiment. 

Bridging the two extreme approaches to inference at Hornsund, hierarchical inference via the partial pooling (PP) approach in panel (c) of Figure~\ref{fig:F_SWE_HS} provides nearly as accurate and precise predictions of the independent SWE in the calibration period as the NP approach. During this period at Hornsund, both NP and PP provide a substantial improvement in mean CRPS over the hierarchical prior of 55\% and 42\%, respectively, unlike CP with a meager 14\% improvement. However, the most marked performance gains with the PP approach are found in the validation period for Hornsund, where the posterior is considerably narrower than the prior while still encompassing the vast majority of independent SWE observations within two standard deviations ($\pm 1$ is shown) of the posterior mean in Figure~\ref{fig:F_SWE_HS}. The corresponding predictions in observation space are shown through the FSCA time series in Figure~S9. Therein, the PP approach is not only able to match predictions to the MODIS FSCA data in the calibration period but also markedly constrain FSCA predictions in the validation period before the start of the MODIS-era in 2000. Improved SWE estimates and constrained FSCA at Hornsund in the pre-MODIS validation period provide a visual demonstration that hierarchical inference via PP has the potential to extend the utility of satellite data well beyond the satellite era into the deeper past as well as the future.

\begin{figure}[ht]
\includegraphics[width=\textwidth]{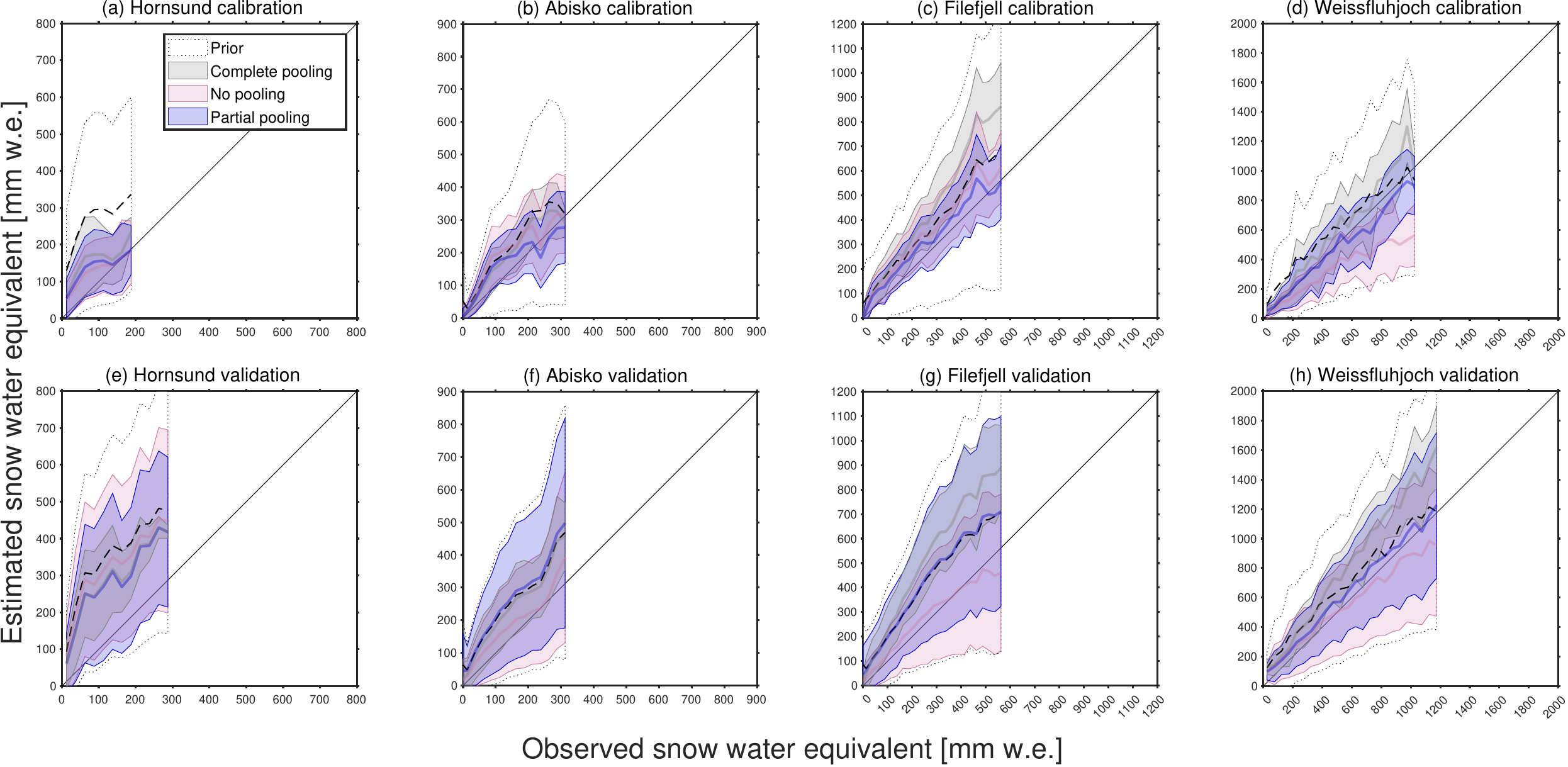}
\caption{As in Figure~\ref{fig:S_scatter} but for the MODIS FSCA snow data assimilation experiment where the in situ SWE data on the $x$-axis serve as fully independent evaluation data. Here the calibration period (top) is 2000-2023 and the remaining water years form the validation period.}
\label{fig:F_scatter}
\end{figure}

% Kris is here
To help gauge the performance of the respective approaches for all snow sites, Figure~\ref{fig:F_scatter} shows binned scatter plots of the respective approaches when assimilating MODIS FSCA. The upper panels show how the PP approach is able to provide samples that are closest to the observations when considering both the mean and spread in the calibration period. The posteriors from all pooling schemes clearly improve on the hierarchical prior for all sites in this period, but both NP and CP tend to sit slightly further off the one-to-one line than PP, except for Hornsund in panel (a), where PP is the runner up to NP. A similar pattern is present in the validation period, where PP performs best with well-calibrated uncertainty for both Hornsund in panel (e) and Weissfluhjoch in panel (h), for which NP is slightly more biased and CP is not only biased but also overconfident. At the other two sites, PP is once more runner up to NP but still outperforms overconfident CP.

The above visual interpretations are confirmed quantitatively by the PP approach, which has either the best or second best CRPS in Table~\ref{tab:crps} for each site in both the validation and calibration periods. The PP approach performs best in the calibration period, coming out on top for three out of four sites, with an overall 37\% mean improvement in CRPS over the hierarchical prior.  The improvements with PP in the calibration are considerable at most sites, with over a 20\% improvement compared to the runner up approach, except for Hornsund. For the validation period, NP and PP each perform best for half the sites, but NP comes out on top overall with a 35\% improvement, followed by PP with 19\%. The fact that NP scores best overall in the validation, where it reverts to the conditional prior, is a reminder not only of the challenges in evaluating $0.01^\circ$ MODIS-based FSCA data assimilation experiments with in situ SWE data, but also that the conditional prior can fortuitously end up being relatively well suited for certain sites, namely Abisko and Filefjell in this case. We see the same behavior in the in situ SWE assimilation, where NP also performed relatively well for these sites in the validation. Still, in both calibration and validation periods, even when PP does not perform best for FSCA assimilation, it is never the worst pooling approach, which was CP in this experiment, and PP always improved on the hierarchical prior.

% ExpF scatter

\subsection{Glacier-wide mass balance data assimilation}
\label{sec:resG}

\begin{figure}[ht]
\includegraphics[width=\textwidth]{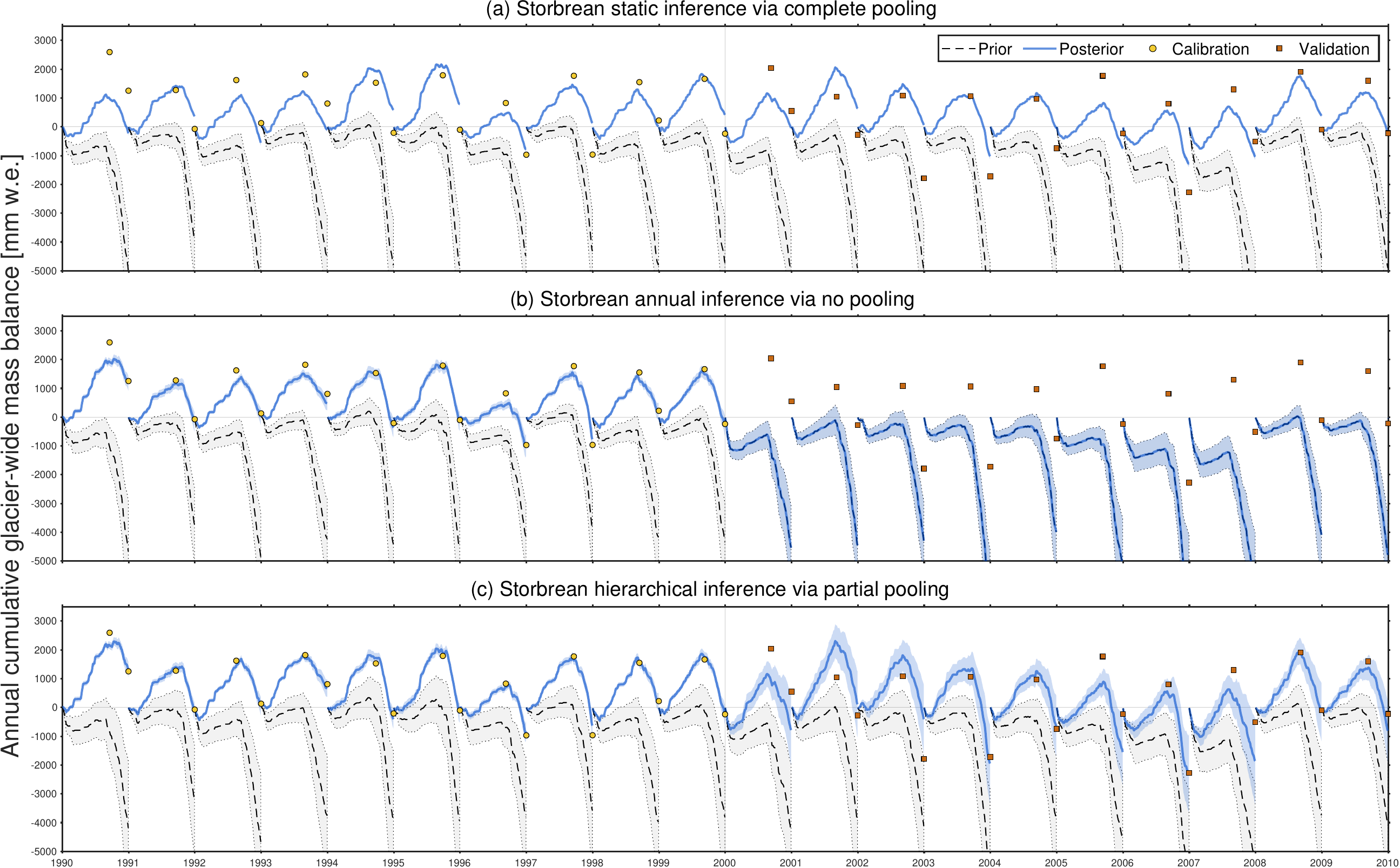}
\caption{As in Figure~\ref{fig:S_SWE_FF} but for the glacier-wide mass balance assimilation experiment at Storbrean, where glacier-wide mass balance data are assimilated (yellow circles) in the pre-2000 calibration period and withheld (orange squares) from 2000 onward in the validation period. The plots show annual cumulative glacier-wide mass balance for each day relative to the start of each balance year indicated as ticks on the $x$-axis.}
\label{fig:G_SB}
\end{figure}

In the glacier-wide mass balance data assimilation experiment, any systematic bias in annual estimates is likely to compound over time in the long-term cumulative glacier mass balance. We thus focus on the performance of the annual cumulative mass balance, i.e., the cumulative mass balance for each day in a balance year. This allows us to evaluate results via comparison to the winter and annual mass balance for each year without the drift that accumulates in multi-decadal cumulative time series. As before, we begin by focusing on the time series for one illustrative glacier, Storbrean, which was particularly challenging for the prior (Figure~\ref{fig:G_SB}). We note that the qualitative findings concerning the respective inference approaches for Storbrean apply to all four glaciers considered (shown in Figures~S13-S16). 

% Kris is here
By inspection of panel (a) in Figure~\ref{fig:G_SB}, the static inference approach via CP performs adequately overall by providing a marked improvement over the conditional prior, which has a strong negative bias. At the same time, the CP approach struggles to capture the nuances of individual balance years, notably years with unusually positive winter mass balance (e.g. 1990) or negative annual mass balance (e.g. 2006), in both the calibration and validation periods. This static approach performs especially poorly in the post-2000 validation period (see also Figure~S15 for 2010-2023), where it is unable to capture the majority of strongly negative annual mass balances. Even if it improves on the prior, the posterior mean from CP provides a generally poor point estimate of the observed mass balance. This applied to both the zoomed in 20 years in Figure~\ref{fig:G_SB} and the entire 75 year simulation period in Figure~S15. Moreover, the spread of the posterior fails to encompass most of the observations, a sign of highly overconfident predictions in both the calibration and validation periods. 

% GMB SB

As in the previous experiments, the reanalysis shifts considerably when using annual inference via NP. Panel (b) in Figure~\ref{fig:G_SB} shows how the no pooling approach at Storbrean is able to perform much better than CP by matching the particularities of individual balance years in the calibration period. For example, the posterior is now able to provide a closer match to the high winter mass balances in 1990 as well as the negative annual mass balances in 1997 (i.e., right before the 1998 tick mark). However, in the validation period, the NP completely reverts back to the conditional prior, yielding strongly negatively biased and uncertain predictions. This is in line with the other experiments and is expected from NP, as this approach has no inferential mechanism to transfer information to unobserved years.

Hierarchical inference via PP at Storbrean, shown in panel (c) of Figure~\ref{fig:G_SB}, once again gets the best of both worlds and outperforms both the non-hierarchical pooling approaches. As with NP, partial pooling provides both more precise and more accurate posterior predictions in the calibration period than CP. In fact, PP also clearly outperforms the NP approach in certain years such as for the winter mass balance in 1992 and 1993. As with CP, PP provides much less negatively biased posterior predictions in the validation period than those of the prior to which NP reverts. Moreover, in this validation period, PP is less overconfident than CP, maintaining a better calibrated posterior spread with improved coverage of the withheld data, as seen in e.g., 2006 and 2009.

\begin{figure}[ht]
\includegraphics[width=\textwidth]{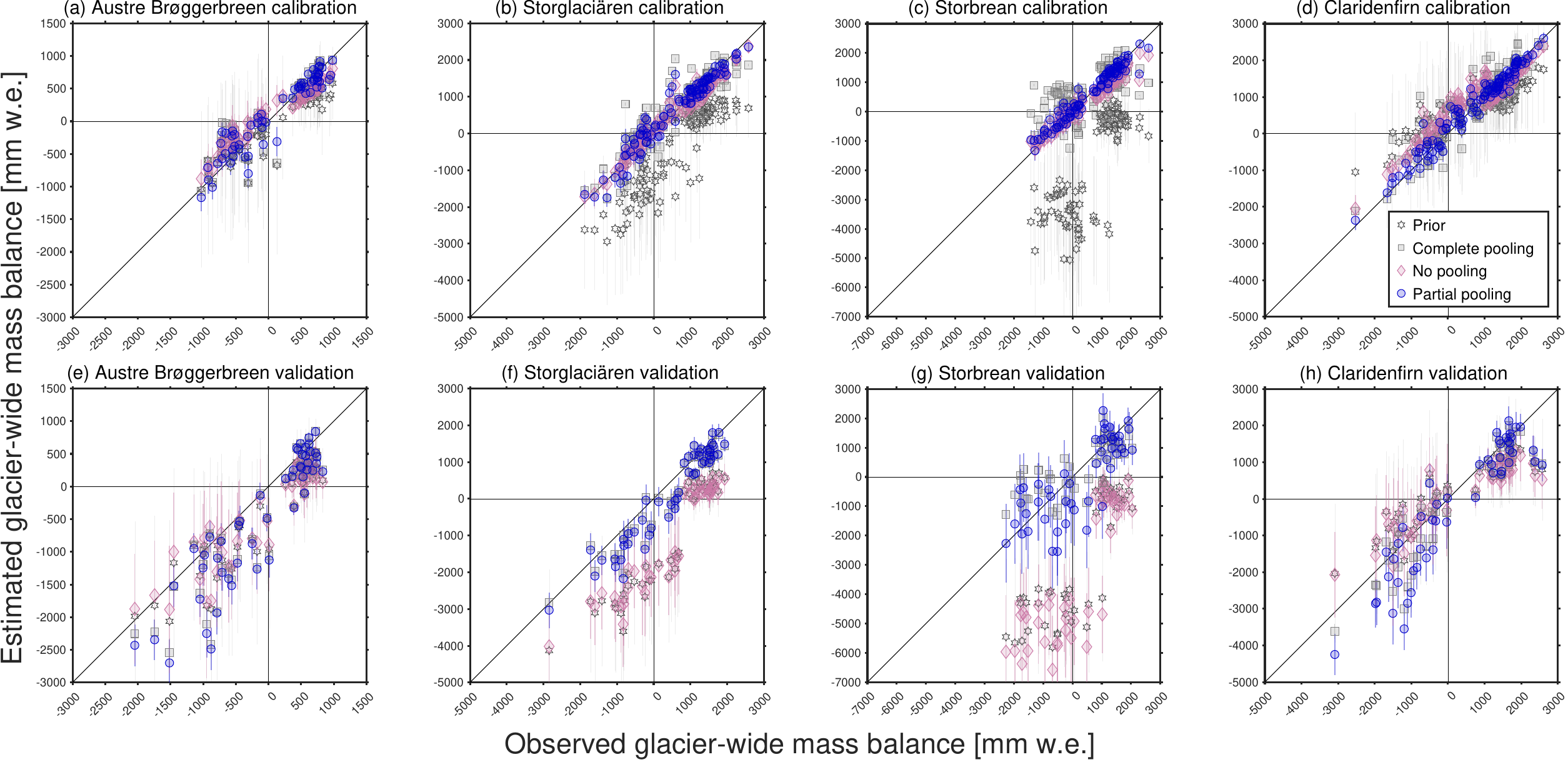}
\caption{Scatter plot showing estimates of glacier-wide winter and annual mass balance from the hierarchical prior (white stars) as well as from the complete pooling (gray squares), no pooling (pink diamonds), and partial pooling (blue circles) posteriors compared to the corresponding observations for each glacier (columns) in the calibration (top) and validation (bottom) periods. The respective markers show the ensemble mean estimate at the observation time, with the associated vertical bars indicating $\pm1$ ensemble standard deviation. The thin solid diagonal line is the 1:1 line with equal square axes for symmetry to cover most of the estimated dynamic range for each glacier. Panels contain both observed winter balances, which are always positive, and annual balances, which are typically negative or near zero.}
\label{fig:G_scatter}
\end{figure}

Across all glaciers in Figure~\ref{fig:G_scatter}, we identify the same general pattern as seen for Storbrean: the PP approach is the only technique that consistently provides accurate posterior mean predictions in both the calibration and validation periods. During calibration (top), static inference via CP (gray) is generally much more scattered about the identity line than NP and PP, which can both adapt to individual balance years. Furthermore, except for Storglaci\"{a}ren where the two are nearly identical, PP (blue) also tends to sit tighter to the 1:1 line than no pooling (red), indicating a further reduction in error. During validation (bottom), the annual inference via NP performs worst, as expected from reversion to a negatively biased prior, as is especially evident at Storglaci\"{a}ren and Storbrean. Both static inference via CP and PP appear visually to perform similarly in this period, effectively correcting for most of the negative bias in the prior. Errors in the prior and NP mean predictions are largest for the mostly negative annual mass balances, which happen to be where CP and PP provide the largest performance gains.  

The generally superior performance of hierarchical inference is quantitatively evident in the probabilistic CRPS metric in Table~\ref{tab:crps}, where PP performs best for three glaciers during calibration and is a close runner up to NP for Storglaci\"{a}ren. For Storbrean and Claridenfirn, the performance gains from partial pooling in the calibration period are particularly notable, with 33\% and 56\% CRPS improvements compared to the runner up NP approach. Averaging across all glaciers during calibration, PP provided a 74\% mean improvement in CRPS relative to the hierarchical prior, followed closely by NP (62\% improvement), and trailed by CP (43\% improvement).

In the validation period, PP provides the lowest CRPS values for three out of four glaciers, and it is the runner-up pooling approach for Austre Brøggerbreen. For the latter, the prior is already performing relatively well in that the mean CRPS of around 300 mm w.e. is less than the assumed observation error standard deviation on annual mass balance ($400$ mm w.e.). The same holds, albeit to a lesser extent, for Claridenfirn. For these two glaciers, the prior is already relatively accurate, so there is not much to gain from PP in the validation period. For the remaining two glaciers where the prior is strongly biased, the performance gains from PP are remarkable. Storbrean, in particular, stands out, with PP providing 76\%, 81\%, and 22\% improvements in CRPS relative to the hierarchical prior, NP, and CP, respectively. At Storglaci\"{a}ren the gains are also considerable, with PP providing corresponding improvements of 70\%, 74\%, and 15\%. Averaging across all glaciers in the validation period, PP provides a 35\% mean improvement in CRPS relative to the hierarchical prior, followed by CP with a 31\% improvement, while NP fails to improve on the hierarchical prior.

\subsection{Hierarchical Bayesian inference schemes}
\label{sec:results_schemes}

In addition to comparing the three approaches to inference through the spectrum of pooling, we benchmark our chosen approach to hierarchical inference via partial pooling using alternative schemes. Recall that the PP heading in Table~\ref{tab:crps} refers to partial pooling using the nested particle smoother workflow in Figure~\ref{fig:workflow}. To put its performance in context, we evaluate two additional hierarchical inference schemes for partial pooling: the fast and approximate EM-based MAP estimate  (Section~\ref{sec:EM}) and the expensive gold-standard PMCMC (Appendix~\ref{app:PMCMC}).

In terms of CRPS in Table~\ref{tab:crps}, the more intensive PMCMC method tends to either match or outperform the nested particle smoothers in the calibration period, while the MAP scheme performs marginally worse. The schemes also perform similarly in the validation period, wherein the nested particle smoothers provide the most consistently robust performance, followed by MAP and the PMCMC scheme. Although the MAP scheme rarely performs best in either the validation or calibration period, it is often close to matching and sometimes outperforms the other hierarchical schemes. In a few cases, such as Filefjell for both snow experiments, the PMCMC benchmark provides poor results during the validation period. This coincides with convergence diagnostics indicating inadequate mixing \citep{Gelman2013}. These cases involve assimilating many observations, which is likely to increase the sampling variance in the particle-based evidence approximation, leading to slow mixing. Resolving this would likely require making the already costly Markov chains much longer. At the opposite extreme in the glacier experiment, with the fewest observations, PMCMC performs best, as expected for this benchmark when the evidence estimate is more reliable. Beyond the poor mixing exceptions, there is generally no systematic difference in the CRPS of the hierarchical schemes when considering both the calibration and validation periods together.

 \subsection{Learning the parameter climatology}
 \label{sec:results_learning}

\begin{figure}[ht]
\includegraphics[width=\textwidth]{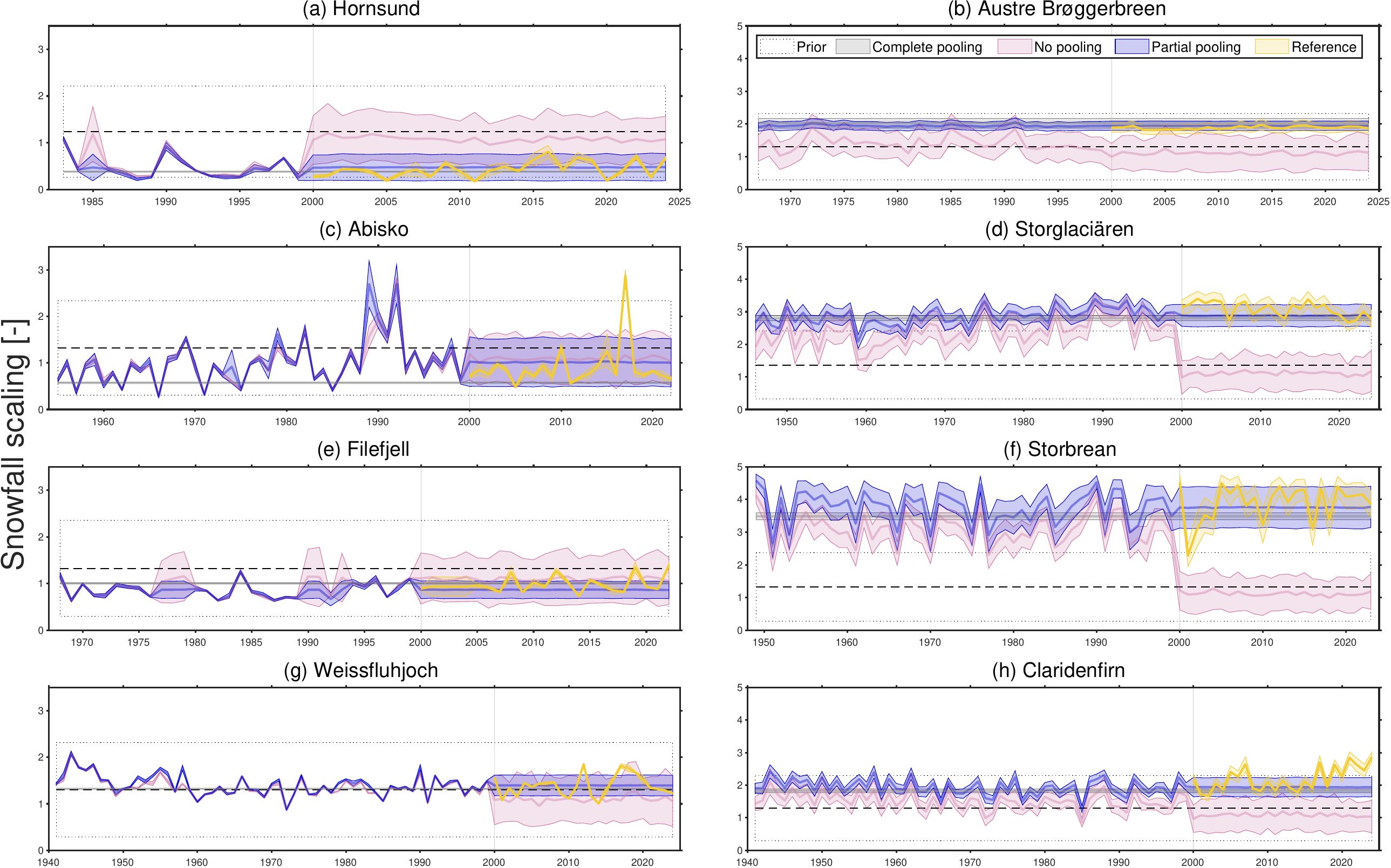}
\caption{Time series of the annual snowfall scaling parameter $c$ distributions from the hierarchical prior (dashed black line, transparent dotted envelope) as well as from the complete pooling (gray line and shading), no pooling (pink line and shading), and partial pooling (blue line and shading) posteriors for each site in the SWE assimilation (left) and glacier mass balance (right) experiments for the calibration (before 2000) and validation (2000 onward) periods. A reference partial pooling (yellow line and shading) run with no data withheld from assimilation is also shown for the validation period. The central lines and envelopes show the ensemble mean $\pm1$ standard deviation.}
\label{fig:learning}
\end{figure}

To help highlight the ability of hierarchical inference via PP to learn the underlying parameter climatology, we reran the SWE and glacier data assimilation experiments without a validation hold-out period (i.e., assimilating all data). This created a reference PP reanalysis that provides `ground truth' parameter values for the withheld validation period, as shown in Figure~\ref{fig:learning}. We could not perform such a reference run for the FSCA data assimilation experiment, as the assimilated FSCA data are absent in the 20$^\text{th}$ century segment of the reanalysis. Nonetheless, we have shown in Section~\ref{sec:resF} how partial pooling of FSCA data can help constrain SWE estimates in the pre-MODIS era validation period. Here, we compare the parameters inferred via the pooling approaches with withheld data during the validation to those in the reference PP reanalysis to show explicitly how such information transfer is achieved by learning the parameter climatology.

In this demonstration, we focus on the snowfall scaling parameter $c$ as visualized in Figure~\ref{fig:learning}, which, together with the air temperature bias $b$, exhibited the largest information gain through hierarchical reanalysis, while the degree day factor $a$ was less constrained. For completeness, plots for all three parameters are also provided in the Supplement via Figures~S17--S19. A lower bound on the information gain can be gauged visually in the validation period by comparing the large spread of the NP reanalysis (pink), which reverts to the conditional prior, to the much more constrained PP spread (blue) across most snow sites and glaciers. The only apparent exception is Abisko, where NP and PP have a similar spread during the validation. Nonetheless, even in this case, there is considerable information gain relative to the much broader hierarchical prior, where the PP posterior becomes narrower and shifts towards lower snowfall scaling than the prior mean.

Comparing the three approaches to pooling, we see that the results in the snowfall scaling parameter space in Figure~\ref{fig:learning} closely mirror those seen in the state space of the respective experiments in Sections~\ref{sec:resS}--\ref{sec:resG}. In particular, static inference via CP (gray) provides very confident predictions for a temporally constant value of the snowfall scaling. Unlike the other approaches, CP is incapable of adapting to the particularities of a given water year and thus fails to capture the parameter inter-annual variability seen in the other approaches during calibration. Moreover, even if CP can often roughly match the temporal mean of the other approaches, it tends to provide overconfident, near degenerate posterior estimates for static snowfall scaling that do not even implicitly encode the actual inter-annual variability in this parameter. Conversely, both annual inference via NP and hierarchical inference via PP capture inter-annual variability in this scaling parameter during the calibration period, tending to agree with one another in mean and variability for all years with observations, especially for the snow sites (left), which are more data rich. We note that this inter-annual variability can be considerable in practice. For example, at both Abisko and Storbrean, the snowfall scaling has a (max-min) range that exceeds $2$. 

During the validation period, the static inference approach via CP sticks to its overconfident, near degenerate estimate, while the annual inference via NP reverts completely back to the uninformed climate-assumed conditional prior. On the other hand, the hierarchical inference via PP successfully learns the parameter climatology that can be transferred to the unseen validation period, capturing not only the mean of the reference PP reanalysis (yellow) but also encoding the inter-annual variability through its uncertainty estimate. In particular, we note that across all sites, the posterior mean $\pm 1$ standard deviation envelope from PP (blue) encompasses $77\pm 15\%$ of the scaling estimates from the fully informed reference PP reanalysis mean (yellow). The fact that PP encompasses most of the reference estimates close to the 68\% coverage that would be expected for a Gaussian is consistent with PP being well calibrated: it is neither overconfident like CP ($12\pm 18\%$ across sites) nor biased yet still uncertain like NP ($39 \pm 47\%$ across sites). Crucially, assuming a stationary climate, this means that PP can provide well-calibrated parameter estimates also in unseen years, both in the deeper past and potentially further into the future. 

\section{Discussion}
\label{sec:discussion}

Here, we focus on the benefits of hierarchical inference, some key remaining challenges, tradeoffs to consider when selecting hierarchical data assimilation schemes, and an outlook for future research. 

\subsection{Benefits of hierarchical inference}
Averaged over the results from all experiments in Sections~\ref{sec:resS}--\ref{sec:resG}, the climate-informed reanalysis using the posterior from hierarchical inference via PP improves on the hierarchical prior CRPS by 65\% in the calibration period. The runner up annual inference NP comes close with a corresponding 54\% improvement, whereas the static calibration via CP falls behind with a 28\% improvement. Performing the same averaging across all experiments in the validation periods, PP also performs best overall with a 36\% improvement, followed by NP with a 22\% improvement, and CP again in last with a 9\% improvement. Averaging these scores across both periods, which are of equal interest for cryospheric reanalysis with sparse data, partial pooling is the clear winner with a 51\% improvement, followed by NP at 38\%, and complete pooling at 19\%. The large difference between NP and CP is partly caused by the fact that the calibration period, where no pooling shines, was much longer than the validation period in most experiments. Furthermore, these improvements are all likely lower bounds due to the lack of spatially representative validation data in the FSCA experiment.

The merits of basing the evaluation on the CRPS are that it is an easily interpretable probabilistic error metric in the same units [mm w.e.] as the quantity of interest, namely SWE or glacier mass balance, tailored to ensemble verification \citep{Gneiting2005}. A major strength of a probabilistic (Bayesian) reanalysis is the provision of both point estimates and uncertainty quantification, where the latter is invaluable for decision making by downstream users and stakeholders. The CRPS helps to quantify how well calibrated such uncertainty estimates are, encoding the idea that it is better to be roughly right than precisely wrong.

Circling back to the CRPS evaluation results, it is noteworthy that hierarchical inference via PP (with $51\%$ improvement over the hierarchical prior) markedly outperforms both annual inference via NP ($38\%$) and static calibration via CP ($19\%$) when combining performance in the calibration and validation periods. This demonstrates that PP can automatically get the best of both worlds of CP and NP by seeing both the forest (climatology) and the trees (inter-annual variability). As such, we obtain a climate-informed reanalysis approach that is able to match data rich years in a similar fashion to annual reanalysis and revert to a constrained parameter climatology in years with little to no data in the spirit of static calibration. The latter is helpful not only for handling sporadic years with data gaps but also for extending reanalyses into the more distant past, before the satellite-era, as well as for informing future forecasts and projections.

What is even more remarkable is that PP also generally manages to beat the two extreme pooling approaches at their own game. That is, for most sites in each experiment, PP outperforms both NP in the calibration period and CP in the validation period (Table~\ref{tab:crps}). On the one hand, inference for observed years in the calibration period gains information myopically from the year in question \emph{and} by pooling information from the climatology. On the other hand, inference in the unobserved validation period gains information by pooling climatological information to constrain the static parameter ensemble \emph{and} by partially retaining spread to encode the inter-annual parameter variability from the observed years. In both cases, the degree of pooling is handled automatically via hierarchical Bayesian inference, which is able to identify the ideal compromise between complete pooling and no pooling.

It may seem somewhat paradoxical that inference for an observed year can benefit from pulling towards the climatology, and that inference in an unobserved year can benefit from injecting interannual variability. Indeed, this is intimately related to Stein's paradox \citep{Efron1977}: joint inference for more than a couple of parameters improves by shrinking towards the mean of all parameters, even if these are seemingly unrelated. In our context, this shrinkage is tantamount to partially pooling information across years in the climatology. As already alluded to by \citet{Efron1977} and later formalized by \citet{Robert2007} and references therein, this Stein effect is not a paradox at all but emerges naturally through the application of hierarchical Bayesian inference and empirical Bayesian approximations thereof. As such, modern probabilistic models often rely heavily on hierarchical Bayesian inference via partial pooling \citep{Gelman2013,McElreath2020}. Our main contribution has been to adapt this hierarchical approach to generating climate-informed cryospheric reanalysis, inspired by applications to training Bayesian neural networks \citep{MacKay1992} and predicting tadpole mortality \citep{McElreath2020}.

\subsection{Learning hyperparameters}

Beyond the Stein effect and the favorable performance of PP, our adoption of hierarchical Bayesian data assimilation was strongly motivated by the very practical concern of hyperparameter selection. Selecting prior distributions (see Figure~\ref{fig:prior}) is often far from trivial and, especially in regimes with sparse and/or very noisy data, inference is sensitive to the choice of prior. This turns out to be a powerful regularizing feature of Bayesian methods and a reminder that one cannot do inference without assumptions that are encoded in the prior \citep{MacKay2003}. To many practitioners, however, such a theoretical justification for the use of subjective priors may offer little comfort. In practice, one generally resorts to using weakly informative priors \citep{Gelman2013} based on scaled back background information, as is becoming standard in cryospheric data assimilation \citep{AlonsoGonzalez2022,Cao2025}. While such an approach can be applicable at the local, basin, or regional scale, it becomes more challenging to entertain at larger scales without invoking a very diffuse prior. For example, building a one size fits all prior to cover the likely range of cryospheric reanalysis parameters for the whole world is a daunting task that likely ends up with an uninformative prior. This, in turn, makes inference challenging for the scarce and noisy data regimes encountered in practical global cryospheric reanalysis settings. 

Hierarchical cryospheric reanalysis can go a long way towards solving the challenging problem of prior specification by learning the prior from the climatology. Strictly speaking, in the fully hierarchical setting, we do not actually learn the prior but the hyperposterior $p(\bpsi|\by_{1:N})$ that weights the usual conditional posterior $p(\btheta_n|\by_n,\bpsi)$ for each year $n$ in the marginalization step in \eqref{eq:nparpost}. Nonetheless, the idea of learning the prior from climatology is instructive: if we approximate the hyperposterior as a Dirac delta $p(\bpsi|\by_{1:N})\simeq \delta(\bpsi-\boldsymbol{\widehat{\psi}})$ centered on the MAP estimate $\boldsymbol{\widehat{\psi}}$, inserting this in \eqref{eq:nparpost} and expanding the usual conditional posterior yields
\begin{equation}
    p(\btheta_n |\by_{1:N}) \simeq p(\by_n|\btheta_n) \int \frac{p(\btheta_n|\bpsi )\delta(\bpsi-\boldsymbol{\widehat{\psi}})}{p(\by_n|\bpsi)} \, \mathrm{d}\bpsi = \frac{p(\by_n|\btheta_n)p(\btheta_n|\widehat{\bpsi})}{p(\by_n|\widehat{\bpsi})} \label{eq:plugin} \, ,
\end{equation}
which is a (type-II) MAP approximation of full hierarchical Bayesian inference \citep{Murphy2023}. Similarly, if $\boldsymbol{\widehat{\psi}}$ maximized the marginal likelihood (evidence), \eqref{eq:plugin} would correspond to the empirical Bayes approximation of hierarchical Bayes. The `plug-in' approximation in \eqref{eq:plugin} is often made implicitly in deep learning, where parameters are generally learned via optimization, which can be seen as a tractable approximation of full hierarchical Bayes \citep{MacKay1992,Murphy2023}. The prior term $p(\btheta_n|\boldsymbol{\widehat{\psi}})$ in \eqref{eq:plugin} now clearly contains the `learned' parameter climatology via the point estimate $\boldsymbol{\widehat{\psi}}$, so the full hierarchical solution involving the hyperposterior is a further extension of this. As such, hierarchical Bayes and approximations thereof provide a mechanism for adapting the hyperparameters to the data at hand. Even if we have tested this method on just a handful of snow sites and glaciers, this hierarchical approach provides a step towards specifying priors in global snow and glacier modeling that can adapt to the local climatology. 

Herein, we have focused on the temporal setting to obtain climate-informed prior hyperparameters. In \citet{Guillet2026}, we show how similar particle methods can be applied to partially pool information in space rather than time to infer regional prior hyperparameters. Recently, \citet{PerezVieites2025} extended hierarchical inference from two (static, slow) to three (static, slow, fast) timescales in general state space modeling, paving the way towards adding additional levels of hyperparameters in cryospheric data assimilation. Beyond the need to consider hyperparameters in the prior, constructing the likelihood often involves equally `subjective' hyperparameter choices. Here too, hierarchical Bayes and its approximations can be used to infer likelihood-related hyperparameters. For example, \citet{Viani2023} and \citet{vanHove2026} both used particle methods to infer the observation error hyperparameters. Beyond strong constraint data assimilation, several studies have also shown how approximate hierarchical methods can be used to infer hyperparameters related to model error in data assimilation \citep{Pulido2018,Lucini2021}. In summary, hierarchical Bayes provides a valuable and general framework for inferring the plethora of uncertain hyperparameters that may arise in geophysical data assimilation.

\subsection{Algorithmic trade-offs}

A major challenge in the application and widespread adoption of hierarchical Bayesian inference in geophysical data assimilation is the development of algorithms that are tractable to implement using forward models that have a non-trivial cost. The knee-jerk response to this challenge would be joint hierarchical inference via state of the art gradient-based MCMC methods, such as Hamiltonian Monte Carlo, that are widely used in machine learning \citep{Neal1996,MacKay2003,Murphy2023}  and statistics \citep{Gelman2013,McElreath2020} as the engine of most modern probabilistic programming languages. Although there have been several successful applications of the more classical gradient free MCMC samplers to cryospheric problems \citep{Kolberg2006,Rounce2020}, these have been mostly limited to relatively simple models and non-hierarchical inference. While our study also falls under the former category of low complexity models, the latter limitation no longer applies.

Beyond the sheer number of forward model evaluations, typically on the order of tens to hundreds of thousands, needed by MCMC, such a joint `all at once' inference approach is poorly suited to state space modeling as it does not leverage their temporal structure \citep{Sarkka2023}. Taking our graphical model for partial pooling in Figure~\ref{fig:pgm} as an example, a joint approach would require simultaneously inferring both the $N_\theta \times 2$ hyperparameter vector $\bpsi$ and a massive $N_\theta \times N$ parameter vector $\btheta_{1:N}$, where we recall that the number of years is on the order of $10^2$ and the number of parameters $N_\theta=3$ or $4$ for glaciers and seasonal snow, respectively (Table~\ref{tab:prior}). While such a relatively large ($>100$) parameter space could still be tractable for gradient-based MCMC with the appropriate reparametrizations \citep{Murphy2023}, it is completely intractable for classical MCMC samplers. Adding parameter memory between years via Markovian transitions, a possible extension of our model, would take even state-of-the-art MCMC samplers off the table. This explains why, in the domain of state space modeling, MCMC is rarely a viable candidate for inference and is replaced by particle methods \citep{Sarkka2023}. The exception to this is in the use of hybrid MCMC-particle methods \citep{Chopin2020}, such as using MCMC to ensure particle diversity in the resample move algorithm \citep{Gilks2001,vanHove2025} or using PMCMC for hierarchical inference \citep{Andrieu2010}. 

The latter PMCMC (Appendix~\ref{app:PMCMC}) approach that we used as a benchmark is the current gold-standard for hierarchical state space models \citep{Chopin2020,Sarkka2023,Murphy2023}. Unlike joint inference with classical gradient-free MCMC, the PMCMC approach uses nested inference to leverage the conditional structure of state space models to make inference tractable. It uses particle methods in the inner parameter loop to infer the conditional posterior $p(\btheta|\by,\bpsi)$ along with the evidence $p(\by|\bpsi)$, where the latter is used as the marginal likelihood together with the hyperprior to update the hyperparameters via MCMC in the outer loop. Even this is computationally very intensive, requiring $10^8$ forward model simulations: arising from $10^5$ iterations in the outer loop multiplied by $10^3$ iterations in the inner loop. The $10^5 $ total number of iterations in the outer loop arises from the number of parallelizable chains ($N_c=10$) multiplied by the number of steps in each chain ($N_m=10^4$). The $10^3$ iterations in the inner loop arise from the number of forward model runs required to run a single generation of MAGPIES, namely $(N_a+1)\times N_e=500$, as described in Section~\ref{sec:MAGPIES}, which is run for \emph{both} the current $\bpsi_i$ and proposed $\bpsi_*$ hyperparameters to approximate the correlated pseudo-marginal method \citep{Deligiannidis2018} (Appendix~\ref{app:PMCMC}) . 

Even if some of the $10^8$ forward model runs are partly parallelizable across both chains and particles, PMCMC remains costly. This was despite the use of a low complexity temperature index model, where running MAGPIES with $5$ iterations of $100$ vectorized ensemble members in the inner loop takes on the order of a second on most sites. In particular, this PMCMC routine takes several hours of wall clock time to complete for a single site running $10$ chains in parallel on $10$ cores and vectorizing the ensemble. So, while it is certainly possible to run our variant of PMCMC for benchmarking at a few sites, scaling this up to regional (let alone global) studies remains prohibitive. Moreover, in our experiments, this gold-standard yet costly PMCMC implementation did not clearly outperform our much more computationally tractable nested particle smoothing approach summarized in Figure~\ref{fig:workflow}. The performance of PMCMC sampling might be improved by further increasing the number of steps in the Markov chains from $N_m=10^4$ up to $10^5$ to handle the poorer mixing at data-rich sites like Filefjell, but that would entail an even more intractable runtime for spatially distributed simulations. We emphasize that these findings are strictly conditional on our PMCMC implementation, which could be refined by adopting the exact conditional pseudo-marginal method as well as gradient-based MCMC. Furthermore, the cost of PMCMC could be reduced considerably by extending likelihood emulation for MCMC \citep{Keetz2025} to evidence emulation for PMCMC. 

% Kris is here
On the other lower end of the computational complexity spectrum for approximate hierarchical inference, we have the type-II MAP method based on particle expectation maximization (Section~\ref{sec:EM}). Using this method, the hyperposterior is replaced by a Dirac delta centered on the maximum a posteriori estimate (MAP) of the hyperparameters $\boldsymbol{\widehat{\psi}}$ as a `plug-in' approximation. As such, we can approximate the full hierarchical solution via partial pooling simply by running the reanalysis with the MAP hyperparameters, as shown in \eqref{eq:plugin}. Thereby, the cost of the MAP approach is only $(G+1)\times(N_a+1)\times N_e=1500$ forward model runs, where $G=2$ is the number of MAGPIES generations (Section~\ref{sec:MAGPIES}), and as before, the ensemble dimension is parallelizable. Strictly, the cost is proportional to $G+1$ rather than $G$, as an ensemble-based reanalysis needs to be rerun with the MAP hyperparameters once these have been estimated via EM. On the one hand, the cost could thus clearly be reduced further by setting the number of MAGPIES generations to $G=1$, whereby the MAP approach would only be twice as expensive as a traditional no pooling cryospheric reanalysis in terms of the number of forward model runs. On the other hand, reducing the number of generations may degrade the accuracy of the particle EM approach. As noted in Section~\ref{sec:results_schemes}, even if the MAP approach performed marginally worse than the nested particle smoother for PP, it often closely matched its performance while also generally outperforming both NP and CP in both the calibration and validation periods. This MAP approach may thus serve as a useful approximation of nested particle smoothing that can be even more tractable at scale, such as for global cryospheric reanalysis. 

Nested particle smoothing nicely bridges the tractability of the MAP method and the theoretically improved performance of the more intensive PMCMC method. Unlike PMCMC, nested particle smoothing currently allows for the implementation of hierarchical cryospheric reanalysis at the regional scale, even with relatively modest computational resources. Just like the MAP approach, nested particle smoothing requires rerunning an ensemble-based reanalysis with the updated hyperparameters. Unlike the MAP approach, however, nested particle smoothing requires reruns for each sample from the hyperposterior rather than just for the MAP estimate. As such, the cost incurred by the nested particle smoothing is first $G\times(N_a+1)\times N_e=2\times5\times100=1000$ forward model runs for a particle approximation of the hyperposterior (Section~\ref{sec:AMIS}), followed by typically $10\leq N_s\leq 100$ unique particles that are then rerun with $N_s\times(N_a+1)\times N_e \leq 50000 $ forward model runs for the posterior predictions. Thus, the total cost of the nested particle smoother is at most on the order of $10^4$ forward model runs, which is four orders of magnitude lower than PMCMC and one order higher than the MAP approach.  

Lastly, it is worth comparing the cost of all the above schemes used for hierarchical reanalysis via PP to the more typical NP and CP approaches for climate-assumed reanalysis and static calibration, respectively. In our experiments, we used the same underlying MAGPIES approach for both no pooling and complete pooling. The main difference is that the hyperparameters were fixed to default climate-assumed values $\boldsymbol{\psi}_0$ corresponding to the hyperprior mean. To avoid giving the hierarchical schemes too much of an advantage in terms of the number of forward model runs they were afforded, we increased the ensemble size from $N_e=100$ to $N_e=1000$ for the CP and NP runs. Thus, NP and CP both incurred a total cost of $(N_a+1)\times N_e=5\times 1000=5000$ forward model runs, which is considerably higher than that of the MAP scheme and within an order of magnitude of the typical cost of nested particle smoothing. The considerable performance gains from hierarchical reanalysis via PP, whether using MAP or nested particle smoothing, compared to the non-hierarchical approaches cannot, therefore, be explained by the number of forward model runs they have been afforded. The difference is probably more fundamental, boiling down to the fact that the `deeper' and more flexible probabilistic graphical model in the case of partial pooling (Figure~\ref{fig:pgm}) is likely a better representation of reality than the non-hierarchical pooling endmembers. 

\subsection{Remaining challenges}

Despite the clear improvements provided by hierarchical inference via PP relative to the non-hierarchical baselines, there are several lingering challenges worth highlighting. While the climate-informed cryospheric reanalysis via PP provides increased flexibility by inferring uncertain climatological hyperparameters as opposed to fixing them a priori, it still assumes stationary hyperparameters. The ability to learn such a static background climatology (Section~\ref{sec:results_learning}), as opposed to assuming that it is known, is a considerable development over previous approaches to reanalysis and calibration. However, the inherent assumption of a stationary parameter climatology is questionable in the face of ongoing climate change \citep{Arias2021}, which is amplified in cold regions \citep{Meredith2019}. Thus, we expect the multi-decadal climatology of the uncertain parameters, especially those related to air temperature ($b$) and snowfall ($c$), to be affected by accelerated warming in recent decades. We see some evidence of this in the later part of the 21$^\text{st}$-century validation period through an increasing discrepancy between the reference (yellow) and PP using only $20^\text{th}$-century calibration data for snowfall at Claridenfirn (Figure~\ref{fig:learning}). 

Although hyperparameter stationarity is currently a limitation in our hierarchical reanalysis, parameter stationarity plagues most, if not all, calibration methods for cryospheric modeling \citep{Rounce2023,Sjursen2025} and beyond \citep{Keetz2025}. Ultimately, we are able to push stationarity up one level by moving from CP to PP. Thereby, with the benefit of hindsight, climate-informed reanalysis gains the advantage of adapting to the particularities of observed years while safely reverting to a well-calibrated climatology in years without observations. Nonetheless, for both the unobserved past and the yet to be observed future, climate-informed reanalysis reverts to the static hyperparameters that have been inferred. A challenge for future work will be to also make the hyperparameters adapt to non-stationary conditions. 

Beyond stationarity, another challenge with hierarchical cryospheric reanalysis via nested particle smoothing is the relatively high complexity of the workflow shown in Figure~\ref{fig:workflow} and described in Section~\ref{sec:method}. As opposed to a more naive yet potentially intractable nested approach using basic particle smoothers without adaptation or guidance, our approach involves a synthesis of several methods: particle-adjusted iterative ensemble Kalman smoothers, expectation maximization, the Laplace approximation, and an amortized AdaPBS scheme. This workflow was devised to minimize the number of forward model runs required to infer an approximate hyperposterior distribution using expectation maximization as a guide and recycling samples by reusing deterministic mixture proposals. The implementation of particle expectation maximization is quite involved, as it requires manually deriving and coding both the Jacobian and Hessian of the energy function used in the gradient-based maximization step (Appendix~\ref{app:grad}). When done manually, this particular step is time consuming, potentially error prone, and challenging to adapt to new models. More generally, at least in its current form, the nested particle smoothing workflow and code implementation can make it challenging to adapt it to higher complexity cryospheric models \citep{Westermann2023} and emerging satellite retrievals \citep{Cluzet2024,Mazzolini2025}. Nonetheless, most of the key schemes used in our workflow, from MAGPIES to AdaPBS, are available to the cryospheric community for further remixing through the MuSA open source snow data assimilation toolbox \citep{AlonsoGonzalez2022,Aalstad2026}.

\subsection{Outlook}

The remaining challenges could be addressed in several ways. Firstly, the current partial pooling model in Figure~\ref{fig:pgm} could be extended. A possible extension would be to add Markovian dynamics to the parameters \citep{Katzfuss2020}, adding conditional dependence (arrows) between the parameters in subsequent years. This is akin to a hierarchical filtering problem in state space modeling \citep{Sarkka2023}, introducing an additional memory hyperparameter for each parameter. Our model is an extreme case of this approach that assumes no short-term memory between years while still sharing information across the climatology via pooling. Just as hierarchical inference automatically sets the degree of pooling, it could also determine the ideal amount of inter-annual memory for each parameter prior. As such, we might be able to identify whether years with a wet/dry (or warm/cold) bias tend to arrive in sequence and leverage this for improved inference, especially in data sparse years. 

Building upon these parameter dynamics, the assumption of static hyperparameters can be relaxed by adding an additional intermediate slower parameter level to the temporal hierarchy. This would yield three parameter levels: a faster (annual), a slower (e.g., decadal), and a static level. A similar three layered approach was recently successfully applied in hierarchical state space modeling by \citet{PerezVieites2025} using hybrid particle and Kalman methods, providing confidence that it would also be possible to extend our model with a new intermediate layer encoding slow dynamics. The hierarchical approach can also be further extended by adding external predictors for higher level parameters, as is commonly done in hierarchical linear models and their generalizations \citep{Gelman2013}. To the best of our knowledge, such a predictive approach has yet to be translated to cryospheric data assimilation with mechanistic models. It would allow parameters to adapt to the specific conditions of unseen years through dynamic predictors, as opposed to reverting to a static climatology, potentially improving reanalyses of the deeper past, near-term forecasts, and even longer-term climate projections. 

A transition to differentiable programming using automatic differentiation \citep{Sapienza2025} has the potential to improve our workflow for hierarchical Bayesian inference in climate-informed cryospheric reanalysis. This would eliminate the time consuming and potentially error-prone exercise of manually deriving and coding hyperparameter gradients that we use for expectation maximization \citep{Neal1998}. It also enables the use of rapid gradient-based Bayesian inference schemes, from the classical Laplace approximation \citep{MacKay1992,MacKay2003} to variational Bayes  \citep{Murphy2023}, that can enhance the Monte Carlo schemes we use \citep{Schillings2020}. Furthermore, differential programming could lead to substantial run time accelerations by facilitating running simulations and performing inference on graphical processing units. Although differentiable programming is gaining traction and showing promise for calibrating cryospheric models \citep{Schmitt2026,Maslov2026}, these efforts have yet to be embedded in a fully Bayesian uncertainty-aware framework. 

Stepping down from these algorithmic clouds, there are several lower hanging fruits that can be picked by adapting the hierarchical cryospheric reanalysis framework. The first step could be to combine our temporal approach with the spatial approach in \citet{Guillet2026} to enable hierarchical spatio-temporal cryospheric reanalysis that can borrow strength in both space and time. Hierarchical cryospheric reanalysis could also benefit from assimilating more satellite retrievals, such as higher resolution FSCA \citep{Willmes2025}, wet snow lines \citep{Cluzet2024}, snow depth \citep{Mazzolini2025,Dunmire2024}, and glacier thickness \citep{Hugonnet2021}. Extending hierarchical Bayesian cryospheric reanalysis to the global scale for snow, glaciers, and permafrost is also becoming an increasingly tractable task, especially with simpler cryospheric models \citep{Rounce2023,Elias2024}. Going global with intermediate complexity \citep{Essery2025}, let alone full complexity \citep{Westermann2023}, cryospheric models will require further exploration of promising techniques such as emulation \citep{Blandini2025} and amortized inference \citep{Arruda2026}. 

\conclusions  %% \conclusions[modified heading if necessary]

In this study, we presented a climate-informed approach to cryospheric reanalysis via hierarchical Bayesian data assimilation. By introducing an additional climatological hyperparameter level of inference above the usual annual (water year) parameter level of inference, this approach jointly infers the historical trajectory of cryospheric state variables \emph{and} the background parameter climatology. Not only can learning the parameter climatology be useful for extending reanalyses further into the sparsely observed past and for future predictions, but it can even improve state estimates for years with dense observations. 

We demonstrated the merits of climate-informed cryospheric reanalysis via partial pooling by comparing it to the current extreme practices of annual cryospheric reanalysis (no pooling) and static parameter calibration (complete pooling). This comparison was carried out in three experiments, assimilating multi-decadal in situ snow water equivalent (SWE), satellite-based fractional snow-covered area (FSCA), and glacier-wide mass balance data spanning four snow sites and four glaciers from the Swiss Alps to the High Arctic. Averaged across all experiments, hierarchical reanalysis improved the continuous ranked probability score (CRPS) by $65\%$ compared to the default prior in the calibration (training) period, beating the runner-up annual reanalysis ($54\%$ improvement). In the more challenging validation (testing) period, hierarchical reanalysis maintained a high $36\%$ score improvement, well above the runner-up annual reanalysis approach ($22\%$). Overall, weighing both periods equally, hierarchical reanalysis provided a $51\%$ overall improvement, exceeding that of annual reanalysis ($38\%$) and static calibration ($19\%$). Given this ranking in terms of improvement in the uncertainty-aware CRPS, our experiments provide empirical evidence that hierarchical inference can markedly enhance cryospheric reanalyses. 

From a Bayesian perspective, the benefit of the hierarchical partial pooling approach to reanalysis is expected from theory, especially in light of Stein's paradox, as it enables learning from climatology. At the same time, we have taken a step towards making hierarchical inference computationally tractable for cryospheric reanalysis by using hybrid nested particle smoothing guided by expectation maximization. Using this nested particle smoothing workflow, we were able to cut computational costs by several orders of magnitude compared to the gold-standard PMCMC method while matching or even improving its performance given the resources at our disposal. We also presented a type-II maximum a posteriori (MAP) hierarchical inference approximation at a fraction of the cost of full nested smoothing that only marginally degrades performance. By making hierarchical inference more tractable, we are further on the path towards advancing Bayesian data assimilation methods for large scale cryospheric reanalysis, forecasting, and projections. 

\appendix

\section{Gradient-based hyperparameter optimization}\label{app:grad}

For the M-step in the EM algorithm, we seek to maximize the objective function $f_i$ with respect to the hyperparameters $\bpsi$. Recall that this is (approximately) equivalent to minimizing the energy function $\phi$ \eqref{eq:energy} based on the ELBO \eqref{eq:ELBO},  whereby $f_i\simeq -\phi$. Although it is not evident how to solve this optimization problem analytically, we have access to the gradients of $f_i$ in closed analytical form, which are related to those of $\phi$ via \eqref{eq:Fisheri}. As such, we are in a position to employ efficient second-order gradient-based numerical optimization methods that can converge rapidly to a local optimum in just a handful of computationally cheap iterations. We use an adaptation of the Levenberg-Marquardt algorithm (LMA) as our second-order gradient-based method, which appears to be a robust choice based on its popularity and efficacy in the inverse problem literature \citep{Pujol2007}. The LMA adjusts a single parameter $\lambda$ that scales a diagonalized Hessian matrix $\mathbf{D}$ to adaptively `interpolate' between the slower but more robust basic gradient descent involving the Jacobian vector $\mathbf{J}=\frac{\partial \phi}{\partial \bpsi}$ for large $\lambda$ and the faster but more unstable second-order Newton method that also involves the Hessian matrix $\mathbf{H}=\frac{\partial^2 \phi}{\partial \bpsi^2}$ for small $\lambda$. Following \citet{Kochenderfer2025}, the second-order gradient-based LMA inspired update step is given by
\begin{equation}
\bpsi_{i+1}^{(\star)}=\bpsi_{i+1}^{(l)}-\left(\mathbf{H}_{i+1}^{(l)}+\lambda \mathbf{D}_{i+1}^{(l)}\right)^{-1}\mathbf{J}_{i+1}^{(l)} \label{eq:LMA} \, ,
\end{equation}
setting $\bpsi_{i+1}^{(l+1)}=\bpsi_{i+1}^{(\star)}$ and $\lambda \leftarrow \max( \lambda/2\Lambda,\lambda_\mathrm{min}) $ only if $f_i(\bpsi_{i+1}^{(\star)})\geq f_i(\bpsi_{i+1}^{(l)})$, i.e., if the fitness encoded in the objective function \eqref{eq:fdef} increases or remains constant. Otherwise, set $\lambda \leftarrow \Lambda \lambda $ and continue to repeat the LMA step above until the fitness increases or $\lambda=\lambda_\mathrm{max}$ before proceeding to the next iteration $l\leftarrow l+1$. Here, we set the $\lambda$ growth factor $\Lambda=10$ and allowed lambda to adapt between $\lambda\in[\lambda_\mathrm{min},\lambda_\mathrm{max}]$ with $\lambda_\mathrm{min}=10^{-7}$ and $\lambda_\mathrm{max}=10^7$, starting with an initial value of $\lambda=1$. The LMA update is carried out sequentially over a set of $l_\mathrm{max}$ iterations indexed by $l=0,\dots,l_\mathrm{max}$, starting with an initial guess for the hyperparameters, for which we use the hyperparameter estimate from the last M-step $\bpsi_{i+1}^{(0)}=\bpsi_{i}$. Note that although one M-step consists of performing $l_\mathrm{max}$ LMA iterations to obtain the updated hyperparameter $\bpsi_{i+1}=\bpsi_{i+1}^{(l_\mathrm{max})}$, these LMA steps are computationally cheap. These steps involve only basic linear algebra operations that do not require rerunning any forward model runs, since they always operate on the same objective function $f_i$ within a given M-step. It was necessary to modify \eqref{eq:LMA} to escape saddle points by ensuring that the Hessian is positive definite. Positive definiteness is readily diagnosed by checking if a Cholesky decomposition is successful. When this decomposition fails, positive definiteness is enforced based on the saddle-free Newton method of \citet{Dauphin2014}, where the Hessian $\mathbf{H}$ in \eqref{eq:LMA} and \eqref{eq:ddiag} is replaced by the matrix $|\mathbf{H}|$ formed by taking the absolute value of the eigenvalues of $\mathbf{H}$. 

What remains is to provide a recipe for computing the gradient-based matrices required by the LMA optimization technique. Specifically, the $N_\psi \times 1$ Jacobian vector $\mathbf{J}$ and the $N_\psi \times N_\psi$ Hessian matrix $\mathbf{H}$ need to be populated by updated gradients after each successful LMA iteration in \eqref{eq:LMA}. Recall that $N_\theta=4$ ($3$) is the number of snow (glacier) parameters in $\btheta$, such that here we have $N_\psi=2N_\theta=8$ ($6$) snow (glacier) model hyperparameters in $\bpsi$, which are the prior mean $\boldsymbol{\mu}$ and the logit-transformed standard deviation $\boldsymbol{\tau}=\mathrm{glt}(\boldsymbol{\sigma})$ of the parameters in $\btheta$. Given that we define the hyperparameter vector as a block vector of the form $\bpsi=[\boldsymbol{\mu}, \,\boldsymbol{\tau}]$, then the Jacobian vector $\mathbf{J}=\frac{\partial \phi}{\partial \bpsi}\simeq -\frac{\partial f_i}{\partial \bpsi}$ becomes a block vector of the form
\begin{equation}
\mathbf{J}=\left[\left(-\frac{\partial f_i}{\partial \boldsymbol{\mu}}\right)^\mathrm{T}, \, \left(-\frac{\partial f_i}{\partial \boldsymbol{\tau}}\right)^\mathrm{T} \right]^\mathrm{T} \, , \label{eq:Jac}
\end{equation}
where the $N_\theta\times1$ column vectors $\frac{\partial f_i}{\partial \boldsymbol{\mu}}$ and $\frac{\partial f_i}{\partial \boldsymbol{\tau}}$ denote the gradient with respect to the prior mean and logit-transformed standard deviation hyperparameters, respectively. When used in the LMA step \eqref{eq:LMA}, our suggestive notation $\mathbf{J}_{i+1}^{(l)}$ is a reminder that this Jacobian is to be evaluated at a particular setting for these hyperparameters, namely $\bpsi=\bpsi_{i+1}^{(l)}$. Combining the definition of $f_i$ \eqref{eq:fdef} and the particle approximation \eqref{eq:parq}, then $f_i$ is
\begin{equation*}
    f_i(\bpsi)=c-\sum_{j=1}^{N_\theta}\left(\frac{(\mu_j-m_j)^2}{2s_j^2}+\frac{(\tau_j-\eta_j)^2}{2\chi_j^2}+\sum_{n=1}^N\sum_{h=1}^{N_h} w_{ni}^{(h)}\left[\frac{(\theta_{nj}^{(h)}-\mu_j)^2}{2\sigma_j^2}+\log(\sigma_j)\right]\right) \, ,
\end{equation*}
where we have used that $\mathrm{get}(\tau_j)=\sigma_j \Leftrightarrow \tau_j = \mathrm{glt}(\sigma_j)$ and that the shorthand term $c$ does not depend on the hyperparameters $\bpsi$, since it consists of the entropy \eqref{eq:entropy} and constants. By judicious application of the chain rule, it is readily shown that the gradients of $f_i$ with respect to the hyperparameters $\mu_k$ and $\tau_k$ in $\bpsi$ are given by
\begin{equation}
    \frac{\partial f_i}{\partial \mu_k}= \frac{m_k-\mu_k}{s_k^2}+\sum_{n=1}^{N}\sum_{h=1}^{N_h}w_{ni}^{(h)}\frac{(\theta_{nk}^{(h)}-\mu_k)}{\sigma_k^2} \, , \label{eq:dfdmu}
\end{equation}
and 
\iffalse
\begin{equation}
    \frac{\partial f_i}{\partial \tau_k}=\frac{\eta_k-\tau_k}{\chi_k^2} + \sum_{n=1}^{N}\sum_{h=1}^{N_h}w_{ni}^{(h)}\left[\frac{(\theta_{nk}^{(h)}-\mu_k)^2}{\exp(2\tau_k)}-1\right] \, ,  \label{eq:dfdtau}
\end{equation}
\fi
\begin{equation}
    \frac{\partial f_i}{\partial \tau_k}=\frac{\eta_k-\tau_k}{\chi_k^2} + \sum_{n=1}^{N}\sum_{h=1}^{N_h}w_{ni}^{(h)}\left[\frac{(\theta_{nk}^{(h)}-\mu_k)^2}{\sigma_k^3}-\frac{1}{\sigma_k}\right]\frac{\partial \sigma_k}{\partial \tau_k} \, ,  \label{eq:dfdtau}
\end{equation}
where here $k=1,\dots,N_\theta$ indexes the $N_\theta=4$ parameters in $\btheta=[\theta_1,\theta_2,\theta_3,\theta_4]^\mathrm{T}=[\alpha,\beta,\gamma,\nu]^\mathrm{T}$ and their associated hyperparameters $\mu_k$ and $\tau_k$ in $\bpsi$. Using the generalized logit ($\mathrm{glt}(\cdot)$) and its inverse transform, the generalized expit ($\mathrm{get}(\cdot)$), it is readily shown that \citep{Keetz2025}
\begin{equation}
 \frac{\partial \sigma_k}{\partial \tau_k}=\frac{(\sigma_k-\mathrm{a}_k)(\mathrm{b}_k-\sigma_k)}{(\mathrm{b}_k-\mathrm{a}_k)} \label{eq:dsdt}
\end{equation}
where $\mathrm{a}_k=0$ and $\mathrm{b}_k=1.4$ are the bounds on $\sigma_k$ given in Table~\ref{tab:prior}. Armed with tractable analytical gradient expressions in \eqref{eq:dfdmu} and \eqref{eq:dfdtau}, we can now rapidly populate the Jacobian vector in \eqref{eq:Jac} at each LMA iteration. 

The $N_\psi\times N_\psi$ Hessian matrix $\mathbf{H}=\frac{\partial^2 \phi}{\partial \bpsi^2} \simeq -\frac{\partial^2 f_i}{\partial \bpsi^2}$ requires computing second-order derivatives of the $f_i$ function. We make the immediate observation that all cross derivatives involving hyperparameters that are tied to different parameters, such as $\frac{\partial^2 f_i}{\partial \mu_l \mu_k}$ for $k\neq l$, will be zero, as can readily be verified by inspection of the  gradients \eqref{eq:dfdmu} and \eqref{eq:dfdtau}. In fact, due in part to the symmetry of second derivatives, we only have to calculate three kinds of second derivatives for each associated parameter ($k=1,\dots,N_\theta$) to populate the Hessian. First in line, using \eqref{eq:dfdmu}, we obtain
\begin{equation}
    \frac{\partial^2 f_i}{\partial\mu_k^2}= -\frac{1}{s_k^2}-\sum_{n=1}^N \sum_{h=1}^{N_h} \frac{w_{ni}^{(h)}}{\sigma_k^2} \, ,  \label{eq:d2fdmu2}
\end{equation}
next, using \eqref{eq:dfdtau}, we obtain
\begin{equation}
\frac{\partial^2 f_i}{\partial \tau_k^2}=-\frac{1}{\chi_k^2}+\sum_{n=1}^{N}\sum_{h=1}^{N_h}w_{ni}^{(h)}\left[\left(\frac{1}{\sigma_k^2}-\frac{3(\theta_{nk}^{(h)}-\mu_k)^2}{\sigma_k^4}\right)\left(\frac{\partial \sigma_k}{\partial \tau_k}\right)^2+\left(\frac{(\theta_{nk}^{(h)}-\mu_k)^2}{\sigma_k^3}-\frac{1}{\sigma_k}\right)\frac{\partial^2 \sigma_k}{\partial \tau_k^2}\right] \, ,  \label{eq:d2fdtau2}
\end{equation}
where using \eqref{eq:dsdt} we have that
\begin{equation*}
    \frac{\partial^2 \sigma_k}{\partial \tau_k^2} = \frac{(\mathrm{b}_k+\mathrm{a}_k-2\sigma_k)}{(\mathrm{b}_k-\mathrm{a}_k)}\frac{\partial \sigma_k}{\partial \tau_k} \, ,
\end{equation*}
and lastly, using both \eqref{eq:dfdmu} and \eqref{eq:dfdtau}, we obtain
\begin{equation}
\frac{\partial^2 f_i}{\partial\tau_k\partial \mu_k}= -2\sum_{n=1}^N \sum_{h=1}^{N_h} w_{ni}^{(h)}\frac{\left(\theta_{nk}^{(h)}-\mu_k\right)}{\sigma_k^3}\frac{\partial \sigma_k}{\partial \tau_k} \, , \label{eq:d2fdtaudmu}
\end{equation}
which together form the three kinds of second derivatives that populate the non-zero entries of our Hessian. So here the $N_\psi \times N_\psi$ Hessian matrix is a block matrix of the form
\begin{equation*}
\mathbf{H}=\begin{bmatrix} \mathbf{H}_{\boldsymbol{\mu}} & \mathbf{H}_{\boldsymbol{\tau}\boldsymbol{\mu}} \\
\mathbf{H}_{\boldsymbol{\tau}\boldsymbol{\mu}} & \mathbf{H}_{\boldsymbol{\tau}} \end{bmatrix} \, ,
    \end{equation*}
where we have defined the $N_\theta \times N_\theta$ diagonal matrices $\mathbf{H}_{\boldsymbol{\mu}}=\mathrm{diag}\left(-\frac{\partial^2 f_i}{\partial \boldsymbol{\mu}^2}\right)$, $\mathbf{H}_{\boldsymbol{\tau}}=\mathrm{diag}\left(-\frac{\partial^2 f_i}{\partial \boldsymbol{\tau}^2}\right)$, and $\mathbf{H}_{\boldsymbol{\tau}\boldsymbol{\mu}}=\mathrm{diag}\left(-\frac{\partial^2 f_i}{\partial \boldsymbol{\tau}\partial\boldsymbol{\mu}}\right)$ that are rapidly populated using the analytical second order derivatives in  \eqref{eq:d2fdmu2}, \eqref{eq:d2fdtau2}, and \eqref{eq:d2fdtaudmu}, respectively. The diagonal matrix $\mathbf{D}$ used in the LMA update \eqref{eq:LMA} is obtained using the diagonal of the Hessian
\begin{equation}  \mathbf{D}=\mathrm{diag}\left(\max\left(\mathrm{diag}\left(\mathbf{H}\right),\boldsymbol{\epsilon}\right)\right) \, , \label{eq:ddiag}
\end{equation}
where $\boldsymbol{\epsilon}=\epsilon \mathbf{1}_{N_\psi}$ with $\mathbf{1}_{N_\psi}$ is a $N_\psi \times 1$ vector of ones and $\epsilon=10^{-6}$ that helps ensure numerical stability \citep{Kochenderfer2025}. Recall that when used in the LMA step \eqref{eq:LMA}, our suggestive notation $\mathbf{H}_{i+1}^{(l)}$ (and $\mathbf{D}_{i+1}^{(l)}$) reminds us that the Hessian (and diagonal) matrix is evaluated for the particular hyperparameter setting $\bpsi=\bpsi_{i+1}^{(l)}$. Lastly, we also compute the Hessian evaluated at the final EM MAP-II hyperparameter estimate $\boldsymbol{\widehat{\psi}}=\bpsi_{\iota}^{(l_\mathrm{max})}$ and denote this as $\widehat{\mathbf{H}}=\mathbf{H}_{\iota}^{(l_\mathrm{max})}$.

\section{AdaPBS algorithm} \label{app:ada}

For hyperposterior inference with AdaPBS: we initialize the current AdaPBS iteration counter $R=1$, set the current sampling proposal to $q^{(R)}(\bpsi)=q^{(1)}(\bpsi)$ in \eqref{eq:defensive}, and proceed as follows:
\begin{enumerate}
\item Generate an ensemble of particles indexed by $i=1\,\dots,N_s$ from the current sampling proposal 
\begin{equation}
\boldsymbol{\bpsi}_i^{(R)}\sim q^{(R)}(\bpsi) \, , \label{eq:AMISsample}
\end{equation}
and append this to the current hyperparameter particle history.
\item For the current hyperparameter particle history, i.e., for all $r=1,\dots,R$ historical AdaPBS iterations and the ensembles of $i=1,\dots,N_s$ particles in each of these iterations, evaluate the hyperparameter DMP $\upsilon^{(R)}(\bpsi)$ for each particle $\bpsi_i^{(r)}$
\begin{equation}
\upsilon^{(R)}(\bpsi_i^{(r)})=\sum_{k=1}^{R} \frac{1}{R}q^{(k)}(\bpsi_i^{(r)}) \, , \label{eq:AMISDMP}
\end{equation} 
where the $i=1,\dots,N_s$ particles $\bpsi_i^{(r)}$ from each iteration $r$ have been previously drawn independently from their respective sampling proposal distributions $q^{(r)}$ through \eqref{eq:AMISsample}. Using the DM approach \citep{Owen2000}, we treat all the particles \emph{as if} they were collectively drawn from the mixture $\upsilon^{(R)}(\bpsi)$, from which we happen to obtain exactly $N_s$ particles from each proposal. 
\item Using the DM in \eqref{eq:AMISDMP} as our effective proposal for importance sampling, compute the unnormalized weights for the current particle history through
\begin{equation*}
\widetilde{\omega}_i^{(r)}=\frac{p(\by\mid \bpsi_i^{(r)})p(\bpsi_i^{(r)})}{\upsilon^{(R)}(\bpsi_i^{(r)})} \, .
\end{equation*}
As in \eqref{eq:MAGPIESjevi}, we can approximate the joint evidence $Z_{i}^{(r)}=\prod_{n=1}^N Z_{ni}^{(r)}\simeq p(\by\mid \bpsi_i^{(r)})$ in the numerator by using the annual factorization in \eqref{eq:jevi} and the MAGPIES estimate \eqref{eq:MAGevi} for the annual evidence $Z_{ni}^{(r)}\simeq p(\by_n\mid \bpsi_i^{(r)})$ to obtain the weights
\begin{equation} 
\widetilde{\omega}_i^{(r)} \simeq \left[\prod_{n=1}^N \frac{1}{N_h}\sum_{k=1}^{N_h} \widetilde{w}_{nir}^{(k)}\right]\frac{p(\bpsi_i^{(r)})}{\upsilon^{(R)}(\bpsi_i^{(r)})}  \, , \label{eq:AMISunw}
\end{equation}
where the extra AdaPBS iteration subscript $r=1,\dots,R$ in $\widetilde{w}_{nir}^{(k)}$ signals that we are evaluating \eqref{eq:MAGevi} at negligible cost to obtain the annual evidence estimates $\widehat{Z}_{ni}^{(r)}$ for hyperparameter setting $\bpsi_i^{(r)}$. For stability, we compute the log of the weights in \eqref{eq:AMISunw} and apply the log-sum-exp trick \citep{Murphy2023,Aalstad2026}. 
\item Using the entire particle history, compute the self-normalized particle weights 
\begin{equation}
\omega_i^{(r)}=\frac{\widetilde{\omega}_i^{(r)}}{N_R \mathcal{Z}_R} \, ,\label{eq:AMISnw}
\end{equation}
where $N_R=R N_s$ is the total number of historical particles at the current AdaPBS iteration $R$, and the normalizing constant in the denominator is
\begin{equation}
\mathcal{Z}_R=\frac{1}{N_R}\sum_{k=1}^{R}\sum_{i=1}^{N_s}\widetilde{\omega}_i^{(k)} \, ,\label{eq:hyperevi}
\end{equation}
is a particle approximation of the hyper evidence $p(\by)$ analogous to that for the evidence in \eqref{eq:ISevi}. We note in passing that the hyperevidence could be used at even higher levels of inference for more general model comparison \citep{Hoge2019,Llorente2023}. 
\item Compute the effective sample size using the $N_R$ historical particle weights \eqref{eq:AMISnw}
\begin{equation}
    N_\mathrm{eff}^{(R)}=\frac{1}{\sum_{k=1}^{R}\sum_{i=1}^{N_s}(\omega_i^{(k)})^{2}} \, , \label{eq:Neff}
\end{equation}
if a high enough effective sample size is obtained $N_\mathrm{eff}^{(R)}\geq \mathcal{I}$ where $\mathcal{I}=\mathrm{round}(\mathcal{D} N_s)$ with a predefined diversity ratio threshold $0<\mathcal{D} \leq 1$, or the maximum  number of iterations is reached, then we flag this $R$ as the final AdaPBS iteration.
\item Unless we have reached the final iteration, use weight clipping to improve the proposal adaptation \citep{Koblents2015} . This clipping first identifies the $\mathcal{I}$-th largest unnormalized weight based on the diversity threshold $\mathcal{D}$. Subsequently, as long as $\widetilde{\omega}_\mathcal{I}>0$, we apply the following weight clipping 
\begin{equation}
    \widetilde{\omega}_i^{(r)}\leftarrow \min\left(\widetilde{\omega}_i^{(r)},\widetilde{\omega}_\mathcal{I}\right) \, , \label{eq:clipping}
\end{equation}
with subsequent normalization via \eqref{eq:AMISnw} to obtain new normalized weights $\omega_i^{(r)}$. Clipping ensures that the $\mathcal{I}$ largest clipped weights will be equal, guaranteeing a clipped effective sample size $\geq \mathcal{I}$, which leads to a more robust and less degenerate sampling proposal adaptation. 
\item Resample an ensemble of $\varrho=1,\dots,N_s$ equally weighted particles $\bpsi_{\varrho}^{(R)}$ from the \emph{entire} current particle history $\bpsi_i^{(r)}$ using all the $N_R$ weights $w_i^{(r)}$ from \eqref{eq:AMISnw}, which will also have been clipped \eqref{eq:clipping} unless we have reached the final iteration. The subscript $\varrho$ (rather than $i$) signals that this is an ensemble of \emph{resampled} generally non-unique (i.e., contains copies) particles which form an equally weighted particle approximation of the hyperposterior 
\begin{equation*}
p(\bpsi\mid \by)\simeq \sum_{\varrho=1}^{N_s}\frac{1}{N_s}\delta(\bpsi-\bpsi_\varrho^{(R)}) \, ,
\end{equation*}
rather than being generally unique weighted particles that were independently drawn from the sampling proposals \eqref{eq:AMISsample}.
\item If we have reached the final iteration, then stop AdaPBS here and use the $N_s$ resampled particles $\bpsi_{\varrho}^{(R)}$ as an equally weighted particle approximation of the hyperposterior distribution. Otherwise, use the resampled particles to adapt the new sampling proposal distribution $q^{(R+1)}(\bpsi)$ for the next iteration. This proposal is constructed as a Gaussian, following \citet{Aalstad2026}, with mean and covariance obtained from the resampled particle ensemble $\bpsi_{\varrho}^{(R)}$. Lastly, update the iteration counter $R\leftarrow R+1$ and return to step 1 with the newly adapted sampling proposal $q^{(R)}$. 
\end{enumerate}

\section{Particle Markov Chain Monte Carlo benchmarks}\label{app:PMCMC}

Particle Markov Chain Monte Carlo \citep[PMCMC;][]{Andrieu2010} is a gold standard approach for hierarchical Bayesian inference in nonlinear state-space models with intractable evidence (marginal likelihood) terms \citep{Chopin2020,Sarkka2023}, such as our climate-informed reanalysis. Unlike the `resample-move' approach, where MCMC is used inside particle methods to help ensure particle diversity \citep{Gilks2001,Chopin2002,vanHove2025}, PMCMC uses particle methods inside MCMC to estimate the intractable evidence, allowing for hierarchical inference in state-space models where standard stand-alone MCMC is either not applicable or scales poorly \citep{Andrieu2010}. Among the many variants of PMCMC, we adopt the Particle Marginal Metropolis Hastings (PMMH) method first proposed by \citet{Andrieu2010} due to its relative ease of implementation and asymptotically robust performance. Therein, we use the Robust Adaptive Metropolis \citep[RAM;][]{Vihola2012} to automate the tuning of MCMC together with the MAGPIES algorithm from Section~\ref{sec:MAGPIES} for efficient particle approximations. As such, our implementation is similar in spirit to that of \citet{Sarkka2015} in terms of using RAM, as well as the recent work of \citet{Katzfuss2020} and \citet{Drovandi2022}
that leverage ensemble Kalman methods in PMCMC-like approaches.

Our PMCMC scheme is initialized by sampling $\bpsi_0$ near the MAP estimate $\widehat{\bpsi}$ from Section~\ref{sec:EM}, and subsequently estimating the associated joint evidence $p(\by\mid \bpsi_0)\simeq \widehat{Z}_0 $ using a single generation (i.e., $G=1$) of MAGPIES via \eqref{eq:MAGPIES}, \eqref{eq:MAGevi}, and \eqref{eq:MAGPIESjevi}. Then, we proceed as follows for $i=0,\dots,(N_m-1)$ sequential Markov chain steps:
\begin{enumerate}
\item Propose a new step in the hyperparameter space via the current Gaussian RAM proposal $q(\bpsi_*\mid \bpsi_i)=\mathcal{N}(\bpsi\mid \bpsi_i,\mathbf{C}_i)$, where $\mathbf{C}_i$ is the $N_\psi \times N_\psi$ proposal covariance matrix at the current iteration. 
\item Estimate the joint evidence (marginal likelihood) for the proposed hyperparameter setting $p(\by\mid \bpsi_*)\simeq \widehat{Z}_* $ using MAGPIES via \eqref{eq:MAGPIES}, \eqref{eq:MAGevi}, and \eqref{eq:MAGPIESjevi}. Since we are not concerned with efficiency in the PMCMC benchmark, we set $G$ in MAGPIES to a maximum of $G_\mathrm{max}=1$ and only keep the last generation of hyperparameter settings consisting of the newly proposed $\bpsi_*$ with the IES in memory. This is tantamount to running a $(N_a+1)N_e$ ensemble of forward models obtained from IES, with hyperparameters set to the proposed $\bpsi_*$ for each step in the Markov chain and using a single generation MAGPIES with the DM proposal to estimate the marginal likelihood \eqref{eq:MAGPIESjevi}.
\item To avoid the well known issue of sticky chains (i.e., poor mixing) with PMCMC \citep{Deligiannidis2018,Drovandi2022}, we approximate the correlated pseudo-marginal method proposed by \citet{Deligiannidis2018}. In practice, we fix the random seed within (but not across) the individual steps of a given Markov chain and re-compute particle approximations of both the proposed $p(\by\mid\bpsi_*)$ and current $p(\by\mid\bpsi_i)$ via single generation MAGPIES runs with completely correlated (i.e., identical) realizations of random variables. This is strictly an approximation of the correlated pseudo-marginal method that combines it with noisy Metropolis Hastings \citep{Medina2016} between steps, which can introduce a slight bias in the sampler but is much more straightforward to implement than the exact correlated pseudo-marginal method. Note that our noisy correlated approach requires running a full IES twice at each step: once again for the current hyperparameters $\bpsi_i$ and once for the proposed hyperparameters $\bpsi_*$, each at a cost of $(N_a+1)N_e$ forward model runs per IES evaluation. We refer the reader to \citet{Deligiannidis2018} for details on the correlated pseudo-marginal method, but note in passing that including this step was key to ensure adequate Markov chain mixing.
\item Estimate the Metropolis Hastings acceptance ratio \citep{Robert2004}
\begin{equation}
A_{i+1} = \frac{p(\by\mid \bpsi_*)p(\bpsi_*)q(\bpsi_i\mid \bpsi_*)}{p(\by\mid \bpsi_i)p(\bpsi_i)q(\bpsi_*\mid \bpsi_i)} \simeq \frac{\widehat{Z}_*p(\bpsi_*)}{\widehat{Z}_i p(\bpsi_i)} \, , \label{eq:aratio}
\end{equation}
where the proposal densities have canceled out due to symmetry $q(\bpsi_*\mid \bpsi_i)=q(\bpsi_i\mid \bpsi_*)$ and in this PMCMC method we use particle approximations of the evidence terms. 
\item The proposed step is probabilistically accepted so that we only move to the proposed point in hyperparameter space $\bpsi_{i+1}=\bpsi_*$ in the next iteration under the condition that $U_{i+1}\leq \mathrm{min}(A_{i+1},1)$ holds, where $U_{i+1}\sim\mathcal{U}(0,1)$ is a realization from the standard uniform distribution. Otherwise, the proposed step is rejected, and the Markov chain remains at the current point $\bpsi_{i+1}=\bpsi_i$ in the next iteration. As such, we are guaranteed to accept proposed points with a higher (estimated) hyperposterior density where $A_{i+1}>1$, while still accepting proposed points with a lower density with probability $A_{i+1}$.  Thereby, the target hyperposterior is the stationary distribution that we sample from asymptotically (i.e., with enough steps) with this Markov chain \citep{MacKay2003,Robert2004}.
\item Use the acceptance ratio $A_{i+1}$ and the current proposal covariance $\mathbf{C}_i$ to obtain the proposal covariance for the next iteration $\mathbf{C}_{i+1}$ by following the RAM covariance adaptation rule \citep{Vihola2012,Sarkka2023}. Lastly, unless $i+1=N_m$, update the iteration counter $i\leftarrow i+1$ and return to step 1 with the adapted proposal covariance $\mathbf{C}_i$. 
\end{enumerate}
In terms of algorithmic settings, we adopt the adaptation parameters suggested by \citet{Vihola2012} and run a total of $N_c=10$ independent Markov chains in parallel, each for a total of $N_m = 10^4$ PMCMC iterations, where the first $50\%$ are discarded as a warm up (or burn in) phase \citep{Gelman2013}. The $N_c=10$ chains are all randomly initialized by sampling $\bpsi_0 \sim \mathrm{N}(\widehat{\bpsi},\mathbf{C}_0)$ independently from a multivariate Gaussian centered on the MAP-II estimate with isotropic covariance $\mathbf{C}_0=\sigma_0^2\mathbf{I}_{N_\psi}$, where $\sigma_0=0.3$. The subsequent proposal covariance matrices $C_i$ are then adapted on the fly via the RAM algorithm \citep{Vihola2012}. From the full collection of samples across all the $N_c=10$ final chains, we then pick $10^3$ hyperparameter samples representing $p(\bpsi\mid \by)$ at random and their associated parameter particle ensembles representing $p(\btheta\mid \by,\bpsi)$. Together these represent the joint posterior $p(\btheta,\bpsi\mid \by)$ from which we obtain the target annual posteriors $p(\btheta_n\mid \by_{1:N})$ via Monte Carlo marginalization. The cost of this PMCMC algorithm is $N_c \times N_m \times 2 \times (N_a+1) \times N_e \simeq 10^8$ mostly non-parallelizable forward model runs over all water years (assimilation windows). This is clearly much higher than the $\simeq 10^3$ model runs cost that we aim for with our proposed nested ensemble-data assimilation method using AMIS and MAGPIES. As such, herein the PMCMC algorithm is used purely as a `gold standard' benchmark in the spirit of \citet{Law2012} and is not (at least in its current form) viable for spatially distributed hierarchical cryospheric reanalysis.

\codedataavailability{ESA Snow\_cci MODIS SCFG satellite retrievals \citep{MODIScci}, based on NASA MODIS Terra radiance data \citep{MOD02}, are available via \url{https://archive.ceda.ac.uk/}.
In situ data from the seasonal snow sites were provided by the Institute of Geophysics, Polish Academy of Sciences, for Hornsund on request, SMHI for Abisko via \url{https://opendata.smhi.se}, NVE for Filefjell on request, and SLF for Weissfluhjoch on request. Glacier-wide mass balance data were provided by NPI for Austre Brøggerbreen via \url{https://mosj.no/en/}, Stockholm University for Storglaciären via \url{https://bolin.su.se/data/v7/}, NVE for Storbrean via \url{https://glacier.nve.no/Glacier/viewer/CI/en/}, GLAMOS for Claridenfirn via \url{https://www.glamos.ch}, and the WGMS for a synthesis through the FoG database via \url{https://doi.org/10.5904/wgms-fog-2025-02b}. Modified Copernicus Sentinel data in Figure~\ref{fig:area} were obtained via Google Earth Engine \citep{Gorelick2017} using the Earth Engine Python API \texttt{ee}. The basemap in Figure~\ref{fig:area} was made with Natural Earth using free raster map data from \url{https://naturalearthdata.com}. We used the \texttt{daft} Python package obtained via \url{https://docs.daft-pgm.org/en/latest/} to render the probabilistic graphical models in Figure~\ref{fig:pgm} and PGF/Ti$k$Z to render the flowchart in Figure~\ref{fig:workflow} in \LaTeX. All experiments carried out in this study were forced by ERA5 reanalysis data \citep{Hersbach2023} obtained from the Copernicus Climate Change Service (C3S) Climate Data Store and topographically downscaled to each study area. These data were generated using modified Copernicus Climate Change Service information [2026]. Neither the European Commission nor ECMWF is responsible for any use that may be made of the Copernicus information or data it contains. All simulations were carried out using basic array programming implemented by the first author in MATLAB \citep{MATLAB}. The code and data needed to reproduce all the results and substantive figures in this study are openly available through \url{https://doi.org/10.5281/zenodo.22967714} \citep{Code} and \url{https://doi.org/10.5281/zenodo.22938686} \citep{Data}, respectively. }

\noappendix       %% use this to mark the end of the appendix section. Otherwise the figures might be numbered incorrectly (e.g. 10 instead of 1).
\appendixfigures  %% needs to be added in front of appendix figures

\appendixtables   %% needs to be added in front of appendix tables

%% Please add \clearpage between each table and/or figure. Further guidelines on figures and tables can be found below.

% KA, EAG, JF, BG, GG, RH, BL, NP, SW, YAY

\authorcontribution{Conceptualization: KA. Data curation: KA, JF, BL, YAY. Formal analysis: KA. Funding acquisition: KA, RH. Investigation: KA. Methodology: KA with contributions from EAG, JF, BG, GG, RH, NP, SW, YAY. Resources: KA, YAY. Software: KA. Validation: KA. Visualization: KA. Writing – original draft: KA. Writing – review and editing: all authors.}%% this section is mandatory

\competinginterests{The contact author has declared that none of the authors has any competing interests.} %% this section is mandatory even if you declare that no competing interests are present

%\disclaimer{TEXT} %% optional section

% Commented for now, fix when submitting (e.g. other projects?)
\financialsupport{KA was financially supported by the European Space Agency Climate Change Initiative (ESA-CCI) Research Fellowship project PATCHES. KA, GG, RH, and YAY were funded by the ERC-2022-ADG project GLACMASS under grant agreement no. 101096057. EAG acknowledges funding from the “Ramon y Cajal” Fellowship RYC2023-044416-I and CSIC Talent Recruitment Program, project AD-For (PIE-20253AT017). JF was funded by the Swiss National Science Foundation (SNSF) grant 179130. BL was funded by the European Union’s Horizon Europe research and innovation programme through the project LIQUIDICE (grant no. 101184962). BG acknowledges funding from the European Union's Horizon Europe program via the project Past2Future under grant agreement no. 101184070. NP was funded by the ERC-2023-StG project ACTIVATE under grant agreement no. 101116083. SW and KA acknowledge funding by the European Space Agency CCI+ Permafrost project (grant no. 4000123681/18/I-NB). This work is a contribution to the strategic research initiative LATICE (UiO/GEO103920), the Center for Biogeochemistry in the Anthropocene (CBA), and the Center for Computational and Data Science (dScience) at the University of Oslo.} 

\begin{acknowledgements}
The authors warmly thank the European Space Agency and the European Research Council for directly supporting this research. We thank all the data providers that supported this study: C3S, ECMWF, ESA, GLAMOS, NASA, NPI, NVE, SLF, SMHI, and WGMS. We gratefully acknowledge the IT services at the Department of Geosciences, University of Oslo, for the provision and maintenance of the \texttt{ivan} and \texttt{mimi} servers that endured many long Markov chains, and the data storage resources (under NS11115K) provided by Sigma2 — the National Infrastructure for High-Performance Computing and Data Storage in Norway. We acknowledge the use of large language models (LLMs; Anthropic Claude and Google Gemini) for proofreading and code review. These LLMs were not used to generate new text, figures, or scientific interpretations. All LLM contributions were independently verified by the corresponding author, who takes full responsibility for this paper.
\end{acknowledgements}

\bibliographystyle{copernicus}
\bibliography{thebib}

\end{document}